%% file: main.tex
\pdfoutput=1
\documentclass[conference,compsoc]{IEEEtran}
\IEEEoverridecommandlockouts

\ifCLASSOPTIONcompsoc
  \usepackage[nocompress]{cite}
\else
  \usepackage{cite}
\fi

\ifCLASSOPTIONcompsoc
    \usepackage[caption=false,font=footnotesize,labelfont=sf,textfont=sf]{subfig}
\else
    \usepackage[caption=false,font=footnotesize]{subfig}
\fi

\usepackage{import}
\import{./lib/}{boilerplate_pdflatex.tex}

\import{./lib/}{colors.tex}

\import{./lib/}{tikzpgfplot.tex}

\import{./lib/}{tikzpgfplot_parulalinestyles.tex}

\import{./lib/}{tikzpgfplot_plotstyles.tex}

\import{./lib/}{tikzpgfplot_blockchainstructures.tex}

\import{./lib/}{tables.tex}

\import{./lib/}{algorithms.tex}

\import{./lib/}{lateplate_pdflatex.tex}

\input{custom_commands}

\title{Optimal Flexible Consensus and its Application to Ethereum}

\newcommand{\gitSourceUrl}[0]{\url{https://github.com/tse-group/flexible-eth}}
\makeatletter
\def\ps@headings{%
\def\@oddhead{\mbox{}\scriptsize\rightmark \hfil \thepage}%
\def\@evenhead{\scriptsize\thepage \hfil \leftmark\mbox{}}}
\makeatother
\begin{document}
\author{%
\IEEEauthorblockN{Joachim Neu}%
\IEEEauthorblockA{Stanford University\\jneu@stanford.edu}%
\and%
\IEEEauthorblockN{Srivatsan Sridhar}%
\IEEEauthorblockA{Stanford University\\svatsan@stanford.edu}%
\and%
\IEEEauthorblockN{Lei Yang}%
\IEEEauthorblockA{Massachusetts Institute of Technology\\leiy@csail.mit.edu}%
\and%
\IEEEauthorblockN{David Tse}%
\IEEEauthorblockA{Stanford University\\dntse@stanford.edu}%
\thanks{JN, SS, LY are listed alphabetically.}%
}

\maketitle
\import{./}{00_abstract.tex}

\import{./}{01_introduction.tex}

\import{./}{02_modelproblem.tex}

\import{./}{03_optflex_general.tex}

\import{./}{04_optflex_streamlet.tex}

\import{./}{05_application_eth.tex}
\import{./}{06_relatedwork.tex}
\import{./}{07_discussion.tex}
\ifCLASSOPTIONcompsoc
  \section*{Acknowledgments}
\else
  \section*{Acknowledgment}
\fi
We thank
Daniel J.\ Aronoff,
Dan Boneh,
Christian Cachin,
Neha Narula,
and 
Ertem Nusret Tas
for fruitful discussions.
JN, SS, LY, and DT are supported by gifts from
the Ethereum Foundation.
JN, SS, and DT are supported by a gift from
Input Output Global.
JN is supported by the Protocol Labs PhD Fellowship.

\bibliographystyle{IEEEtran}
\bibliography{IEEEabrv,references}

\appendices
\crefalias{section}{appendix}
\crefalias{subsection}{subappendix}
\crefalias{subsubsection}{subsubappendix}

\import{./}{A01_streamlet_proofs.tex}
\import{./}{A02_strongflex_converse.tex}

\end{document}

%% file: lib/boilerplate_pdflatex.tex
\usepackage[utf8]{inputenc}

\input{boilerplate_base.tex}

%% file: lib/boilerplate_base.tex
\PassOptionsToPackage{hyphens}{url}%
\usepackage{hyperref}
\usepackage{graphicx}

\usepackage{amsmath}
\usepackage{amsthm}
\usepackage{amssymb}
\usepackage{amsfonts}
\usepackage{mathtools}

\usepackage{shortsym}

\allowdisplaybreaks

\usepackage[shortlabels,inline]{enumitem}
\setlist[itemize]{leftmargin=1.25em}
\setlist[enumerate]{leftmargin=1.75em}

\usepackage{xstring}   %
\usepackage{import}
\subimport{.}{boilerplate_base_hyphenations.tex}

\subimport{.}{boilerplate_base_mathenvironments.tex}

\subimport{.}{boilerplate_base_math.tex}
\subimport{.}{boilerplate_base_symbols.tex}

\subimport{.}{boilerplate_base_textdecorations.tex}
\subimport{.}{boilerplate_base_textabbreviations.tex}

%% file: lib/boilerplate_base_hyphenations.tex
\hyphenation{a-vail-a-bil-i-ty}

\hyphenation{ac-count-a-bil-i-ty}
\hyphenation{ac-count-a-ble}
\hyphenation{ac-count-a-bly}

\hyphenation{led-ger}

\hyphenation{Na-ka-mo-to}

\hyphenation{white-pa-per}

\hyphenation{equi-vo-ca-tion}
\hyphenation{equi-vo-ca-ting}

\hyphenation{con-sen-sus}
\hyphenation{pro-to-col}

\hyphenation{block-chain}
\hyphenation{block-chains}

\hyphenation{dem-on-strate}

%% file: lib/boilerplate_base_mathenvironments.tex
\theoremstyle{plain}
\newtheorem{theorem}{Theorem}
\newtheorem*{theorem*}{Theorem}

\newtheorem{lemma}{Lemma}

\theoremstyle{definition}
\newtheorem{definition}{Definition}
\makeatletter
\renewcommand*\env@matrix[1][*\c@MaxMatrixCols c]{%
    \hskip -\arraycolsep
    \let\@ifnextchar\new@ifnextchar
    \array{#1}}
\makeatother

%% file: lib/boilerplate_base_math.tex
\DeclarePairedDelimiter\abs{\lvert}{\rvert}
\DeclarePairedDelimiter\len{\lvert}{\rvert}
\DeclarePairedDelimiter\norm{\lVert}{\rVert}

\makeatletter
\let\oldabs\abs
\def\abs{\@ifstar{\oldabs}{\oldabs*}}
\let\oldlen\len
\def\len{\@ifstar{\oldlen}{\oldlen*}}
\let\oldnorm\norm
\def\norm{\@ifstar{\oldnorm}{\oldnorm*}}
\makeatother

\newcommand{\scriptspacing}[0]{\medmuskip=0mu\thickmuskip=0mu}

%% file: lib/boilerplate_base_symbols.tex
\usepackage{pifont}

\usepackage{dsfont}

\DeclareRobustCommand{\svdots}{%
    \vcenter{%
        \offinterlineskip%
        \hbox{.}%
        \vskip0.25\normalbaselineskip%
        \hbox{.}%
        \vskip0.25\normalbaselineskip%
        \hbox{.}%
    }%
}

%% file: lib/boilerplate_base_textabbreviations.tex
\usepackage{xspace}

\newcommand{\cf}[0]{cf.\xspace}
\newcommand{\ie}[0]{\emph{i.e.}\xspace}
\newcommand{\eg}[0]{\emph{e.g.}\xspace}

%% file: lib/colors.tex
\usepackage{xcolor}

\definecolor{jnSUCardinalRed}{HTML}{8c1515}
\definecolor{jnSUCardinalRedLight}{HTML}{B83A4B}
\definecolor{jnSUCardinalRedDark}{HTML}{820000}
\definecolor{jnSUWhite}{HTML}{ffffff}
\definecolor{jnSUCoolGrey}{HTML}{53565A}
\definecolor{jnSUBlack}{HTML}{2e2d29}
\definecolor{jnSUBlack100}{HTML}{2e2d29}
\definecolor{jnSUBlack90}{HTML}{43423E}
\definecolor{jnSUBlack80}{HTML}{585754}
\definecolor{jnSUBlack70}{HTML}{6D6C69}
\definecolor{jnSUBlack60}{HTML}{767674}
\definecolor{jnSUBlack50}{HTML}{979694}
\definecolor{jnSUBlack40}{HTML}{ABABA9}
\definecolor{jnSUBlack30}{HTML}{C0C0BF}
\definecolor{jnSUBlack20}{HTML}{D5D5D4}
\definecolor{jnSUBlack10}{HTML}{EAEAEA}

\definecolor{jnSUPaloAlto}{HTML}{175E54}
\definecolor{jnSUPaloAltoLight}{HTML}{2D716F}
\definecolor{jnSUPaloAltoDark}{HTML}{014240}
\definecolor{jnSUPaloVerde}{HTML}{279989}
\definecolor{jnSUPaloVerdeLight}{HTML}{59B3A9}
\definecolor{jnSUPaloVerdeDark}{HTML}{017E7C}
\definecolor{jnSUOlive}{HTML}{8F993E}
\definecolor{jnSUOliveLight}{HTML}{A6B168}
\definecolor{jnSUOliveDark}{HTML}{7A863B}
\definecolor{jnSUBay}{HTML}{6FA287}
\definecolor{jnSUBayLight}{HTML}{8AB8A7}
\definecolor{jnSUBayDark}{HTML}{417865}
\definecolor{jnSUSky}{HTML}{4298B5}
\definecolor{jnSUSkyLight}{HTML}{67AFD2}
\definecolor{jnSUSkyDark}{HTML}{016895}
\definecolor{jnSULagunita}{HTML}{007C92}
\definecolor{jnSULagunitaLight}{HTML}{009AB4}
\definecolor{jnSULagunitaDark}{HTML}{006B81}
\definecolor{jnSUPoppy}{HTML}{E98300}
\definecolor{jnSUPoppyLight}{HTML}{F9A44A}
\definecolor{jnSUPoppyDark}{HTML}{D1660F}
\definecolor{jnSUSpirited}{HTML}{E04F39}
\definecolor{jnSUSpiritedLight}{HTML}{F4795B}
\definecolor{jnSUSpiritedDark}{HTML}{C74632}
\definecolor{jnSUIlluminating}{HTML}{FEDD5C}
\definecolor{jnSUIlluminatingLight}{HTML}{FFE781}
\definecolor{jnSUIlluminatingDark}{HTML}{FEC51D}
\definecolor{jnSUPlum}{HTML}{620059}
\definecolor{jnSUPlumLight}{HTML}{734675}
\definecolor{jnSUPlumDark}{HTML}{350D36}
\definecolor{jnSUBrick}{HTML}{651C32}
\definecolor{jnSUBrickLight}{HTML}{7F2D48}
\definecolor{jnSUBrickDark}{HTML}{42081B}
\definecolor{jnSUArchway}{HTML}{5D4B3C}
\definecolor{jnSUArchwayLight}{HTML}{766253}
\definecolor{jnSUArchwayDark}{HTML}{2F2424}
\definecolor{jnSUStone}{HTML}{7F7776}
\definecolor{jnSUStoneLight}{HTML}{D4D1D1}
\definecolor{jnSUStoneDark}{HTML}{544948}
\definecolor{jnSUFog}{HTML}{DAD7CB}
\definecolor{jnSUFogLight}{HTML}{F4F4F4}
\definecolor{jnSUFogDark}{HTML}{B6B1A9}

\definecolor{jnSUDigitalRed}{HTML}{B1040E}
\definecolor{jnSUDigitalRedLight}{HTML}{E50808}
\definecolor{jnSUDigitalRedDark}{HTML}{820000}
\definecolor{jnSUDigitalBlue}{HTML}{006CB8}
\definecolor{jnSUDigitalBlueLight}{HTML}{6FC3FF}
\definecolor{jnSUDigitalBlueDark}{HTML}{00548f}
\definecolor{jnSUDigitalGreen}{HTML}{008566}
\definecolor{jnSUDigitalGreenLight}{HTML}{1AECBA}
\definecolor{jnSUDigitalGreenDark}{HTML}{006F54}

\definecolor{myParula01Blue}{RGB}{0,114,189}
\definecolor{myParula02Orange}{RGB}{217,83,25}
\definecolor{myParula03Yellow}{RGB}{237,177,32}
\definecolor{myParula04Purple}{RGB}{126,47,142}
\definecolor{myParula05Green}{RGB}{119,172,48}
\definecolor{myParula06LightBlue}{RGB}{77,190,238}
\definecolor{myParula07Red}{RGB}{162,20,47}

%% file: lib/tikzpgfplot.tex
\usepackage{tikz}
\usetikzlibrary{calc}
\usetikzlibrary{arrows}
\usetikzlibrary{arrows.meta}
\usetikzlibrary{patterns}
\usetikzlibrary{positioning}
\usetikzlibrary{decorations.pathreplacing}
\usetikzlibrary{calligraphy}
\usetikzlibrary{shapes.misc}
\usetikzlibrary{shapes.symbols}
\usetikzlibrary{shapes.arrows}
\usetikzlibrary{shapes.geometric}
\usetikzlibrary{shapes.callouts}
\usetikzlibrary{spy}

\usepackage{pgfplots}
\pgfplotsset{compat=1.17}
\usepgfplotslibrary{fillbetween}

\pgfplotsset{
    discard if not/.style 2 args={
            x filter/.code={
                    \edef\tempa{\thisrow{#1}}
                    \edef\tempb{#2}
                    \ifx\tempa\tempb
                    \else
                        \def\pgfmathresult{inf}
                    \fi
                }
        },
}

%% file: lib/tikzpgfplot_parulalinestyles.tex
\tikzset{myparula11/.style={color=myParula01Blue,solid,mark=+,mark options={solid}}}
\tikzset{myparula12/.style={color=myParula01Blue,densely dashed,mark=x,mark options={solid}}}
\tikzset{myparula13/.style={color=myParula01Blue,densely dotted,mark=o,mark options={solid}}}
\tikzset{myparula14/.style={color=myParula01Blue,dashdotted,mark=triangle,mark options={solid}}}
\tikzset{myparula15/.style={color=myParula01Blue,dashdotdotted,mark=square,mark options={solid}}}

\tikzset{myparula21/.style={color=myParula02Orange,solid,mark=+,mark options={solid}}}
\tikzset{myparula22/.style={color=myParula02Orange,densely dashed,mark=x,mark options={solid}}}
\tikzset{myparula23/.style={color=myParula02Orange,densely dotted,mark=o,mark options={solid}}}
\tikzset{myparula24/.style={color=myParula02Orange,dashdotted,mark=triangle,mark options={solid}}}
\tikzset{myparula25/.style={color=myParula02Orange,dashdotdotted,mark=square,mark options={solid}}}

\tikzset{myparula31/.style={color=myParula03Yellow,solid,mark=+,mark options={solid}}}
\tikzset{myparula32/.style={color=myParula03Yellow,densely dashed,mark=x,mark options={solid}}}
\tikzset{myparula33/.style={color=myParula03Yellow,densely dotted,mark=o,mark options={solid}}}
\tikzset{myparula34/.style={color=myParula03Yellow,dashdotted,mark=triangle,mark options={solid}}}
\tikzset{myparula35/.style={color=myParula03Yellow,dashdotdotted,mark=square,mark options={solid}}}

\tikzset{myparula41/.style={color=myParula04Purple,solid,mark=+,mark options={solid}}}
\tikzset{myparula42/.style={color=myParula04Purple,densely dashed,mark=x,mark options={solid}}}
\tikzset{myparula43/.style={color=myParula04Purple,densely dotted,mark=o,mark options={solid}}}
\tikzset{myparula44/.style={color=myParula04Purple,dashdotted,mark=triangle,mark options={solid}}}
\tikzset{myparula45/.style={color=myParula04Purple,dashdotdotted,mark=square,mark options={solid}}}

\tikzset{myparula51/.style={color=myParula05Green,solid,mark=+,mark options={solid}}}
\tikzset{myparula52/.style={color=myParula05Green,densely dashed,mark=x,mark options={solid}}}
\tikzset{myparula53/.style={color=myParula05Green,densely dotted,mark=o,mark options={solid}}}
\tikzset{myparula54/.style={color=myParula05Green,dashdotted,mark=triangle,mark options={solid}}}
\tikzset{myparula55/.style={color=myParula05Green,dashdotdotted,mark=square,mark options={solid}}}

\tikzset{myparula61/.style={color=myParula06LightBlue,solid,mark=+,mark options={solid}}}
\tikzset{myparula62/.style={color=myParula06LightBlue,densely dashed,mark=x,mark options={solid}}}
\tikzset{myparula63/.style={color=myParula06LightBlue,densely dotted,mark=o,mark options={solid}}}
\tikzset{myparula64/.style={color=myParula06LightBlue,dashdotted,mark=triangle,mark options={solid}}}
\tikzset{myparula65/.style={color=myParula06LightBlue,dashdotdotted,mark=square,mark options={solid}}}

\tikzset{myparula71/.style={color=myParula07Red,solid,mark=+,mark options={solid}}}
\tikzset{myparula72/.style={color=myParula07Red,densely dashed,mark=x,mark options={solid}}}
\tikzset{myparula73/.style={color=myParula07Red,densely dotted,mark=o,mark options={solid}}}
\tikzset{myparula74/.style={color=myParula07Red,dashdotted,mark=triangle,mark options={solid}}}
\tikzset{myparula75/.style={color=myParula07Red,dashdotdotted,mark=square,mark options={solid}}}

%% file: lib/tikzpgfplot_plotstyles.tex
\pgfplotsset{
    mysimpleplot/.style = {
            width=1.05\linewidth,
            height=0.75\linewidth,
            title style={font=\footnotesize,align=center},
            legend cell align=left,
            legend style={font=\footnotesize},
            legend columns=3,
            legend style={
                    at={(0.5,1)},
                    yshift=0.3em,
                    anchor=south,
                    draw=none,
                    /tikz/every even column/.append style={
                            column sep=0.3em
                        },
                    cells={
                            align=left
                        }
                },
            grid=both,
            minor tick num=4,
            major grid style={solid,very thin,draw=gray!50},
            minor grid style={solid,ultra thin,draw=gray!20},
            label style={font=\footnotesize,align=center},
            tick label style={font=\footnotesize,align=center},
        },
}

\pgfplotsset{
    mysimpleresilienceplot01/.style = {
            mysimpleplot,
            ylabel={Adversary resilience $\beta$},
            height=0.45\linewidth,
            width=\linewidth,
            ymin=0,ymax=0.5,
            ytick={0,0.1,0.2,0.3,0.4,0.5},
            xmin=1e-5,xmax=1e2,
            grid=major,
        },
}

%% file: lib/tikzpgfplot_blockchainstructures.tex
\tikzset{blockchainold/.style={
            x=1.5cm,
            y=0.6cm,
            node distance=0.5cm,
            block/.style = {
                    minimum width=0.25cm,
                    minimum height=0.25cm,
                    draw,
                    shade,
                    top color=white,
                    bottom color=black!10,
                },
            block-adv1/.style = {
                    block,
                    bottom color=myParula01Blue!50,
                    draw=myParula01Blue!50!black
                },
            block-adv2/.style = {
                    block,
                    bottom color=myParula07Red!50,
                    draw=myParula07Red!50!black,
                },
            block-adv3/.style = {
                    block,
                    bottom color=myParula05Green!50,
                    draw=myParula05Green!50!black,
                },
            block-green/.style = {
                    block,
                    bottom color=myParula05Green!50,
                    draw=myParula05Green!50!black,
                },
            block-red/.style = {
                    block,
                    bottom color=myParula07Red!50,
                    draw=myParula07Red!70!black,
                },
            block-gray/.style = {
                    block,
                    bottom color=black!30,
                },
            block-big/.style = {
                    minimum width=0.7cm,
                    minimum height=0.7cm,
                    draw,
                    shade,
                    top color=white,
                    bottom color=black!10,
                },
            branch/.style = {
                    minimum width=0.1cm,
                    minimum height=0.1cm,
                    draw,
                    circle,
                    inner sep=0,
                    fill=black,
                },
            link/.style = {
                    -latex,
                },
            hiddenlink/.style = {
                    dashed,
                },
            hiddenlink-adv1/.style = {
                    hiddenlink,
                    draw=myParula01Blue!50!black,
                },
            hiddenlink-adv2/.style = {
                    hiddenlink,
                    draw=myParula07Red!50!black,
                },
            hiddenlink-adv3/.style = {
                    hiddenlink,
                    draw=myParula05Green!50!black,
                },
            label/.style = {
                },
            label-adv1/.style = {
                    label,
                    text=myParula01Blue!50!black,
                },
            label-adv2/.style = {
                    label,
                    text=myParula07Red!50!black,
                },
            label-adv3/.style = {
                    label,
                    text=myParula05Green!50!black,
                },
        }
}

\tikzset{blockchain/.style={
            x=0.5cm,
            y=0.55cm,
            node distance=0.5cm,
            block/.style = {
                    minimum width=0.5cm,
                    minimum height=0.5cm,
                    draw,
                    shade,
                    top color=white,
                    bottom color=black!10,
                    inner sep=0,
                },
            block-adv/.style = {
                    block,
                    bottom color=myParula07Red!50,
                    draw=myParula07Red!50!black,
                },
            block-hon/.style = {
                    block,
                    bottom color=myParula05Green!50,
                    draw=myParula05Green!50!black,
                },
            block-blank/.style = {
                    minimum width=0.3cm,
                    minimum height=0.3cm,
                    rounded corners,
                    inner sep=0,
                },
            vote/.style = {
                    minimum width=0.5cm,
                    minimum height=0.5cm,
                    draw,
                    shade,
                    top color=white,
                    bottom color=black!10,
                    inner sep=0,
                },
            link/.style = {
                    -latex,
                },
            link-adv/.style = {
                    link,
                },
            link-hon/.style = {
                    link,
                },
        }
}

%% file: lib/tables.tex
\usepackage{multirow,booktabs}
\usepackage{makecell}

%% file: lib/algorithms.tex
\usepackage{algorithm}
\usepackage{algorithmicx}
\usepackage[noend]{algpseudocode}

\algnewcommand{\algorithmicswitch}{\textbf{switch}}
\algdef{SE}[SWITCH]{Switch}{EndSwitch}[1]{\algorithmicswitch\ #1\ \algorithmicdo}{\algorithmicend\ \algorithmicswitch}%
\algtext*{EndSwitch}%

\algnewcommand{\algorithmiccase}{\textbf{case}}
\algdef{SE}[CASE]{Case}{EndCase}[1]{\algorithmiccase\ #1}{\algorithmicend\ \algorithmiccase}%
\algtext*{EndCase}%

\algnewcommand{\algorithmicon}{\textbf{on}}
\algdef{SE}[ON]{On}{EndOn}[1]{\algorithmicon\ #1\ \algorithmicdo}{\algorithmicend\ \algorithmicon}%
\algtext*{EndOn}%

\algnewcommand{\algorithmicat}{\textbf{at}}
\algdef{SE}[AT]{At}{EndAt}[1]{\algorithmicat\ #1\ \algorithmicdo}{\algorithmicend\ \algorithmicat}%
\algtext*{EndAt}%

\algnewcommand{\algorithmicrealfunction}{\textbf{function}}
\algdef{SE}[REALFUNCTION]{RealFunction}{EndRealFunction}[1]{\algorithmicrealfunction\ #1\ \algorithmicdo}{\algorithmicend\ \algorithmicrealfunction}%
\algtext*{EndRealFunction}%

\algnewcommand{\algorithmicthroughout}{\textbf{do throughout}}
\algdef{SE}[Throughout]{Throughout}{EndThroughout}[1]{\algorithmicthroughout\ #1\ \algorithmicdo}{\algorithmicend\ \algorithmicthroughout}%
\algtext*{EndThroughout}%

\algnewcommand{\algorithmicforever}{\textbf{forever}}
\algdef{SE}[Forever]{Forever}{EndForever}[0]{\algorithmicforever\ \algorithmicdo}{\algorithmicend\ \algorithmicforever}%
\algtext*{EndForever}%

\algrenewcommand{\algorithmicdo}{}
\algrenewcommand{\algorithmicthen}{}

\algnewcommand{\algorithmicgoto}{\textbf{goto}}%
\algnewcommand{\Goto}[1]{\algorithmicgoto~\ref{#1}}%

\algnewcommand{\algorithmicassert}{\textbf{assert}}%
\algnewcommand{\Assert}[1]{\algorithmicassert~{#1}}%

\algnewcommand{\algorithmicbreak}{\textbf{break}}%
\algnewcommand{\Break}[0]{\algorithmicbreak}%

\algnewcommand{\algorithmicwaiton}{\textbf{wait on}}%
\algnewcommand{\WaitOn}[1]{\algorithmicwaiton~{#1}}%

\algnewcommand{\LineComment}[1]{\State \(\triangleright\) \textit{#1}}
\algnewcommand{\InlineRequire}[1]{\textbf{require} {#1}}

%% file: lib/lateplate_pdflatex.tex
\input{lateplate_base.tex}

%% file: lib/lateplate_base.tex
\usepackage[sort&compress,capitalize,nameinlink]{cleveref}

\crefname{figure}{Fig.}{Figs.}
\Crefname{figure}{Fig.}{Figs.}

\crefname{table}{Tab.}{Tabs.}
\Crefname{table}{Tab.}{Tabs.}

\crefname{section}{Sec.}{Secs.}
\Crefname{section}{Sec.}{Secs.}
\crefname{appendix}{App.}{Apps.}
\Crefname{appendix}{App.}{Apps.}

\crefname{algorithm}{Alg.}{Algs.}
\Crefname{algorithm}{Alg.}{Algs.}
\crefname{line}{l.}{ll.}
\Crefname{line}{l.}{ll.}

\crefname{proposition}{Prop.}{Props.}
\Crefname{proposition}{Prop.}{Props.}
\crefname{lemma}{Lem.}{Lems.}
\Crefname{lemma}{Lem.}{Lems.}
\crefname{theorem}{Thm.}{Thms.}
\Crefname{theorem}{Thm.}{Thms.}
\crefname{corollary}{Cor.}{Cors.}
\Crefname{corollary}{Cor.}{Cors.}
\crefname{definition}{Def.}{Defs.}
\Crefname{definition}{Def.}{Defs.}

%% file: custom_commands.tex
\usepackage[normalem]{ulem}
\usepackage{noto-emoji-easy}
\usepackage{tango-icons}

\usetikzlibrary{spy}

\newcommand*\DiagramStep[1]{\tikz[baseline=(char.south)]{
                \node[shape=circle,draw=none,fill=black,text=white,inner sep=0.75pt] (char) {\tiny\textbf{#1}}; }}

\newcommand{\tL}[0]{\ensuremath{t^{\mathrm{L}}}}
\newcommand{\tS}[0]{\ensuremath{t^{\mathrm{S}}}}

\newcommand{\tLrel}[0]{\ensuremath{\mathfrak{t}^{\mathrm{L}}}}
\newcommand{\tSrel}[0]{\ensuremath{\mathfrak{t}^{\mathrm{S}}}}
\newcommand{\frel}[0]{\ensuremath{\mathfrak{f}}}
\newcommand{\qrel}[0]{\ensuremath{\mathfrak{q}}}

\newcommand{\FBFT}[0]{FBFT\xspace}
\newcommand{\FBFTtwo}[0]{SFT-DiemBFT\xspace}

\newcommand{\Oflex}[0]{OFlex\xspace}

\newcommand{\olock}[0]{perma-lock\xspace}
\newcommand{\olocks}[0]{perma-locks\xspace}
\newcommand{\olocked}[0]{perma-locked\xspace}
\newcommand{\olocking}[0]{perma-locking\xspace}
\newcommand{\Olock}[0]{Perma-lock\xspace}

\newcommand{\ovote}[0]{post-vote\xspace}
\newcommand{\ovotes}[0]{post-votes\xspace}
\newcommand{\ovoted}[0]{post-voted\xspace}
\newcommand{\ovoting}[0]{post-voting\xspace}
\newcommand{\Ovote}[0]{Post-vote\xspace}

\newcommand{\pipeliningchaining}[0]{streamlining\xspace}
\newcommand{\pipelinechain}[0]{streamline\xspace}
\newcommand{\pipelinedchained}[0]{streamlined\xspace}
\newcommand{\Pipelinedchained}[0]{Streamlined\xspace}

\newcommand{\LOG}[2]{\ensuremath{\LOGempty_{#1}^{#2}}}
\newcommand{\LOGempty}[0]{\ensuremath{\mathsf{LOG}}}

\newcommand{\PI}[0]{\ensuremath{\BPi}}
\newcommand{\PiR}[0]{\ensuremath{\Pi}}
\newcommand{\PiC}[0]{\ensuremath{\CC}}
\newcommand{\PIOpart}[0]{\ensuremath{\BPi^{\textrm{O}}}}
\newcommand{\PiROpart}[0]{\ensuremath{\Pi^{\textrm{O}}}}
\newcommand{\PiCOpart}[0]{\ensuremath{\CC^{\textrm{O}}}}
\newcommand{\PIO}[0]{\ensuremath{\BPi^*}}

\newcommand{\PImod}[0]{\ensuremath{\BPi'}}
\newcommand{\PiRmod}[0]{\ensuremath{\Pi'}}
\newcommand{\PiCmod}[0]{\ensuremath{\CC'}}

\newcommand{\tx}[0]{\ensuremath{\mathsf{tx}}}
\newcommand{\txs}[0]{\ensuremath{\mathsf{txs}}}

\newcommand{\GST}[0]{\ensuremath{\mathsf{GST}}}

\newcommand{\blockLabel}[1]{#1}

\newcommand{\algolock}[0]{\ensuremath{\mathsf{permalock}}}
\newcommand{\algovote}[0]{\ensuremath{\mathsf{postvote}}}
\newcommand{\algovotes}[0]{\ensuremath{\mathsf{postvotes}}}

%% file: 00_abstract.tex
\begin{abstract}
    Classic 
    BFT 
    consensus protocols
    guarantee safety and liveness for \emph{all} clients
    if fewer than one-third of replicas are faulty.
    However,
    in
    applications such as
    high-value payments,
    some clients may 
    want to prioritize safety over liveness.
    \emph{Flexible} consensus allows
    \emph{each} client to
    opt for
    higher safety resilience,
    albeit at the expense of reduced liveness resilience.
    We present the first construction that allows \emph{optimal safety--liveness tradeoff for every client simultaneously}.
    This construction is modular and is realized as an add-on applied on top of an existing consensus protocol.
    The add-on consists of 
    an \emph{additional} round of voting and \emph{permanent} locking
    done by the replicas,
    to sidestep
    a sub-optimal quorum-intersection-based constraint
    present in previous solutions.
    We
    adapt our
    construction to 
    the existing Ethereum protocol
    to
    derive
    optimal
    flexible confirmation rules
    that
    clients can adopt
    unilaterally
    \emph{without} requiring 
    system-wide changes.
    This is possible because existing Ethereum protocol features
    can double as
    the extra voting and locking.
    We 
    show
    an implementation
    using Ethereum's consensus API.
\end{abstract}

%% file: 01_introduction.tex
\section{Introduction}
\label{sec:intro}

\subsection{Flexible Consensus}
\label{sec:intro-flexible}

A state-machine replication (SMR) \emph{consensus protocol}
has two groups of participants: \emph{clients} and \emph{replicas}.
Continuously,
clients input \emph{transactions} to the replicas.
The replicas partake in a distributed algorithm to develop a common
ordering among the transactions,
and report back to the clients.
At any time, each client outputs a \emph{log} 
which is the ordered sequence of transactions
it deems \emph{confirmed} so far.
Two desiderata for consensus protocols are:
\emph{safety}, meaning that logs are consistent across clients and across time;
and
\emph{liveness}, meaning that transactions submitted to all replicas
eventually appear in all clients' logs.
Byzantine-fault tolerant (BFT) consensus protocols guarantee these properties
if the fraction $\frel$ of \emph{adversary} replicas
out of all $n$ replicas
is not too large,
where adversary replicas
may deviate from the protocol in any arbitrary and coordinated fashion,
and non-adversary \emph{honest} replicas follow the protocol as specified.%
\footnote{Since no guarantee can be given for logs produced by adversary clients following arbitrary logic,
the focus is on clients that follow the protocol.}
Based on terminology of~\cite{mtbft,mtrbc,mtrbc2,aa,basilic},
a protocol's
\emph{safety resilience} $\tSrel$
is defined as the maximum fraction of adversary replicas the protocol can tolerate
while still guaranteeing safety.
Similarly, the \emph{liveness resilience} $\tLrel$ is the maximum fraction of adversary replicas the protocol can tolerate while guaranteeing liveness.
Classical \emph{PBFT-style}
protocols
\cite{pbft,sbft,tendermint,casper,hotstuff,streamlet}
provide
balanced safety and liveness resiliences
$\tLrel = \tSrel = \frac{1}{3}$~\cite{pbft},
which is the highest that both the resiliences can simultaneously be in the partially-synchronous setting~\cite{dls84} of interest here.

\import{./figures/}{safety-liveness-tradeoffs.tex}

However,
users (\ie, clients)
of a system may not
desire
equal resilience to safety and liveness faults~\cite{aa,mtbft,mtrbc,mtrbc2,basilic}.
In fact,
different clients 
may prefer
different
(liveness-/safety-)\emph{resilience pairs} altogether~\cite{fbft,fbft2}!\footnote{Note that 
like in~\cite{fbft},
`clients' are 
a logical abstraction for whenever physical users require different resilience pairs.
Thus, one physical user
can achieve different resilience pairs
at different points in time or to confirm different transactions,
and would then manifest as multiple clients.}
For instance,
for a
distributed ledger such as a cryptocurrency,
clients
who perform high-value transfers
may prefer a higher safety resilience,%
\footnote{For example, in Ethereum
as of 06/2023,
a single organization (Lido)
controls more than one-third
of replicas,
causing concerns about 
a possible coordinated failure,
\eg, from technical malfunction or social engineering,
exceeding Ethereum's one-third resilience~\cite{social_danny,social_danny2,social_reddit,social_superphiz,danny_github}.}
even at the expense of lower liveness resilience,
because business lost during a system downtime
may cause less harm than a 
double spend.
In the same vein,
by definition,
liveness only breaks if transaction inclusion
is
denied \emph{forever}~\cite{safetyliveness}.
Thus, liveness can be `restored' relatively easily,
\eg,
through exogenous reconfiguration and removal of adversary replicas.
In contrast,
safety already breaks if inconsistency occurs \emph{ever}.
These
considerations
suggest that
guaranteeing consensus in the
\emph{high-safety regime}, where $\tSrel \geq \tLrel$,
is
of particular interest
to 
some 
\emph{safety-favoring} clients.
At the same time,
other \emph{liveness-preserving}
clients may want to retain $\tLrel=\frac{1}{3}$,
even at the expense of not increasing $\tSrel$ beyond $\frac{1}{3}$.%
\footnote{In contrast, the regime $\tLrel > \tSrel$ is
subsequently disregarded because its `live but inconsistent' logs are of questionable utility~\cite{mtbft,fbft2}.}

This motivates \emph{flexible consensus}~\cite{fbft}
(also called
\emph{strengthened fault tolerance}~\cite{fbft2}),
where each client $k$ chooses a
resilience pair
$(\tLrel_k, \tSrel_k)$ with $\tSrel_k \geq \tLrel_k$,
and
the consensus properties are guaranteed
for all clients who have chosen adequate resiliences.
That is,
for all clients $k, k'$ 
if the adversary fraction 
$\frel \leq \tSrel_{k}$ and  $\frel \leq \tSrel_{k'}$,
their logs are guaranteed to be consistent across time (safety);
and transactions submitted to all replicas eventually appear in
the log of every client $k$ with $\frel \leq \tLrel_k$ (liveness).

\subsection{Quest for Optimal Flexible Consensus}
\label{sec:intro-flexiblerelated}

Given the flexible consensus problem,
it is natural to ask:
\emph{What is the `maximum' flexibility a protocol can provide?}
The classic impossibility result $2 \tLrel_k + \tSrel_k \leq 1$~\cite{dls84,aav1} shows that
we cannot hope to do better than
for
each client to achieve a
resilience pair $(\tLrel_{k},\tSrel_{k})$ on the straight line between $(0,1)$ and $(\frac{1}{3},\frac{1}{3})$, shown in 
\cref{fig:safety-liveness-tradeoffs} 
(\tikz[baseline=0.15em,x=0.9em,y=0.9em]{ \draw [pattern=north east lines,pattern color=red!30,draw=none] (0,0) rectangle (1,1); }).
\emph{But is
there a protocol which allows clients to achieve all such pairs simultaneously, or are there some further limits to flexibility?}
In typical PBFT-style
protocols,
all clients have the same pair of resiliences $\tLrel_{k} = \tSrel_{k} = \frac{1}{3}$.
In that sense, these protocols are a degenerate case of flexible consensus where the set of resilience pairs $(\tLrel_{k},\tSrel_{k})$ that clients can simultaneously `choose' from is the singleton $\{ (\frac{1}{3}, \frac{1}{3}) \}$.
In these protocols, one could vary the quorum size $\qrel$ used in the protocol such that for any given quorum $\qrel$, all clients achieve a single resilience pair $(\tLrel,\tSrel)$ that is different from $(\frac{1}{3},\frac{1}{3})$.
\Cref{fig:safety-liveness-tradeoffs-classic} shows one such achievable pair.
However,
these protocols still have \emph{no flexibility} to support
liveness-preserving
and safety-favoring clients simultaneously.

To add some flexibility,
for any fixed system-wide \emph{replica quorum} $\qrel$,
\FBFT~\cite{fbft} and \FBFTtwo~\cite{fbft2}
(\cref{fig:safety-liveness-tradeoffs-fbft})
allow every client $k$ to opt for higher
$\tSrel_k$
by using a higher client-specific \emph{confirmation quorum} $\qrel_k \in [\qrel,1]$
when confirming its output log.
This construction results in
$\tSrel_k = \qrel_k + \qrel - 1$ and
$\tLrel_k = 1-\qrel_k$,
allowing the trade-off exemplified in \cref{fig:safety-liveness-tradeoffs-fbft},
between
$(0,\qrel)$ for the most safety-favoring client with $\qrel_k=1$,
and
$(1-\qrel,2\qrel-1)$ for the 
liveness-preserving
client with $\qrel_k=\qrel$.
By tuning the replica quorum $\qrel$,
it is
possible to run \emph{different instances}
of the protocol which support \emph{different regions of resilience pairs},
but no single instance can simultaneously support a
maximally safety-favoring client at $(0,1)$
and a 
liveness-preserving
client at $(\frac{1}{3},\frac{1}{3})$.

\subsection{Contribution: \Oflex Family of Resilience-Optimal Flexible Consensus Protocols}
\label{sec:intro-contribution}

\import{./figures/}{tab-oflex-constructions.tex}

We present the \Oflex family of flexible consensus protocols, 
the first
\emph{resilience-optimal}
flexible consensus protocols. Each protocol in this family can 
support all clients on the straight line
between $(0,1)$ and $(\frac{1}{3},\frac{1}{3})$
simultaneously
(\cref{fig:safety-liveness-tradeoffs-optflex}). 
Each protocol in this family is obtained by applying a modification to an existing consensus protocol, and our constructions are applicable to a broad class of existing protocols.

Our constructions extend the approach from \FBFT and \FBFTtwo \cite{fbft,fbft2} to decouple the consensus protocol $\PI$ into two phases: an interactive replica logic $\PiR$ that depends only on a system-wide quorum $\frac{2}{3}$ (unaware of clients' choices of resilience pairs), and a confirmation rule $\PiC$ run locally by clients that additionally depends on client-specific quorums $\qrel_k \geq \frac{2}{3}$.
The key new idea in \Oflex is to augment the replica logic with an additional round of voting (called \emph{\ovoting}) and permanent locking (\emph{\olocking}).
Replicas hold on to their \olock \emph{permanently} and never \olock or \ovote anything inconsistent with their \olock, regardless of how powerful the adversary is. 
In contrast to FBFT, where safety is limited by a quorum-intersection-based constraint between a large client quorum and a small replica quorum ($\tSrel_k \leq \qrel_k + \qrel - 1$), \Oflex due to the \olock and \ovote only needs a constraint between two large client quorums ($\tSrel_k \leq 2\qrel_k - 1$). This enhances the safety resilience of \Oflex.
Even though the \olock appears to restrict a replica's \ovotes, \Oflex in fact preserves \FBFT's liveness resilience of $\tLrel_k = 1 - \qrel_k$.
This is because for all resilience pairs of interest (\cref{fig:safety-liveness-tradeoffs-optflex}), the liveness resilience $\tLrel_k \leq \frac{1}{3}$,
and when the adversary fraction is smaller than this,
all replicas' \olocks are consistent with each other, so honest replicas continue voting 
in a manner that ensures liveness.
We present
three \Oflex protocol constructions,
that differ in 
how they implement 
\olocking and \ovoting
(see \cref{tab:oflex-constructions} for a summary):
\begin{itemize}
    \item 
        \import{./figures/}{blockdiagrams-oflex-generic.tex}
        The \emph{generic \Oflex construction (\cref{sec:optflex}, \cref{fig:blockdiagrams-oflex-generic})}
        adds both \olocking and \ovoting
        as separate replica-side logic
        that acts on the output log of any
        $(\frac{1}{3},\frac{1}{3})$-resilient consensus protocol for partial synchrony
        to make it optimally flexible in a closed-box manner.
        \emph{This is a modular addition to the replica logic.}

    \item %
        To \emph{modify \emph{textbook} PBFT-style protocols for \Oflex
        (\cref{sec:optflex-streamlet})},
        \ovoting requires no replica-side changes,
        because existing votes can be
        `reused' as \ovotes (\cf~\pipeliningchaining~\cite{hotstuff,casper,streamlet}).
        Only \olocking needs to be added by means of an additional constraint for replicas while voting.
        We demonstrate the application of \Oflex to Streamlet~\cite{streamlet}
to obtain \Oflex-Streamlet in \cref{sec:optflex-streamlet}, and the technique readily carries over to other PBFT-style protocols.
    \emph{This non-modular modification of the replica logic preserves the amount of communication of the original PBFT-style protocol.} 

    \item 
        \import{./figures/}{eth-experiment-intro-latency.tex}
        Many \emph{implementations} of PBFT-style protocols
        employ a
        performance optimization
        that de-facto already implements the \olocking mechanism.
        For \Oflex variants of such protocols,
        \emph{no replica-side changes are needed}.
        An example is Ethereum
        (which is based on Casper~\cite{casper}, a PBFT-style protocol),
        for which we derive
    optimally
    flexible confirmation rules
    (\cref{sec:eth})
    that
    clients can adopt 
    without requiring any system-wide changes.
    We demonstrate an implementation
    using Ethereum's consensus API,
    and study the practical safety vs.\ confirmation latency trade-offs, as summarized in \cref{fig:eth-experiment-intro-latency}.
\end{itemize}

Our \Oflex protocols also have other practically relevant qualities such as 
stronger consistency guarantees between clients with different safety resiliences, accountable safety, and the possibility of preserving safety even after external repair (`social consensus'). We discuss these in \cref{sec:discussion}.%

\subsection{Methods}
\label{sec:intro-methods}

\subsubsection{Generic \Oflex Construction (\cref{sec:optflex})}
\label{sec:intro-methods-oflex-generic}

Our generic \Oflex construction best highlights the key ideas
that are used in all three of our constructions.
The generic construction has two stages (\cref{fig:blockdiagrams-oflex-generic}).
First,
any off-the-shelf $(\frac{1}{3},\frac{1}{3})$-resilient consensus
protocol
is invoked in a closed-box manner to sequence incoming transactions
into a log.
In the second stage,
whenever the log output by the first stage
changes,
lock \emph{permanently}
(\emph{\olock})
on 
the new log
(%
only if the
new log
extends the
earlier \olock).
They then cast a \ovote vote 
for
the latest \olocked log.
Finally, clients use local quorums $\qrel_k\geq \frac{2}{3}$
among the second stage's \ovote votes
to decide which log entries can be confirmed.
Unlike in \FBFT,
the
client-local confirmation rules
are not applied to votes 
preexistent in PBFT-style protocols,
but rather to votes from an \emph{additional}
round of
\ovotes
which respect the \olock
(\cref{fig:blockdiagrams-oflex-generic}).
This is crucial to achieve the optimal quorum-interesection constraint $\tSrel_k = 2\qrel_k - 1$ because an honest replica will never \ovote two inconsistent logs, \emph{no matter what the first stage outputs}.
This generic construction applies to any $(\frac{1}{3},\frac{1}{3})$-resilient consensus protocol such as~\cite{tendermint,dumbo,dumbong,dispersed,honeybadger,pace}, including DAG-based protocols~\cite{narwhaltusk,bullshark,dashingstar,dagrider}.
While it also applies to \pipelinedchained PBFT-style protocols, we also provide a specific construction for them.

\subsubsection{\Oflex PBFT-Style Protocols (\cref{sec:optflex-streamlet})}
\label{sec:intro-methods-oflex-specific}
With the insight that optimal flexibility
can be achieved with an extra round of \olock and \ovote,
we
revisit
PBFT-style protocols 
and modify them
for optimal flexibility.
To this end, recall that PBFT proceeds in views,
where in each view a batch of new transactions is proposed,
and after meeting quorum in multiple voting phases, the proposal is confirmed.
Some recent PBFT-style protocols~\cite{casper,hotstuff,streamlet}
\pipelinechain these phases into a \pipelinedchained protocol,
where in each round a proposal and voting takes place.
Then,
votes simultaneously serve for different voting phases with respect
to different proposals~\cite[Sec.~5]{hotstuff}.
It is easy to see that
the additional round of voting (\ovotes)
required for \Oflex is easily integrated into 
such \pipelinedchained
PBFT-style protocols
using subsequent votes.
Besides the system-wide replica quorum $\qrel=2/3$, clients impose client-local
confirmation quorums $\qrel_k$ for the additional voting round.

On the other hand, a mechanism like \olock
is not readily found in the \emph{textbook version}
of PBFT-style protocols.
But the generic \Oflex construction suggests the following modification
which turns out to be sufficient:
in the place where the original PBFT-style protocol would have confirmed a block (using a system-wide confirmation rule),
replicas \olock the respective block (and never vote for anything inconsistent
with their \olock).

\subsubsection{\Oflex Confirmation Rules for Ethereum (\cref{sec:eth})}
\label{sec:intro-methods-oflex-eth}

While a mechanism like \olock is not commonly found
in the pseudo-code of popular PBFT-style protocols,
it
is, however, already present
in many \emph{implementations} of PBFT-style protocols
in the form of an unrelated performance optimization.
For instance, Ethereum replicas (called `validators') will forever lock on a block they view as `finalized' according to the rules of Casper~\cite{casper,gasper}.
They will then
refuse to consider any conflicting block going forward,
even if the conflicting block has also received enough
votes to become finalized
(such conflicting finalizations can only occur
if more than $1/3$ of replicas are adversary).
This `pruning' of the block tree
improves validators' computational efficiency.

Since Ethereum has already implemented an equivalent of
\olocking,
and \ovoting is easily accommodated in a PBFT-style protocol like Casper,
as discussed above,
applying the \Oflex construction to Ethereum
requires
no system-wide changes to the replica logic.
Client-side changes to the confirmation logic suffice.

\subsection{Outline}
\label{sec:intro-outline}

After recapitulating the
model and problem formulation
in \cref{sec:modprob},
we describe and prove secure the generic 
\Oflex construction
in \cref{sec:optflex}.
We show in \cref{sec:optflex-streamlet}
how to use
\Oflex's key insight
to modify PBFT-style protocols to provide
optimal flexibility.
In \cref{sec:eth}, we apply \Oflex to obtain
optimal-resilience
flexible-consensus confirmation-rules
for Ethereum, and report on an implementation thereof.
We conclude with a comparison to related work in \cref{sec:relatedwork},
and a discussion of extensions of \Oflex in \cref{sec:discussion}.

%% file: figures/safety-liveness-tradeoffs.tex
\tikzset{fig_safety_liveness_tradeoff/.style={
            x=4cm,
            y=4cm,
            bboxfix/.style={
                    fill=none,
                    draw=none,
                    southwest/.style={
                        },
                    northeast/.style={
                        },
                },
            converselabel/.style={
                    red!40,
                    anchor=south east,
                    inner sep=0,
                    yshift=2pt,
                },
            conversearea/.style={
                    pattern=north east lines,
                    pattern color=red!30,
                },
        }}
\begin{figure}[tb]%
    \centering%
    \subfloat[PBFT-style prot.\label{fig:safety-liveness-tradeoffs-classic}]{%
        \begin{tikzpicture}[fig_safety_liveness_tradeoff]%
            \footnotesize%
            \scriptspacing%

            \pgfdeclarelayer{background}
            \pgfdeclarelayer{foreground}
            \pgfsetlayers{background,main,foreground}

            \draw [bboxfix] ([southwest]0,0.333) rectangle ([northeast]0.4,1.0);

            \fill [conversearea] (0,1.05) -- (0,1) -- (0.333,0.333) -- (0.4,0.333) -- (0.4,1.05) -- cycle;
            \node [converselabel] at (0.4,1.05) {\footnotesize\textsc{Impossible}};

            \draw [-latex] (0,0.333) -- (0.45,0.333) node [below] {$\tLrel_k$};
            \draw [-latex] (0,0.333) -- (0,1.1) node [left] {$\tSrel_k$};

            \draw (0,0.333) -- ++([yshift=2pt]0,0) -- ++([yshift=-4pt]0,0) node [below] {$0$};
            \draw (0.333,0.333) -- ++([yshift=2pt]0,0) -- ++([yshift=-4pt]0,0) node [below] {$1/3$};

            \draw (0,0.333) -- ++([xshift=2pt]0,0) -- ++([xshift=-4pt]0,0) node [left] {$\frac{1}{3}$};
            \draw (0,0.667) -- ++([xshift=2pt]0,0) -- ++([xshift=-4pt]0,0) node [left] {$\frac{2}{3}$};
            \draw (0,1.000) -- ++([xshift=2pt]0,0) -- ++([xshift=-4pt]0,0) node [left] {$1$};

            \draw [line width=2pt,myParula01Blue,name path=linePsync,line cap=round] (0.200,0.600) -- (0.200,0.600);
            \node [inner sep=1.5pt,circle,draw=myParula01Blue] at (0.200,0.600) {};
            \node [myParula01Blue,align=left,anchor=south west] at (0.200,0.600) {All clients};

            \begin{pgfonlayer}{background}
                \draw [draw=none,fill=myParula01Blue,fill opacity=0.3] (0.200,0.600) -- (0.200,0.333) -- (0,0.333) -- (0,0.600) -- cycle;
            \end{pgfonlayer}

        \end{tikzpicture}%
    }%
    \hfill%
    \subfloat[\FBFT, \FBFTtwo\label{fig:safety-liveness-tradeoffs-fbft}]{%
        \begin{tikzpicture}[fig_safety_liveness_tradeoff]%
            \footnotesize%
            \scriptspacing%

            \pgfdeclarelayer{background}
            \pgfdeclarelayer{foreground}
            \pgfsetlayers{background,main,foreground}

            \draw [bboxfix] ([southwest]0,0.333) rectangle ([northeast]0.4,1.0);

            \fill [conversearea] (0,1.05) -- (0,1) -- (0.333,0.333) -- (0.4,0.333) -- (0.4,1.05) -- cycle;
            \node [converselabel] at (0.4,1.05) {\footnotesize\textsc{Impossible}};

            \draw [-latex] (0,0.333) -- (0.45,0.333) node [below] {$\tLrel_k$};
            \draw [-latex] (0,0.333) -- (0,1.1) node [left] {$\tSrel_k$};

            \draw (0,0.333) -- ++([yshift=2pt]0,0) -- ++([yshift=-4pt]0,0) node [below] {$0$};
            \draw (0.333,0.333) -- ++([yshift=2pt]0,0) -- ++([yshift=-4pt]0,0) node [below] {$1/3$};

            \draw (0,0.333) -- ++([xshift=2pt]0,0) -- ++([xshift=-4pt]0,0) node [left] {$\frac{1}{3}$};
            \draw (0,0.667) -- ++([xshift=2pt]0,0) -- ++([xshift=-4pt]0,0) node [left] {$\frac{2}{3}$};
            \draw (0,1.000) -- ++([xshift=2pt]0,0) -- ++([xshift=-4pt]0,0) node [left] {$1$};

            \draw [line width=2pt,myParula05Green,name path=lineFlex2,line cap=round] (0.200,0.600) -- (0,0.800);
            \node [inner sep=1.5pt,circle,draw=myParula05Green] at (0.200,0.600) {};
            \node [myParula05Green,align=left,anchor=west,xshift=2pt] at (0.200,0.600) {Liveness-\\preserv.}; %
            \node [inner sep=1.5pt,circle,draw=myParula05Green] at (0,0.800) {};
            \node [align=left,anchor=south west,myParula05Green] at (0.0,0.800) {Safety-\\favor.}; %

            \begin{pgfonlayer}{background}
                \draw [draw=none,fill=myParula05Green,fill opacity=0.3] (0.200,0.600) -- (0,0.800) -- (0,0.333) -- (0.200,0.333) -- cycle;
            \end{pgfonlayer}

        \end{tikzpicture}%
    }%
    \hfill%
    \subfloat[Our \Oflex prot.\label{fig:safety-liveness-tradeoffs-optflex}]{%
        \begin{tikzpicture}[fig_safety_liveness_tradeoff]%
            \footnotesize%
            \scriptspacing%

            \pgfdeclarelayer{background}
            \pgfdeclarelayer{foreground}
            \pgfsetlayers{background,main,foreground}

            \draw [bboxfix] ([southwest]0,0.333) rectangle ([northeast]0.4,1.0);

            \fill [conversearea] (0,1.05) -- (0,1) -- (0.333,0.333) -- (0.4,0.333) -- (0.4,1.05) -- cycle;
            \node [converselabel] at (0.4,1.05) {\footnotesize\textsc{Impossible}};

            \draw [-latex] (0,0.333) -- (0.45,0.333) node [below] {$\tLrel_k$};
            \draw [-latex] (0,0.333) -- (0,1.1) node [left] {$\tSrel_k$};

            \draw (0,0.333) -- ++([yshift=2pt]0,0) -- ++([yshift=-4pt]0,0) node [below] {$0$};
            \draw (0.333,0.333) -- ++([yshift=2pt]0,0) -- ++([yshift=-4pt]0,0) node [below] {$1/3$};

            \draw (0,0.333) -- ++([xshift=2pt]0,0) -- ++([xshift=-4pt]0,0) node [left] {$\frac{1}{3}$};
            \draw (0,0.667) -- ++([xshift=2pt]0,0) -- ++([xshift=-4pt]0,0) node [left] {$\frac{2}{3}$};
            \draw (0,1.000) -- ++([xshift=2pt]0,0) -- ++([xshift=-4pt]0,0) node [left] {$1$};

            \draw [line width=2pt,myParula04Purple,name path=lineNew,line cap=round] (0.333,0.333) -- (0,1);
            \node [inner sep=1.5pt,circle,draw=myParula04Purple] at (0.333,0.333) {};
            \node [myParula04Purple,align=left,anchor=south west,xshift=-0.7em] at (0.333,0.333) {Liveness-\\\hspace{0.7em}preserv.}; %
            \node [inner sep=1.5pt,circle,draw=myParula04Purple] at (0,1) {};
            \node [align=left,anchor=west,xshift=2pt,myParula04Purple,yshift=-5pt] at (0,1) {Safety-\\\hspace{0.7em}favor.}; %

            \begin{pgfonlayer}{background}
                \draw [draw=none,fill=myParula04Purple,fill opacity=0.3] (0.333,0.333) -- (0,1) -- (0,0.333) -- cycle;
            \end{pgfonlayer}

        \end{tikzpicture}%
    }%
    \caption[]{%
        Pareto front
        (\tikz[baseline=-0.3em,x=1.5em]{ \draw [line width=2pt,myParula01Blue,line cap=round] (0,0) -- (0,0); },
        \tikz[baseline=-0.3em,x=1.5em]{ \draw [line width=2pt,myParula05Green,line cap=round] (0,0) -- (1,0); },
        \tikz[baseline=-0.3em,x=1.5em]{ \draw [line width=2pt,myParula04Purple,line cap=round] (0,0) -- (1,0); })
        and
        region
        (\tikz[baseline=0.15em,x=0.9em,y=0.9em]{ \draw [draw=none,fill=myParula01Blue,fill opacity=0.3] (0,0) rectangle (1,1); },
        \tikz[baseline=0.15em,x=0.9em,y=0.9em]{ \draw [draw=none,fill=myParula05Green,fill opacity=0.3] (0,0) rectangle (1,1); },
        \tikz[baseline=0.15em,x=0.9em,y=0.9em]{ \draw [draw=none,fill=myParula04Purple,fill opacity=0.3] (0,0) rectangle (1,1); })
        of
        liveness-/safety-resilience pairs $(\tLrel_k, \tSrel_k)$
        that each client $k$ can choose from,
        for three flexible consensus protocols
        under partial synchrony.
        Recall the classic impossibility result
        $2\tLrel_k + \tSrel_k \leq 1$
        (\tikz[baseline=0.15em,x=0.9em,y=0.9em]{ \draw [pattern=north east lines,pattern color=red!30,draw=none] (0,0) rectangle (1,1); })
        \cite{pbft,aav1}.
        \subref{fig:safety-liveness-tradeoffs-classic}~%
        PBFT-style protocols
        \cite{pbft,hotstuff,casper,tendermint,streamlet}
        use a system-wide replica quorum $\qrel$.
        For any fixed $\qrel$ ($=\frac{4}{5}$ here),
        these protocols
        have no flexibility
        (\tikz[baseline=-0.3em,x=1.5em]{ \draw [line width=2pt,myParula01Blue,line cap=round] (0,0) -- (0,0); })
        to support liveness-preserving clients and safety-favoring
        clients \emph{simultaneously};
        all clients operate at
        $(1-\qrel,2\qrel-1)$.
        \subref{fig:safety-liveness-tradeoffs-fbft}~%
        \FBFT \cite{fbft}
        and \FBFTtwo \cite{fbft2}
        additionally use
        client-side confirmation quorums $\qrel_k \in [\qrel,1]$
        to
        allow for some flexibility
        (\tikz[baseline=-0.3em,x=1.5em]{ \draw [line width=2pt,myParula05Green,line cap=round] (0,0) -- (1,0); })
        between the 
        liveness-preserving
        client
        at $(1-\qrel,2\qrel-1)$ for $\qrel_k=\qrel$
        and the most safety-favoring client
        at $(0,\qrel)$ for $\qrel_k=1$.
        \subref{fig:safety-liveness-tradeoffs-optflex}~%
        Our \Oflex protocols
        with only client-side confirmation quorums $\qrel_k \in [\frac{2}{3},1]$
        provide \emph{optimal} flexibility
        (\tikz[baseline=-0.3em,x=1.5em]{ \draw [line width=2pt,myParula04Purple,line cap=round] (0,0) -- (1,0); } vs.\ \tikz[baseline=0.15em,x=0.9em,y=0.9em]{ \draw [pattern=north east lines,pattern color=red!30,draw=none] (0,0) rectangle (1,1); })
        between the 
        liveness-preserving
        client
        at $(\frac{1}{3},\frac{1}{3})$ for $\qrel_k=\frac{2}{3}$
        and the most safety-favoring client
        at $(0,1)$ for $\qrel_k=1$.
    }%
    \label{fig:safety-liveness-tradeoffs}%
\end{figure}%

%% file: figures/tab-oflex-constructions.tex
\begin{table}[tb]
    \centering
    \caption{%
        Realization of key mechanisms in different protocols of the \Oflex family (\cref{sec:intro-contribution}):\\
        \textcolor{jnSUDigitalRed}{\textbf{addition}} of logic
        vs.\ re-use of \textcolor{jnSUDigitalGreen}{\textbf{existing}} logic.%
    }
    \setlength{\tabcolsep}{3pt}
    \begin{tabular}{@{}>{\raggedright\arraybackslash}p{0.23\columnwidth}>{\raggedright\arraybackslash}p{0.20\columnwidth}>{\raggedright\arraybackslash}p{0.20\columnwidth}>{\raggedright\arraybackslash}p{0.29\columnwidth}@{}}%
        \toprule
         & \multicolumn{2}{l}{\textbf{Replica logic (system-wide)}}
         & \multicolumn{1}{l}{\textbf{Client logic (local)}}
        \\ \cmidrule(){2-4}
        \textbf{Construction}
         & \multicolumn{1}{l}{\textbf{\Ovote}}
         & \multicolumn{1}{l}{\textbf{\Olock}}
         & \multicolumn{1}{l}{\textbf{Confirmation rule}}
        \\ \midrule
        Generic (add-on to any protocol)
         & \textcolor{jnSUDigitalRed}{\textbf{add}} \ovotes external to the base protocol
         & \textcolor{jnSUDigitalRed}{\textbf{add}} constraint on \ovotes
         & \textcolor{jnSUDigitalRed}{\textbf{new}} (quorum $\qrel_{k}$ fraction of \ovotes)
        \\\addlinespace[2pt]%
        \Pipelinedchained PBFT-style (\eg, \Oflex-Streamlet)
         & \textcolor{jnSUDigitalGreen}{\textbf{existing}} votes (\pipeliningchaining)
         & \textcolor{jnSUDigitalRed}{\textbf{add}} constraint on votes
         & \textcolor{jnSUDigitalRed}{\textbf{new}} (quorum $\qrel_{k}$ fraction of votes)
        \\\addlinespace[2pt]%
        \Oflex conf.\ rules for Ethereum
         & \textcolor{jnSUDigitalGreen}{\textbf{existing}} votes (\pipeliningchaining)
         & \textcolor{jnSUDigitalGreen}{\textbf{existing}} performance optimization
         & \textcolor{jnSUDigitalRed}{\textbf{new}} (quorum $\qrel_{k}$ fraction of votes)
        \\
        \bottomrule
    \end{tabular}
    \label{tab:oflex-constructions}
\end{table}

%% file: figures/blockdiagrams-oflex-generic.tex
\tikzset{fig_blockdiagrams/.style={
            x=2.6cm,
            y=1.6cm,
            pir/.style={
                    fill=white,
                    draw,
                    inner sep=0pt,
                    minimum height=1.5cm,
                    minimum width=1.5cm,
                    align=center,
                },
            pic/.style={
                    fill=white,
                    draw,
                    inner sep=0pt,
                    shape=rounded rectangle,
                    minimum width=1.4cm,
                    minimum height=0.5cm,
                    xshift=1.4cm,
                    align=center,
                },
            bboxfix/.style={
                    fill=none,
                    draw=none,
                    cornerSW/.style={
                        xshift=-1.6cm,
                        yshift=-1em,
                    },
                    cornerNE/.style={
                        xshift=7.125cm,
                        yshift=1em,
                    },
                },
            boxOverall/.style={
                densely dotted,
                cornerSW/.style={
                    xshift=-0.8em,
                    yshift=-1em,
                },
                cornerNE/.style={
                    xshift=0.8em,
                    yshift=2.75em,
                },
            }
        }}
\begin{figure}[tb]%
    \centering%
    \begin{tikzpicture}[fig_blockdiagrams]%
        \footnotesize%
        \scriptspacing%

        \draw [bboxfix] ([cornerSW]0,0) rectangle ([cornerNE]0,0);

        \begin{scope}[xshift=0cm]
            \coordinate (txs) at (-0.60,0);

            \node [pic] (PiCmin) at (0,0) {$\PiC$};
            \node [pir] (PiR) at (0,0) {$\PiR$};

            \draw [-latex] (txs) -- (PiR) node [above,pos=0.25] {$\txs$};
        \end{scope}

        \begin{scope}[xshift=3.3cm]
            \coordinate (LOGmin_oflex) at ([yshift=0.5cm]1.1,0);
            \coordinate (LOGmax_oflex) at ([yshift=-0.5cm]1.1,0);

            \node [pic] (PiCmin_oflex) at ([yshift=0.5cm]0,0) {$\PiCOpart_{\qrel_k=\frac{2}{3}}$};
            \node [pic] (PiCmax_oflex) at ([yshift=-0.5cm]0,0) {$\PiCOpart_{\qrel_k=1}$};
            \node [xshift=1.35cm] at (0,0) {$\svdots$};
            \node [pir] (PiR_oflex) at (0,0) {$\PiROpart$\\[2pt]\olock\\\& \ovote};

            \draw [-latex] (PiCmin) -- (PiR_oflex);
            \draw [-latex] (PiCmin_oflex) -- (LOGmin_oflex) node [pos=1,anchor=north west,xshift=-1.5em,yshift=1.5em] {$\LOG{(\frac{1}{3},\frac{1}{3})}{}$};
            \draw [draw=none] (LOGmin_oflex) -- (LOGmax_oflex) node [midway,yshift=3pt] {$\svdots$};
            \draw [-latex] (PiCmax_oflex) -- (LOGmax_oflex) node [pos=1,anchor=north west,xshift=-1.5em,yshift=1.5em] {$\LOG{(0,1)}{}$};
        \end{scope}

        \draw [boxOverall] ([cornerSW]PiR.south west) rectangle ([cornerNE]PiCmin_oflex.east |- PiR_oflex.north) coordinate [pos=0] (ProtoBoxSW) coordinate [pos=1] (ProtoBoxNE);

        \draw [decorate,decoration={calligraphic brace}] ([yshift=0.75em,xshift=1pt]PiR.north west) -- ([yshift=0.75em,xshift=-1pt]PiR_oflex.north east) node [midway,above,anchor=base,yshift=5pt] {Replicas};
        \draw [decorate,decoration={calligraphic brace}] ([yshift=0.75em,xshift=1pt]PiR_oflex.north east) -- ([yshift=0.75em,xshift=-1pt]PiCmin_oflex.east |- PiR_oflex.north east) node [midway,above,anchor=base,yshift=5pt] {Clients};

    \end{tikzpicture}%
    \caption[]{%
        Block diagram of our
        generic \Oflex construction 
        (\cref{sec:optflex}).
        Starting from an \emph{unmodified}
        consensus protocol
        with
        replica logic $\PiR$
        and confirmation rule $\PiC$
        with resiliences $\tLrel=\tSrel=\frac{1}{3}$,
        the construction applies
        a round of permanently locking (\emph{\olocking}) and voting (\emph{\ovoting})
        by replicas
        ($\PiROpart$)
        on the log output by
        $(\PiR, \PiC)$,
        the unmodified protocol,
        followed by client-specific confirmation quorums $\qrel_k$
        on the \ovote messages
        for confirmation
        ($\PiCOpart$).
    }%
    \label{fig:blockdiagrams-oflex-generic}%
\end{figure}%

%% file: figures/eth-experiment-intro-latency.tex
\begin{figure}[tb]
    \centering
    \begin{tikzpicture}[]%
        \footnotesize
        \begin{axis}[
            mysimpleplot,
            name=plot1,
            xlabel={Safety resilience $\tSrel_k$},
            ylabel={Latency [minutes]},
            xmin=0.3, xmax=1,
            ymin=0, ymax=30,
            height=0.5\linewidth,
            width=\linewidth,
            legend columns=2,
        ]

            \addplot+[myparula21,thick,mark=square*,only marks] 
            table[ignore chars={(,)},col sep=comma, x expr=1.0-(1.0-\thisrowno{0})*2,y expr=\thisrowno{1}*12/60.0] {figures/eth-experiment-intro-latency/finalization-avg-latency.txt};
            \addlegendentry{Casper finality};

            \addplot+[myparula11,thick,mark=*] 
            table[ignore chars={(,)},col sep=comma, x expr=1.0-(1.0-\thisrowno{0})*2,y expr=\thisrowno{1}*12/60.0] {figures/eth-experiment-intro-latency/avg-latency.txt};
            \addlegendentry{\Oflex rule};
            
        \end{axis}
    \end{tikzpicture}%
    \caption[]{%
        Average latency to confirm Ethereum mainnet blocks between slots $5{,}970{,}000$ and $6{,}970{,}000$
        (\ie, most recent $1{,}000{,}000$ slots as of this writing, corresponding to March 10, 2023 to July 27, 2023) by Casper finality~\cite{casper} (the confirmation rule as per Ethereum's specifications)
        and by \Oflex
        confirmation rules (\cref{sec:eth})
        with different safety levels. 
        See \cref{sec:eth-experiments} for details of the setup.
        \Oflex incurs approx.\ $4{.}25\,\text{min}$ ($21\,\text{slots}$)
        increase in latency over Casper finality,
        even at the same safety resilience
        $\tSrel_k = \frac{1}{3}$, 
        due to \ovoting.
        However, beyond that,
        \Oflex achieves extremely high safety resilience while incurring only a modest increase in latency.
        Each client can operate on any point along the blue line 
        to match its desired trade-off between safety and latency.
        }
    \label{fig:eth-experiment-intro-latency}
\end{figure}
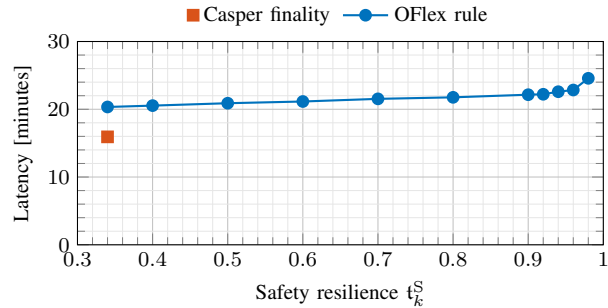

%% file: 02_modelproblem.tex
\section{Model and Problem Formulation}
\label{sec:modprob}

SMR consensus is run by $n$ \emph{replicas}
and a group of \emph{clients}.
Out of these,
$f$ replicas are \emph{adversary}%
\footnote{To state results with precision, we henceforth denominate $f,q,\tL,\tS$ in number of replicas (instead of fractions $\frel,\qrel,\tLrel,\tSrel$ as in \cref{sec:intro}).}
(`Byzantine faults';
assumed to be chosen
at the start of the execution before global randomness is drawn),
and the remaining replicas are \emph{honest}
and follow the specified \emph{replica logic} $\PiR$
to receive transactions from the environment,
interact with other replicas, and send updates
to clients.
At all times $\tau$,
clients (indexed by $k$)
use a \emph{confirmation rule},
which only involves local computation on the updates received from replicas,
to confirm and output a log $\LOG{k}{\tau}$ of transactions.
In classical consensus, there is one confirmation rule $\PiC$.
In flexible consensus,
each client $k$ may 
use a different confirmation rule $\PiC_k$. 
The replica protocol and the confirmation rule(s) together are called the \emph{consensus protocol} $\PI$.

Communication among replicas and from replicas to clients is authenticated
using digital signatures.
We assume the \emph{eventual-synchrony} variant of the
\emph{partially-synchronous} network model~\cite{pbft},
\ie,
there is an adversarially chosen
\emph{global stabilization time} ($\GST$),
\emph{unknown} to the protocol,
before and after which the network is \emph{asynchronous} and \emph{synchronous}, respectively.
During asynchrony, the adversary can delay messages arbitrarily.
During synchrony, the adversary can delay messages up to a delay bound $\Delta$,
\emph{known} to the replicas and clients.
Protocols may instruct replicas to \emph{gossip} received messages,
so that after $\GST$ every message received by any honest
replica by $\tau$ is received by all honest replicas by $\tau + \Delta$.

\smallskip\noindent\textbf{Notation:}~%
Let $\LOGempty \preceq \LOGempty'$ denote that $\LOGempty$ is a prefix of or identical to $\LOGempty'$.
Two logs are \emph{consistent} iff $\LOGempty \preceq \LOGempty'$ or $\LOGempty' \preceq \LOGempty$.

\smallskip
We now formally define the flexible consensus problem.\footnote{As is common practice,
we prove which security properties hold when the resiliences are satisfied \emph{throughout the execution}.
In \cref{sec:discussion},
we discuss how security may be recovered after resiliences are violated.}
\begin{definition}%
    \label{def:flex-consensus}
    Protocol $\PI = (\PiR, \{\PiC_k\})$
    provides \emph{flexible consensus}
    with resilience pairs
    $\left\{(\tL_k,\tS_k)\right\}$ iff:
    \begin{itemize}
        \item
              \textbf{Liveness:}
              For every
              $\tx$
              input to all honest replicas,
              eventually,
              for all clients $k$ with $f \leq \tL_k$,
              $\tx \in \LOG{k}{}$.

        \item
              \textbf{Safety:}
              For all clients $k, k'$ with $f \leq \min\{\tS_k, \tS_{k'}\}$,
              for all times $\tau, \tau'$,
              $\LOG{k}{\tau}$ and $\LOG{k'}{\tau'}$ are consistent.
    \end{itemize}
\end{definition}

We use three special cases of the above definition.
We refer to the case where
for all $k$, $\tL_{k}=\tL$, $\tS_{k}=\tS$, as \emph{$(\tL,\tS)$-consensus}.
We call the further specialized case
$\tL=n/3-1$, $\tS=n/3$
\emph{classical consensus}.
The third case is the following:
\begin{definition}
    \label{def:optimal-flexible}
    A flexible consensus protocol
    provides \emph{optimal flexible consensus} if it
    supports
    any set $\left\{(\tL_k,\tS_k)\right\}$
    of resilience pairs
    with for all $k$, $2\tL_{k} + \tS_{k} < n$ and $\tL_{k} \leq \tS_{k}$.
\end{definition}

%% file: 03_optflex_general.tex
\section{Generic \Oflex Construction}
\label{sec:optflex}

We 
describe how to construct an optimal high-safety flexible consensus protocol from any classical consensus protocol in a generic closed-box manner.
The key ingredient is new replica logic we add, which we call an \emph{optimal-flexibility (\Oflex) gadget}: 
an additional round of voting
(called `\ovote' to distinguish from votes in the classical protocol)
in which replicas follow a permanent lock (`\olock') constraint.

\subsection{Construction}
\label{sec:optflex-protocol}

\import{./figures/}{alg-optflex.tex}

\import{./figures/}{protocol-views-oflex-generic.tex}

A high-level block diagram of this construction is shown in \cref{fig:blockdiagrams-oflex-generic}.
We start with any protocol $\PI=(\PiR,\PiC)$ that provides $(n/3-1,n/3)$-consensus. We call $\PI$ the \emph{base protocol}.
Replicas running the replica protocol $\PiR$ additionally run the \Oflex gadget protocol $\PiROpart$ (which implements `\olock and \ovote'). Clients obtain their flexible logs by running new confirmation rules 
$\PiCOpart_k$
on messages received as part of $\PiROpart$ from replicas.
The entire construction, encompassing $\PI$, $\PiROpart$,
and the family of confirmation rules $\{\PiC_k\}$,
is denoted $\PIO(\PI)$.

Pseudocode of $\PiROpart$ and $\PiCOpart_k$ is given in \cref{alg:optflex}.
The replicas' role in $\PiROpart$ is illustrated in \cref{fig:protocol-views-oflex-generic}.
Each replica 
uses the confirmation rule of the base protocol $\PiC$ to obtain a confirmed log at each time step.
Each replica maintains a \olocked log.
Whenever the replica sees an update to 
the output log of the base protocol
according to the confirmation rule $\PiC$,
the replica checks if the new log \emph{strictly extends} its \olocked log
(\cref{alg:optflex}~\cref{loc:optflex-check-permalock}).
Only if so, the replica updates its \olock to the new log
(\cref{fig:protocol-views-oflex-generic}~\DiagramStep{1}),
and broadcasts a `\ovote' message for the new log
(\cref{fig:protocol-views-oflex-generic}~\DiagramStep{2}~\DiagramStep{3}).

The client's confirmation rule $\PiCOpart_k$ is parameterized by 
a client-specific confirmation quorum $q_{k}$,
chosen as $(n+\tS_{k})/2 < q_{k} \leq n - \tL_{k}$ to satisfy the client's resilience pair $(\tL_k, \tS_k)$.
The client confirms a log $\LOG{}{}$ if at least $q_{k}$ replicas \ovote $\LOG{}{}$ or a log that extends $\LOG{}{}$ (\cref{alg:optflex}~\cref{loc:optflex-confirmation-rule}).

\subsection{Security Analysis}
\label{sec:optflex-analysis}

We will prove that $\PIO(\PI)$,
\ie, our \Oflex gadget
applied to
any classical consensus protocol $\PI$,
provides optimal high-safety flexible consensus.
To build intuition,
consider the simple example of a client who chooses 
a confirmation quorum $q_{k} = n$.
\emph{Safety} of $\PIO(\PI)$
depends purely on the 
\Oflex gadget,
and does not rely on safety or liveness of the base protocol.
When an honest replica \ovotes a log
(\cref{fig:protocol-views-oflex-generic}~\DiagramStep{2}), it signals to clients that it has \olocked that log
(\cref{fig:protocol-views-oflex-generic}~\DiagramStep{1}); hence, it has never \ovoted and will never \ovote an inconsistent log
(\cref{fig:protocol-views-oflex-generic}~\DiagramStep{3}).
Thus, when the client confirms a log which all $q_k = n$ replicas \ovote, 
then 
no client $k'$ with $q_{k'} = n$
will ever confirm an inconsistent log, unless all replicas are adversary,
thus achieving $\tS_{k} = \tS_{k'} = n-1$.
\emph{Liveness} of $\PIO(\PI)$, on the other hand, depends on both liveness and safety of the base protocol $\PI$,
and liveness of the \Oflex gadget.
However, we require $\tS_{k} \geq \tL_{k}$ and we know that it is impossible to achieve both resiliences $\geq n/3$.
Therefore,
clients can only expect $\tL_{k} < n/3$,
and
the base protocol is guaranteed to be live and safe in this regime.
Then, all honest replicas obtain live logs from $\PI$, and also eventually \ovote them because they never see inconsistent logs confirmed by $\PI$.
Therefore, if all replicas are honest, the client in our example eventually sees $q_k = n$ \ovotes for logs that were obtained by honest replicas from $\PI$, and as a result this client also confirms the live output log of $\PI$, thus achieving $\tL_{k} = 0$.
Simultaneously, another client that chooses a confirmation quorum $q_{k'} = 2n/3 + 1$ obtains resilience guarantees $\tS_{k'} = n/3$, $\tL_{k'} = n/3-1$ (the classical parameters).
Thus, our \Oflex gadget simultaneously supports clients with resiliences $(0,n-1)$ and $(n/3-1,n/3)$ (see~\cref{fig:safety-liveness-tradeoffs-optflex}),
while 
earlier \FBFT and \FBFTtwo
could not
support both simultaneously (\cref{fig:safety-liveness-tradeoffs-fbft}).

\begin{theorem}
    \label{thm:generic-oflex-security}
    If $\PI = (\PiR,\PiC)$ provides $(n/3-1,n/3)$-consensus, then the construction $\PIO(\PI) = (\PiROpart \circ \PiC \circ \PiR, \{\PiCOpart_k\})$, where the \Oflex gadget is applied to $\PI$,
    provides optimal high-safety flexible consensus.
\end{theorem}
\begin{proof}
    We show that all clients with $f \leq \tS_k$ and confirmation quorums $q_{k} > (n+\tS_{k})/2$ have safety,
    and all clients with $f \leq \tL_k \leq \tS_k$ and $q_{k} \leq n - \tL_{k}$ have liveness.
    Therefore, by changing the quorum size $q_{k}$, clients can achieve any pair of resiliences that satisfies $(n+\tS_{k})/2 < n - \tL_{k}$, \ie, $2\tL_{k} + \tS_{k} < n$, and $\tL_k \leq \tS_k$.

    \emph{Safety:}
    Suppose, for contradiction, that for two clients $k,l$,
    $f \leq \min\{\tS_{k},\tS_{k'}\}$,
    and they confirm inconsistent logs $\LOG{k}{\tau}$ and $\LOG{k'}{\tau'}$.
    Consider any honest replica that \ovoted $\LOG{k}{\tau}$ or an extension thereof
    (\cf \cref{fig:protocol-views-oflex-generic}~\DiagramStep{2}).
    (The quorum intersection argument below shows that such a replica exists.)
    Then this replica can never \ovote a log $\LOGempty'$ that is inconsistent with $\LOG{k}{\tau}$
    (\cf \cref{fig:protocol-views-oflex-generic}~\DiagramStep{3}).
    It does not do so after it \ovoted $\LOG{k}{\tau}$ because it has already \olocked $\LOG{k}{\tau}$ or an extension thereof
    (\cf \cref{fig:protocol-views-oflex-generic}~\DiagramStep{1}).
    It could not have done so before it \ovoted $\LOG{k}{\tau}$ because then it must have \olocked $\LOGempty'$ and hence would not have \ovoted $\LOG{k}{\tau}$.
    
    For confirmation by client $k$, at least $q_{k}$ replicas \ovoted $\LOG{k}{\tau}$ or an extension thereof. Similarly, at least $q_{k'}$ replicas \ovoted $\LOG{k'}{\tau'}$ or an extension thereof.
    Therefore, at least $q_{k} + q_{k'} - n$ replicas \ovoted two inconsistent logs.
    Due to the preceding argument, these must all be adversary replicas, so $f \geq q_{k} + q_{k'} - n$.
    Due to the choice of quorums $q_{k} > (n+\tS_{k})/2$ and $q_{k'} > (n+\tS_{k'})/2$, safety violation between clients $k,l$ requires $f \geq (\tS_{k} + \tS_{k'})/2 + 1$ adversary replicas.
    This is a contradiction to the assumption that $f \leq \min\{\tS_{k},\tS_{k'}\}$.

    \emph{Liveness:}
    Suppose that $f \leq \tL_{k}$.
    Note that since $\tL_{k} \leq \tS_{k}$, the feasibility condition $2\tL_{k} + \tS_{k} < n$ also implies that $\tL_{k} < n/3$.
    Therefore, $f < n/3$,
    which means that $\PI$ is safe and live.
    Due to the safety of $\PI$, the log obtained by a replica is always consistent with the logs it obtained in the past, and hence consistent with the replica's \olock.
    Therefore, whenever a replica updates its log from $\PI$, it will either \ovote the new log, or has already \ovoted that log or an extension thereof.
    Thus,
    an honest replica \ovotes every log that it obtains from $\PI$.
    Furthermore, since $\PI$ is live, transactions eventually appear
    in every honest replica's log of $\PI$.
    Since $f \leq \tL_{k}$, 
    client $k$ eventually sees enough \ovotes to confirm every log output by $\PI$, since \ovotes from all honest replicas are enough to satisfy the quorum $q_{k} \leq n - \tL_{k}$.
\end{proof}

This construction comes with a communication overhead of
one additional round of voting (\ovotes)
per confirmed batch of transactions.
However, in \pipelinedchained 
PBFT-style protocols 
such as 
Casper~\cite{casper},
HotStuff~\cite{hotstuff},
or
Streamlet~\cite{streamlet},  
we can re-use existing successive rounds of votes in the base protocol to also serve as \ovotes, and this eliminates the communication overhead.
We describe this modification for Streamlet
in \cref{sec:optflex-streamlet},
and for Ethereum's implementation of Casper in \cref{sec:eth}.

%% file: figures/alg-optflex.tex
\begin{algorithm}[tb]
    \caption{Optimal-flexibility (\Oflex) gadget $\PIOpart$}
    \label{alg:optflex}
    \begin{algorithmic}[1]
        \footnotesize
        \LineComment{Replica-side logic $\PiROpart$}%
        \On{\Call{init}{\null}}
            \State $\PiC.\Call{init}{\null}$
                \Comment{Replica runs base protocol confirmation rule}
            \State $\algolock \gets [\,]$ \Comment{Initialize \olock to the empty log}
        \EndOn
    
        \Forever
            \State $\LOGempty \gets \PiC.\Call{getLog}{\null}$
            \If{$\algolock \prec \LOGempty$}%
                \Comment{$\LOGempty$ (strictly) extends \olock}\label{loc:optflex-check-permalock}
                \State $\algolock \gets \LOGempty$ \Comment{\Olock}
                \State Broadcast $\langle\algovote, \LOGempty\rangle$ to all clients
                    \Comment{\Ovote $\LOGempty$}
            \EndIf
        \EndForever
        \medskip
        \LineComment{Client-side confirmation rule $\PiCOpart$ for resilience pair $(\tL_k, \tS_k)$}
        \On{\Call{init}{$q_{k}$} where $q_k$ shall satisfy $(n+\tS_{k})/2 < q_{k} \leq n - \tL_{k}$}
            \State Set local quorum size $q_{k}$
            \State $\algovotes \gets \emptyset$
                \Comment{Set of received \ovotes}
        \EndOn
        \On{receiving $\langle\algovote,\LOGempty\rangle$ from replica $i$}
            \State $\algovotes \gets \algovotes \cup \{ (\LOGempty, i) \}$
                \Comment{Record \ovote} %
        \EndOn
        \On{\Call{getLog}{\null}}
            \LineComment{Confirm $\LOG{}{*}$ if $q_k$ replicas \ovoted $\LOG{}{*}$ or an extension} 
            \State $\LOG{}{*} \gets \arg\max_{\LOGempty} |\LOGempty|$
            subject to
            $\algovotes$ contains at least $q_k$ pairs $(\LOG{i}{},i)$ for distinct $i$ with some $\LOG{i}{} \succeq \LOGempty$ \label{loc:optflex-confirmation-rule}
            \State \textbf{return} $\LOG{}{*}$
        \EndOn
    \end{algorithmic}
\end{algorithm}

%% file: figures/protocol-views-oflex-generic.tex
\import{./}{protocol-views-_utils.tex}%
\begin{figure}[tb]
    \centering
    \begin{tikzpicture}[fig_protocol_views,x=0.8cm]%
        \footnotesize%
        \scriptspacing%

        \pgfdeclarelayer{background}
        \pgfdeclarelayer{foreground}
        \pgfsetlayers{background,main,foreground}

        \node (b0) at (0,0) {$\tx$};
        \draw [ultra thick,densely dotted] (b0) -- ++([xshift=-0.65cm]0,0);

        \node (bA) at (1,+1) {$\tx'$};
        \draw [ultra thick] (bA) -- (b0);

        \node (bC) at (3,+1) {$\tx''$};
        \draw [ultra thick] (bC) -- (bA);

        \coordinate (bE) at (1,-1);

        \node (bF) at (2,-1) {$\tx'''$};
        \draw [ultra thick] (bF) -- (bE) -- (b0);

        \begin{pgfonlayer}{background}
            \draw [fill=red,fill opacity=0.2,draw=none] ([yshift=1.1em,xshift=-0.75cm]b0.center) -- ([yshift=1.1em]b0.center) -- ([yshift=1.1em]bA.center) -- ([yshift=1.1em,xshift=0.4cm]bC.center) -- ([yshift=-1.1em,xshift=0.4cm]bC.center) -- ([yshift=-1.1em]bA.center) -- ([yshift=-1.1em]b0.center) -- ([yshift=-1.1em,xshift=-0.75cm]b0.center);
            \draw [fill=blue,fill opacity=0.2,draw=none] ([yshift=0.9em,xshift=-0.75cm]b0.center) -- ([yshift=0.9em]b0.center) -- ([yshift=0.9em]bE.center) -- ([yshift=0.9em,xshift=0.4cm]bF.center) -- ([yshift=-0.9em,xshift=0.4cm]bF.center) -- ([yshift=-0.9em]bE.center) -- ([yshift=-0.9em]b0.center) -- ([yshift=-0.9em,xshift=-0.75cm]b0.center);
        \end{pgfonlayer}

        \node (lockRed) at ([yshift=-1.1em,xshift=0.4cm]bC.center) {\notoemoji[height=0.4cm]{lock}};

        \node (validator) at (5.5,0) {\notoemoji[height=0.6cm]{technologist}};
        \draw [-latex,dashed] (validator) -- ++(0.75,1.5) node [pos=1,right,anchor=west] {\DiagramStep{2} $\langle\algovote,\tikz[x=0.6em,y=0.25em]{ \draw [fill=red,fill opacity=0.2,draw=none] (-1,-1) rectangle (1,1); }\rangle$};
        \draw [-latex,dashed] (validator) -- ++(0.75,-1.5) node [pos=1,right,anchor=west] {\DiagramStep{3} $\langle\algovote,\tikz[x=0.6em,y=0.25em]{ \draw [fill=blue,fill opacity=0.2,draw=none] (-1,-1) rectangle (1,1); }\rangle$};
        \draw [-latex,dashed,shorten >=-3pt] (validator) -- (lockRed) node [pos=0.2] {\DiagramStep{1}};
        \node [cloud callout,cloud puffs=15,aspect=2.5,cloud puff arc=120,draw,callout absolute pointer={(6,0)},xshift=1.3cm] at (validator.east) {\DiagramStep{1} \notoemoji[height=0.9em]{lock}\tikz[x=0.6em,y=0.25em]{ \draw [fill=red,fill opacity=0.2,draw=none] (-1,-1) rectangle (1,1); }};

        \node at (7.5,-1.5) {\notoemoji[height=1.3em]{prohibited}};

        \coordinate (LOGredS) at ([xshift=2.3cm,yshift=0.85cm]validator.center);
        \coordinate (LOGredT) at ([yshift=1.1em,xshift=0.4cm]bC.center);
        \draw [-latex,bend right=15,red,densely dotted] (LOGredS) to (LOGredT);

        \coordinate (LOGblueS) at ([xshift=2.3cm,yshift=-0.85cm]validator.center);
        \coordinate (LOGblueT) at ([yshift=-0.9em,xshift=0.4cm]bF.center);
        \draw [-latex,bend left=15,blue,densely dotted] (LOGblueS) to (LOGblueT);

        \draw [dashed] ([xshift=-0.9cm,yshift=-1cm]0,0) rectangle ([xshift=3.3cm,yshift=1cm]0,0);
        \node [anchor=south west] at ([xshift=-0.9cm,yshift=1cm]0,0) {$\PI$};

        \draw [dashed] ([xshift=-0.9cm,yshift=-1cm]validator.center) rectangle ([xshift=2.8cm,yshift=1cm]validator.center);
        \node [anchor=south east] at ([xshift=2.8cm,yshift=1cm]validator.center) {$\PiROpart$};

        \draw [dashed] ([xshift=-1cm,yshift=-1.1cm]0,0) rectangle ([xshift=2.9cm,yshift=1.45cm]validator.center) node [midway,yshift=1.45cm] {$\PIO(\PI)$};

    \end{tikzpicture}%
    \caption[]{%
        Illustration of a replica's role in \Oflex gadget $\PiROpart$ (\cref{alg:optflex}), which can augment any classical consensus protocol $\PI$ to make an optimal flexible consensus protocol $\PIO(\PI)$.
        * \emph{Protocol rules:}
        \DiagramStep{1}~Honest replicas \olock the log they obtain from $\PI$ iff it extends their previous \olock,
        \DiagramStep{2}~\ovote the log that they \olock,
        and
        \DiagramStep{3}~never \ovote any log that does not extend their \olocked log.
        * \emph{Safety intuition:}
        A client $k$
        with $q_k=n$ confirms a log only
        when it sees $n$ \ovotes for the log.
        Due to steps \DiagramStep{1} to \DiagramStep{3},
        no conflicting log can receive $n$ \ovotes
        and be confirmed by another client $k'$ with
        $q_{k'}=n$,
        \emph{unless all replicas are adversary},
        implying $\tS_k = \tS_{k'} = n-1$.
    }
    \label{fig:protocol-views-oflex-generic}
\end{figure}

%% file: 04_optflex_streamlet.tex
\section{\Oflex PBFT-Style Protocols}
\label{sec:optflex-streamlet}

In this section, we use Streamlet~\cite{streamlet}
as an example
to show how the extra round of \olocking and \ovoting for \Oflex
can be realized in
PBFT-style protocols.
Specifically,
we show how \emph{existing} votes can be reused to double as \ovotes
(so no extra votes are added),
and how \olocking is implemented as an additional constraint
in the protocol's voting rule.
The construction carries over rather straightforwardly
to other PBFT-style protocols like
Hotstuff~\cite{hotstuff}, Casper~\cite{casper}, Tendermint~\cite{tendermint}.

We first recap Streamlet~\cite{streamlet} (\cref{sec:optflex-streamlet-recap}),
then describe the modifications required for \Oflex (\cref{sec:optflex-streamlet-protocol}).
We call the resulting protocol \Oflex-Streamlet.
We prove optimal flexibility of \Oflex-Streamlet in \cref{sec:optflex-streamlet-analysis}.
To contrast \Oflex and \FBFT in \cref{sec:optflex-streamlet-fbft-comparison},
we compare \Oflex-Streamlet
with an adaptation of \FBFT~\cite{fbft} to Streamlet (`\FBFT-Streamlet').

\subsection{Recap of PBFT-Style Consensus: Streamlet}
\label{sec:optflex-streamlet-recap}

Streamlet~\cite{streamlet} (\cref{alg:streamlet})
proceeds in \emph{epochs} of duration $2\Delta$,
where $\Delta$ is the bound on message delays after $\GST$.
There are two message types: \emph{blocks} and \emph{votes}.
Every message is signed by the replica that creates it.
A block consists of
a hash pointer to its \emph{parent} block
(transitively forming a \emph{chain} back to a commonly known \emph{genesis block}),
an epoch number,
and transaction payload.
A vote consists of a hash pointer to a block.
Based on the relation of a block and its parent,
`ancestor', `descendant', and `child' are defined in the canonical way.
Two blocks are \emph{adjacent} if one is the other's parent.
A block is \emph{notarized} in the view of a replica/client
when it has
seen votes for that block from a quorum of at least $q = 2n/3 + 1$ replicas.
A chain is notarized if every block in the chain is notarized.
Throughout, replicas echo every message they receive.
The \emph{replica logic $\PiR$} consists of two steps:
\emph{1) Propose:}
Each epoch has a \emph{leader} replica elected using public randomness.
At the start of the epoch, the leader broadcasts a new block
containing unconfirmed transactions and the current epoch;
the parent of the new block
is the tip of
any one of the longest notarized chains in the \emph{leader's} view.
\emph{2) Vote:}
During the epoch,
each replica broadcasts a vote for the first block received from the epoch's leader,
\emph{if} the block's parent is the tip of
any one of the longest notarized chains in the \emph{replica's} view.

\emph{Confirmation rule $\PiC$}:
If a client sees, in a \emph{notarized} chain, three
\emph{adjacent} blocks $A,B,C$
from \emph{consecutive} epochs,
then it \emph{confirms} $B$, \ie,
sets its log to the sequence of transactions
as ordered in the chain from the genesis block to $B$.

\subsection{\Oflex-Streamlet Protocol}
\label{sec:optflex-streamlet-protocol}

\import{./figures/}{alg-oflex-streamlet.tex}

\import{./figures/}{protocol-views-oflex-streamlet.tex}

We modify Streamlet
(\cref{alg:streamlet})
to \Oflex-Streamlet
(\cref{alg:oflex-streamlet}).
Changes are highlighted \textcolor{jnSUDigitalGreen}{green} in \cref{alg:oflex-streamlet}.

To implement \olocking,
we extend the replica logic with a $\algolock$,
initialized to the genesis block.
Whenever a replica sees a block $B$ extending its $\algolock$
such that a client would have confirmed $B$
using Streamlet's original confirmation rule $\PiC$,
then the replica updates its $\algolock$ to $B$
(\cref{fig:protocol-views-oflex-streamlet}~\DiagramStep{1}, \cref{alg:oflex-streamlet}~\cref{loc:oflex-streamlet-olock}).
A replica's voting is constrained to blocks that are
consistent with the $\algolock$
(\cref{fig:protocol-views-oflex-streamlet}~\DiagramStep{2}~\DiagramStep{3}).

In \Oflex-Streamlet's confirmation rule $\PiCmod(q_k)$,
clients require $q_k$ votes for a block $D$ which is a descendant of the
three blocks $A,B,C$ required by Streamlet's rule $\PiC$ (\cref{loc:oflex-streamlet-conf}).
These $q_k$ votes serve as \ovotes
(\cref{fig:protocol-views-oflex-streamlet}~\DiagramStep{2}), and allow
clients to reason that honest replicas voting for $D$ are \olocked on $B$
(\cref{fig:protocol-views-oflex-streamlet}~\DiagramStep{1})
and will not vote for anything conflicting with $B$
(\cref{fig:protocol-views-oflex-streamlet}~\DiagramStep{3}).
This reprises the safety argument of
\Oflex in \cref{sec:optflex},
and
is key to \Oflex-Streamlet's
safety argument and improved resilience
compared to \FBFT and \FBFTtwo.

Note that we did not have to modify Streamlet's proposing rule,
which is in line with the generic \Oflex construction.

\subsection{Security Analysis}
\label{sec:optflex-streamlet-analysis}

To build intuition, in \cref{fig:protocol-views-oflex-streamlet}, we show an example of how \Oflex-Streamlet provides safety
with
$\tS_{k} = n-1$ 
for clients that confirm according to the rule $\PiCmod(q_{k}=n)$.
Such a client confirms the block $B$ 
after seeing the blocks $A,B,C,D$
with the required votes.
When the client sees a replica vote on the block $D$
(\cref{fig:protocol-views-oflex-streamlet}~\DiagramStep{2}), it knows that this replica, if honest, must have seen the blocks $A,B,C$ notarized (honest replicas only vote for blocks whose parent chain they have seen notarized), and therefore must have \olocked the block $B$
(\cref{fig:protocol-views-oflex-streamlet}~\DiagramStep{1}), and will thus never vote for a block that is inconsistent with $B$
(\cref{fig:protocol-views-oflex-streamlet}~\DiagramStep{3}).
Thus,
if a client sees all $n$ replicas vote for block $D$,
then it knows that unless all these replicas are adversary, a block inconsistent with $B$ cannot obtain votes from all replicas.
Thus, clients that confirm blocks with $q_k=n$
remain safe even when $f = n-1$ replicas
are adversary.

To intuit liveness,
recall that the requirement $\tS_{k} \geq \tL_{k}$ implies $\tL_{k} < n/3$.
Thus, clients can expect liveness only when $f < n/3$.
Tracing Streamlet's safety analysis
in this regime shows that every replica's $\algolock$
is a prefix of every longest notarized chain
in its view at all times.
As a result, when $f < n/3$,
the $\algolock$ constraint in \Oflex-Streamlet is 
inactive (\cf \cref{loc:oflex-streamlet-vote}),
and \Oflex-Streamlet `behaves like' Streamlet.
Therefore, for a client with $q_{k} \leq n - \tL_{k}$, if $f \leq \tL_{k} < n/3$,
after sufficiently many successive honest leaders,
the client eventually sees blocks with $q_{k}$ votes, in particular, blocks that satisfy the confirmation rule.

\begin{theorem}
    \label{thm:oflex-streamlet-security}
    \Oflex-Streamlet provides optimal flexible consensus.
\end{theorem}
\begin{proof}
    We prove safety and liveness in \cref{thm:oflex-streamlet-safety,thm:oflex-streamlet-liveness}, respectively, for clients who choose their confirmation quorums appropriately.
    For any client $k$ choosing resilience pair $(\tL_{k},\tS_{k})$ such that $2\tL_{k} + \tS_{k} < n$ and $\tL_{k} \leq \tS_{k}$,
    there exists a confirmation quorum
    $q_{k}$
    that satisfies the requirements of \cref{thm:oflex-streamlet-safety,thm:oflex-streamlet-liveness}, \ie,
    $(n+\tS_{k})/2 < q_{k} \leq n - \tL_{k}$.
    Thus, by changing the confirmation quorum $q_{k}$, clients can achieve any $(\tL_{k},\tS_{k})$ that satisfies $2\tL_{k} + \tS_{k} < n$ and $\tL_k \leq \tS_k$.
\end{proof}

\begin{lemma}[Safety]
    \label{thm:oflex-streamlet-safety}
    For all clients $k, k'$ such that
    $q_{k} > (n+\tS_{k})/2$
    and $q_{k'} > (n+\tS_{k'})/2$,
    if $f \leq \min\{\tS_{k}, \tS_{k'}\}$,
    then for all times $\tau$, $\tau'$,
    $\LOG{k}{\tau}$ and $\LOG{k'}{\tau'}$ are consistent.
\end{lemma}
\begin{proof}
    Suppose, for contradiction, that there exist two clients $k,k'$ and two time instants $\tau,\tau'$ such that $\LOG{k}{\tau}$ and $\LOG{k'}{\tau'}$ are not consistent.
    Then, there must exist two notarized chains that satisfy the respective confirmation rules of the clients.
    So in the first chain, there
    are four blocks $A,B,C,D$
    from
    epochs $e-x-2, e-x-1, e-x, e$ for some $x\geq 1$,
    where $A, B, C$ are adjacent,
    and $D$ is a descendant of $C$
    and has received $q_{k}$ votes in client $k$'s view.
    In the second chain, there are four blocks $H,I,J,K$
    from
    epochs $e'-y-2, e'-y-1, e'-y, e'$ for some $y\geq 1$,
    where $H, I, J$ are adjacent,
    and $K$ is a descendant of $J$
    and has received $q_{k'}$ votes in client $k'$'s view.
    Furthermore, $B$ and $I$ are inconsistent.
    See \cref{fig:protocol-views-oflex-streamlet} for illustration.

    Without loss of generality, we may assume that $e' \geq e$.
    Honest replicas only vote for a block if they have seen its parent chain notarized.
    Therefore, an honest replica who voted for the block $D$ in epoch $e$
    (\cf \cref{fig:protocol-views-oflex-streamlet}~\DiagramStep{2})
    must have seen blocks $A,B,C$ notarized.
    Furthermore, an honest replica only votes for the block $D$ if it is also a descendant of the replica's \olock.
    Therefore, 
    either the block $B$ is a descendant of the replica's \olock, or the replica must have already \olocked $B$. In either case, the replica will have \olocked the block $B$ by the end of epoch $e$
    (\cf \cref{fig:protocol-views-oflex-streamlet}~\DiagramStep{1}).
    An honest replica that votes for the block $D$ in epoch $e$ does not vote for any other block in epoch $e$. Furthermore, it does not vote for any block that is inconsistent with block $B$ in epochs $>e$, due to its \olock on $B$
    (\cf \cref{fig:protocol-views-oflex-streamlet}~\DiagramStep{3}).
    Since $K$ is a descendant of $I$, but $I$ and $B$ are inconsistent, we conclude that no honest replica votes for both blocks $D$ and $K$.

    Since blocks $D$ and $K$ have received at least $q_{k}$ and $q_{k'}$ votes respectively, at least $q_{k} + q_{k'} - n$ replicas must have voted for both blocks.
    Due to the preceding argument, these must all be adversary replicas, so $f \geq q_{k} + q_{k'} - n$.
    Due to the choice of quorums $q_{k} > (n+\tS_{k})/2$ and $q_{k'} > (n+\tS_{k'})/2$, safety violation between clients $k,k'$ requires $f \geq (\tS_{k} + \tS_{k'})/2 + 1$ adversary replicas.
    This is a contradiction to the assumption that $f \leq \min\{\tS_{k},\tS_{k'}\}$.
\end{proof}

The key new step towards proving liveness of \Oflex-Streamlet is proving that
when $f < n/3$
(recall that we only require liveness for clients with $f \leq \tL_{k} < n/3$), the additional constraint on voting of \Oflex-Streamlet is never active (\cref{lem:voting}).
This follows from a safety property of Streamlet (which also holds in \Oflex-Streamlet) that once a block is confirmed according to Streamlet (or \olocked according to \Oflex-Streamlet), no block inconsistent with it ever gets notarized (\cref{lem:no-conflict-notarize}).
Thus, under the regime of interest for liveness ($f < n/3$), the voting rule of \Oflex-Streamlet behaves exactly like that of Streamlet,
and the rest of the proof for liveness follows using techniques from~\cite{streamlet}, except that \Oflex-Streamlet requires one additional epoch to confirm blocks compared to Streamlet (\cref{lem:liveness-intermediate}).
Liveness (\cref{thm:oflex-streamlet-liveness}) follows immediately.
Proofs of \Cref{lem:no-conflict-notarize,lem:liveness-intermediate} are given in \cref{sec:oflex-streamlet-proofs} as they mostly follow steps from~\cite{streamlet}.

\begin{lemma}[Liveness]
    \label{thm:oflex-streamlet-liveness}
    For every $\tx$ input to all honest replicas, eventually, for all clients $k$
    such that $q_{k} \leq n - \tL_{k}$ (quorum choice), $2\tL_{k} + \tS_{k} < n$, $\tL_{k} \leq \tS_{k}$ (optimal flexible resiliences), and $f \leq \tL_{k}$,
    $\tx \in \LOG{k}{}$.
\end{lemma}

The following lemmas build up to the proof of \cref{thm:oflex-streamlet-liveness}.

\begin{lemma}[{\cf \cite[Lem.~2]{streamlet}}]
    \label{lem:no-conflict-notarize}
    If $f < n/3$, then if some honest replica sees three adjacent blocks $A,B,C$ with consecutive epoch numbers on a notarized blockchain, then there cannot be a block $F \neq B$ at the same height as $B$ that is also notarized in an honest replica's view.
\end{lemma}

Proof is given in \cref{sec:oflex-streamlet-proofs}.

\begin{lemma}
    \label{lem:voting}
    If $f < n/3$,
    then if a block $B$'s parent's chain is one of the longest notarized chains in a replica's view, then $B$ is also a descendant of that replica's \olock.
\end{lemma}

\begin{proof}
    Suppose that in epoch $e$, a block $B$ is proposed whose parent chain is one of the longest notarized chains seen by an honest replica at epoch $e$.
    Let $B'$ be this honest replica's \olock as of epoch $e$.
    If $B'$ is the genesis block, then $B$ must be its descendant.
    Otherwise, at the end of some epoch $e' < e$, this replica saw three adjacent blocks $A',B',C'$ with consecutive epoch numbers on a notarized blockchain.
    For a contradiction, suppose that $B$ is not a descendant of $B'$.
    Since the replica saw block $C'$ notarized at epoch $e'$, it must be that
    $B$ is at a greater height than $B'$.
    Then there must be a block
    on the parent chain of $B$ at the same height as $B'$
    that is also notarized. Due to \cref{lem:no-conflict-notarize}, this is a contradiction.
\end{proof}

\begin{lemma}[{\cf \cite[Thm.~4]{streamlet}}]
    \label{lem:liveness-intermediate}
    After $\GST$, suppose that there are $6$ consecutive epochs $e, e+1, ..., e+5$ with honest leaders. Then by the beginning of epoch $e+6$,
    every client $k$ such that $f \leq \tL_{k} < n/3$ and $q_{k} \leq n - \tL_{k}$
    will have confirmed a new block which it had not confirmed at the beginning of epoch $e$. Moreover, this new block was proposed by an honest leader.
\end{lemma}

Proof is given in \cref{sec:oflex-streamlet-proofs}.

\begin{proof}[Proof of \cref{thm:oflex-streamlet-liveness}]
    The conditions $2\tL_{k} + \tS_{k} < n$ and $\tL_{k} \leq \tS_{k}$ imply $\tL_{k} < 1/3$.
    Then, \cref{lem:liveness-intermediate} directly implies liveness.
    The conditions required in \cref{lem:liveness-intermediate} occur eventually, \ie, after $\GST$ and after $\left(\frac{n}{n-f}\right)^6$ epochs in expectation.
    Due to the honest replicas' propose rule, the new block by an honest leader (referred to in \cref{lem:liveness-intermediate}) includes any $\tx$ input to all honest replicas, that were not already in its parent chain.
    This block and its parent chain are confirmed by all clients $k$ for which $q_k \leq n - \tL_{k}$.
\end{proof}

\subsection{Comparison with \FBFT}
\label{sec:optflex-streamlet-fbft-comparison}

\import{./figures/}{protocol-views-fbft.tex}%

We briefly describe
a straightforward
and truthful
adaptation\footnote{While the \FBFT protocol of~\cite{fbft} is inspired by HotStuff, the similarity between Streamlet and HotStuff~\cite{hotstuff,fbft2,streamlethotstuff} enables this adaptation.}
of \FBFT~\cite{fbft} to Streamlet
(which we call \FBFT-Streamlet)
to facilitate a clear comparison of
\FBFT's and \Oflex's paradigms.
FBFT-Streamlet is identical to Streamlet, except for its confirmation rule:
``When client $k$ sees three adjacent blocks $A,B,C$ from consecutive epochs in a notarized chain,
\uline{each with $q_{k}$ votes},
then it confirms $B$.''
Note that this is Streamlet's original confirmation rule, except with $q_k$ instead of $q$ votes for each block.

First,
for liveness,
observe that with $q_k \geq q$,
both \FBFT-Streamlet and \Oflex-Streamlet
require
$f \leq \tL_k \leq n - q_k$.
For safety, however,
\FBFT-Streamlet requires
$f \leq \tS_k < q_{k} + q - n$,
while \Oflex-Streamlet only requires
$f \leq \tS_k < 2q_{k} - n$.
This enables \Oflex-Streamlet's optimal flexibility.

To understand where \FBFT-Streamlet's sub-optimal bound
on $\tS_k$ comes from,
consider
how \FBFT-Streamlet's safety breaks
in the example of $q = \frac{2n}{3}+1$, clients with $q_{k} = n$,
and $f$ just exceeding the clients' safety resilience in FBFT-Streamlet,
\ie, $f = q_{k} + q - n = \frac{2n}{3}+1$
(illustrated in \cref{fig:protocol-views-fbft}):
Suppose blocks $A,B,C$ received votes from all replicas, leading a client $k$ with $q_k = n$ to confirm block $B$.
Other clients may not hear of these blocks and votes for a while (before $\GST$).
Since the adversary controls $\frac{2n}{3}+1$ replicas,
it can, by its own efforts,
notarize inconsistent blocks $E,F,G$.
In the next slot, the leader proposes block $H$ with parent $G$. This indeed follows the protocol rules because $G$ is `the tip of one of the longest notarized chains'. When $H$ is proposed, all honest replicas will vote for $H$ because $H$ is the only block proposed by the leader and its parent $G$ is the tip of one of the longest notarized chains
(\cf \cref{sec:optflex-streamlet-recap}).
The adversary replicas also vote for $H$ which results in $n$ votes for $H$ in total. Similarly, leaders propose blocks $I,J$ and they too gather $n$ votes each. Ultimately, the block $I$ satisfies the confirmation rule for $q_{k'} = n$ in another client $k'$ (who has not heard of $B$). This is a safety violation
between $k$ and $k'$ because $B$ and $I$ are inconsistent.

Key issue at hand:
the adversary was able to
by-pass the heavy quorum $q_k$ on $A,B,C$
with a light quorum $q$ on $E,F,G$,
and honest replicas subsequently helped the adversary
achieve a conflicting heavy quorum $q_{k'}$
on $H,I,J$.

To rule this out, \FBFT-Streamlet
imposes $f \leq \tS_k < q_{k} + q - n$,
which (by a standard quorum intersection argument)
ensures that no block $F$ can reach quorum $q$
on the same height as $B$ when $A, B, C$ reach quorum $q_k$
(red dashed in \cref{fig:protocol-views-fbft}).
Thus, no by-passing, and no conflicting heavy quorum.
But the constraint turns out to be sub-optimal.

Instead, \Oflex-Streamlet addresses the key issue
with \olocking and \ovoting,
leaving the adversary the power to reach light quorum $q$
on a block $F$ conflicting with $B$,
but preventing honest replicas from contributing to the conflicting
heavy quorum $q_{k'}$ that would be necessary for
a conflicting confirmation.
See \cref{fig:protocol-views-oflex-streamlet} for \Oflex-Streamlet in a situation similar to \cref{fig:protocol-views-fbft}, where, however, honest replicas do not vote for blocks $H,I,J$.
First, when honest replicas see blocks $A,B,C$ notarized ($q = \frac{2n}{3}+1$ votes are enough for this), they \olock the block $B$
(\cref{fig:protocol-views-oflex-streamlet}~\DiagramStep{1}).
But, we also require the client to know that honest replicas have \olocked $B$, so that the client can safely confirm $B$,
knowing honest replicas will not contribute
votes in favor of a conflicting block
(\cref{fig:protocol-views-oflex-streamlet}~\DiagramStep{3}).
This is why in \Oflex-Streamlet, the confirmation rule requires the client to see an additional block $D$ with $n$ votes
(\cref{fig:protocol-views-oflex-streamlet}~\DiagramStep{2}),
from which the client infers that all honest replicas must have \olocked block $B$.
Thus, it would require $n$ adversary replicas to confirm an inconsistent block such as $I$.
In \Oflex-Streamlet, only $\frac{2n}{3}+1$ votes are required for the blocks $A,B,C$ and $q_{k}$ votes are required only for block $D$ (\cref{fig:protocol-views-oflex-streamlet}), which suffice for the client to infer replicas' \olocks.

%% file: figures/alg-oflex-streamlet.tex
\begin{algorithm}[tb]
    \caption{\Oflex-Streamlet protocol $\PImod$ (changes relative to \cref{alg:streamlet} are highlighted \textcolor{jnSUDigitalGreen}{green})}
    \label{alg:oflex-streamlet}
    \begin{algorithmic}[1]
        \footnotesize
        \LineComment{\textcolor{jnSUDigitalGreen}{$(\algolock \prec B) \triangleq$ chain of $\algolock$ is a prefix of chain of $B$}}%
        \medskip%
        \LineComment{Replica-side logic \textcolor{jnSUDigitalGreen}{$\PiRmod$}}
        \smallskip
        \On{\Call{init}{\null}}
            \State $\CB, \CV \gets \{ B_0 \}, \{ \}$
                \Comment{Background task: receive blocks and votes into $\CB$ and $\CV$, respectively, subject to canonical validation (\cref{alg:streamlet})}
            \State \textcolor{jnSUDigitalGreen}{$\algolock \gets B_0$}
                \Comment{\textcolor{jnSUDigitalGreen}{Initialize \olock to genesis block $B_0$}}
        \EndOn
        \smallskip
        \For{each epoch $e=1,2,3,...$}
            \LineComment{\textbf{Propose} (done by epoch leader at the start of the epoch)}
            \State $B' \gets$ tip of any one longest notarized chain in $(\CB, \CV)$
            \State $h \gets \operatorname{Hash}(B')$
            \State $\txs \gets$ transactions not present in chain of $B'$
            \State Sign and broadcast block $(h, e, \txs)$
            \smallskip
            \LineComment{\textbf{Vote} (done by all replicas \emph{once} during the epoch)}
            \State $B \gets$ first block from epoch $e$ in $\CB$ signed by epoch leader
            \State $B' \gets$ parent block of $B$ in $\CB$
            \If{$B'$ is tip of any longest notarized chain in $(\CB, \CV)$}
                \label{loc:oflex-streamlet-vote}
                \If{\textcolor{jnSUDigitalGreen}{$\algolock \prec B'$}}
                    \State $h \gets \operatorname{Hash}(B)$
                    \State Sign and broadcast vote $h$
                \EndIf
            \EndIf
            \smallskip
            \LineComment{\textcolor{jnSUDigitalGreen}{\textbf{\Olock} (done by all replicas throughout the epoch)}}
            \label{loc:oflex-streamlet-olock}
            \If{\textcolor{jnSUDigitalGreen}{$(\CB, \CV)$ contains a notarized chain with three adjacent blocks $A, B, C$ from consecutive epochs, and $\algolock \prec B$}}
                \State \textcolor{jnSUDigitalGreen}{$\algolock \gets B$}
            \EndIf
        \EndFor
        \medskip
        \LineComment{Client-side confirmation rule \textcolor{jnSUDigitalGreen}{$\PiCmod$} \textcolor{jnSUDigitalGreen}{for resilience pair $(\tL_k, \tS_k)$}}
        \smallskip
        \On{\Call{init}{\textcolor{jnSUDigitalGreen}{$q_{k}$}} \textcolor{jnSUDigitalGreen}{where $q_k$ shall satisfy $(n+\tS_{k})/2 < q_{k} \leq n - \tL_{k}$}}
            \State \textcolor{jnSUDigitalGreen}{Set local quorum size $q_{k}$}
            \State $\CB, \CV \gets \{ B_0 \}, \{ \}$
                \Comment{Background task: receive blocks and votes into $\CB$ and $\CV$, respectively, subject to canonical validation (\cref{alg:streamlet})}
        \EndOn
        \smallskip
        \LineComment{\textbf{Confirmation}}
        \label{loc:oflex-streamlet-conf}
        \If{$(\CB, \CV)$ contains a notarized chain with three adjacent blocks $A, B, C$ from consecutive epochs, \textcolor{jnSUDigitalGreen}{and $(\CB, \CV)$ contains a block $D$ such that $C \prec D$ and $D$ has received $q_k$ votes}}
            \State Choose $A, B, C$ as such blocks with maximum height
            \State $\LOG{}{} \gets$ sequence of transactions as ordered in chain of $B$
        \EndIf
    \end{algorithmic}
\end{algorithm}

%% file: figures/protocol-views-oflex-streamlet.tex
\import{./}{protocol-views-_utils.tex}%
\begin{figure}[tb]
    \centering
    \begin{tikzpicture}[fig_protocol_views]%
        \footnotesize%
        \scriptspacing%

        \node [block] (b0) at (0,0) {};
        \draw [link,densely dotted] (b0) -- ++([xshift=-0.65cm]0,0);

        \node [block] (bA) at (1,+1) {\blockLabel{A}};
        \draw [link] (bA) -- (b0);

        \node [vote] at ([pos11]1,+3) {\tangoicon[height=0.3cm]{mimetypes_application-certificate}};
        \node [vote,opacity=0.2] at ([pos12]1,+3) {\tangoicon[height=0.3cm]{mimetypes_application-certificate}};
        \node [vote,opacity=0.2] at ([pos13]1,+3) {\tangoicon[height=0.3cm]{mimetypes_application-certificate}};
        \node [vote] at ([pos21]1,+3) {\tangoicon[height=0.3cm]{mimetypes_application-certificate}};
        \node [vote] at ([pos22]1,+3) {\tangoicon[height=0.3cm]{mimetypes_application-certificate}};
        \node [vote] at ([pos23]1,+3) {\tangoicon[height=0.3cm]{mimetypes_application-certificate}};
        \node [vote] at ([pos31]1,+3) {\tangoicon[height=0.3cm]{mimetypes_application-certificate}};
        \node [vote] at ([pos32]1,+3) {\tangoicon[height=0.3cm]{mimetypes_application-certificate}};
        \node [vote] at ([pos33]1,+3) {\tangoicon[height=0.3cm]{mimetypes_application-certificate}};
        \node [votelabel,above] at (1,+3) {$\frac{2n}{3}+1$};

        \node [block] (bB) at (2,+1) {\blockLabel{B}};
        \draw [link] (bB) -- (bA);

        \node [vote] at ([pos11]2,+3) {\tangoicon[height=0.3cm]{mimetypes_application-certificate}};
        \node [vote,opacity=0.2] at ([pos12]2,+3) {\tangoicon[height=0.3cm]{mimetypes_application-certificate}};
        \node [vote,opacity=0.2] at ([pos13]2,+3) {\tangoicon[height=0.3cm]{mimetypes_application-certificate}};
        \node [vote] at ([pos21]2,+3) {\tangoicon[height=0.3cm]{mimetypes_application-certificate}};
        \node [vote] at ([pos22]2,+3) {\tangoicon[height=0.3cm]{mimetypes_application-certificate}};
        \node [vote] at ([pos23]2,+3) {\tangoicon[height=0.3cm]{mimetypes_application-certificate}};
        \node [vote] at ([pos31]2,+3) {\tangoicon[height=0.3cm]{mimetypes_application-certificate}};
        \node [vote] at ([pos32]2,+3) {\tangoicon[height=0.3cm]{mimetypes_application-certificate}};
        \node [vote] at ([pos33]2,+3) {\tangoicon[height=0.3cm]{mimetypes_application-certificate}};
        \node [votelabel,above] at (2,+3) {$\frac{2n}{3}+1$};

        \node [block] (bC) at (3,+1) {\blockLabel{C}};
        \draw [link] (bC) -- (bB);

        \node [vote] at ([pos11]3,+3) {\tangoicon[height=0.3cm]{mimetypes_application-certificate}};
        \node [vote,opacity=0.2] at ([pos12]3,+3) {\tangoicon[height=0.3cm]{mimetypes_application-certificate}};
        \node [vote,opacity=0.2] at ([pos13]3,+3) {\tangoicon[height=0.3cm]{mimetypes_application-certificate}};
        \node [vote] at ([pos21]3,+3) {\tangoicon[height=0.3cm]{mimetypes_application-certificate}};
        \node [vote] at ([pos22]3,+3) {\tangoicon[height=0.3cm]{mimetypes_application-certificate}};
        \node [vote] at ([pos23]3,+3) {\tangoicon[height=0.3cm]{mimetypes_application-certificate}};
        \node [vote] at ([pos31]3,+3) {\tangoicon[height=0.3cm]{mimetypes_application-certificate}};
        \node [vote] at ([pos32]3,+3) {\tangoicon[height=0.3cm]{mimetypes_application-certificate}};
        \node [vote] at ([pos33]3,+3) {\tangoicon[height=0.3cm]{mimetypes_application-certificate}};
        \node [votelabel,above] at (3,+3) {$\frac{2n}{3}+1$};

        \node [block] (bD) at (4,+1) {\blockLabel{D}};
        \draw [link] (bD) -- (bC);

        \node [vote] at ([pos11]4,+3) {\tangoicon[height=0.3cm]{mimetypes_application-certificate}};
        \node [vote] at ([pos12]4,+3) {\tangoicon[height=0.3cm]{mimetypes_application-certificate}};
        \node [vote] (voteYes) at ([pos13]4,+3) {\tangoicon[height=0.3cm]{mimetypes_application-certificate}};
        \node [vote] at ([pos21]4,+3) {\tangoicon[height=0.3cm]{mimetypes_application-certificate}};
        \node [vote] at ([pos22]4,+3) {\tangoicon[height=0.3cm]{mimetypes_application-certificate}};
        \node [vote] at ([pos23]4,+3) {\tangoicon[height=0.3cm]{mimetypes_application-certificate}};
        \node [vote] at ([pos31]4,+3) {\tangoicon[height=0.3cm]{mimetypes_application-certificate}};
        \node [vote] at ([pos32]4,+3) {\tangoicon[height=0.3cm]{mimetypes_application-certificate}};
        \node [vote] at ([pos33]4,+3) {\tangoicon[height=0.3cm]{mimetypes_application-certificate}};
        \node [votelabel,above] at (4,+3) {$n$};

        \coordinate (bE) at (1,-1);

        \node [block] (bF) at (2,-1) {\blockLabel{F}};
        \draw [link,densely dotted] (bF) -- (bE) -- (b0);

        \node [vote] at ([pos11]2,-3) {\tangoicon[height=0.3cm]{mimetypes_application-certificate}};
        \node [vote] at ([pos12]2,-3) {\tangoicon[height=0.3cm]{mimetypes_application-certificate}};
        \node [vote] at ([pos13]2,-3) {\tangoicon[height=0.3cm]{mimetypes_application-certificate}};
        \node [vote] at ([pos21]2,-3) {\tangoicon[height=0.3cm]{mimetypes_application-certificate}};
        \node [vote] at ([pos22]2,-3) {\tangoicon[height=0.3cm]{mimetypes_application-certificate}};
        \node [vote] at ([pos23]2,-3) {\tangoicon[height=0.3cm]{mimetypes_application-certificate}};
        \node [vote] at ([pos31]2,-3) {\tangoicon[height=0.3cm]{mimetypes_application-certificate}};
        \node [vote,opacity=0.2] at ([pos32]2,-3) {\tangoicon[height=0.3cm]{mimetypes_application-certificate}};
        \node [vote,opacity=0.2] at ([pos33]2,-3) {\tangoicon[height=0.3cm]{mimetypes_application-certificate}};
        \node [votelabel,below] at (2,-3) {$\frac{2n}{3}+1$};

        \coordinate (bG) at (3,-1);

        \node [block] (bH) at (4,-1) {\blockLabel{H}};
        \draw [link,densely dotted] (bH) -- (bG) -- (bF);

        \node [vote] at ([pos11]4,-3) {\tangoicon[height=0.3cm]{mimetypes_application-certificate}};
        \node [vote] at ([pos12]4,-3) {\tangoicon[height=0.3cm]{mimetypes_application-certificate}};
        \node [vote] at ([pos13]4,-3) {\tangoicon[height=0.3cm]{mimetypes_application-certificate}};
        \node [vote] at ([pos21]4,-3) {\tangoicon[height=0.3cm]{mimetypes_application-certificate}};
        \node [vote] at ([pos22]4,-3) {\tangoicon[height=0.3cm]{mimetypes_application-certificate}};
        \node [vote] at ([pos23]4,-3) {\tangoicon[height=0.3cm]{mimetypes_application-certificate}};
        \node [vote] at ([pos31]4,-3) {\tangoicon[height=0.3cm]{mimetypes_application-certificate}};
        \node [vote,opacity=0.2] at ([pos32]4,-3) {\tangoicon[height=0.3cm]{mimetypes_application-certificate}};
        \node [vote,opacity=0.2] at ([pos33]4,-3) {\tangoicon[height=0.3cm]{mimetypes_application-certificate}};
        \node [votelabel,below] at (4,-3) {$\frac{2n}{3}+1$};

        \node [block] (bI) at (5,-1) {\blockLabel{I}};
        \draw [link] (bI) -- (bH);

        \node [vote] at ([pos11]5,-3) {\tangoicon[height=0.3cm]{mimetypes_application-certificate}};
        \node [vote] at ([pos12]5,-3) {\tangoicon[height=0.3cm]{mimetypes_application-certificate}};
        \node [vote] at ([pos13]5,-3) {\tangoicon[height=0.3cm]{mimetypes_application-certificate}};
        \node [vote] at ([pos21]5,-3) {\tangoicon[height=0.3cm]{mimetypes_application-certificate}};
        \node [vote] at ([pos22]5,-3) {\tangoicon[height=0.3cm]{mimetypes_application-certificate}};
        \node [vote] at ([pos23]5,-3) {\tangoicon[height=0.3cm]{mimetypes_application-certificate}};
        \node [vote] at ([pos31]5,-3) {\tangoicon[height=0.3cm]{mimetypes_application-certificate}};
        \node [vote,opacity=0.2] at ([pos32]5,-3) {\tangoicon[height=0.3cm]{mimetypes_application-certificate}};
        \node [vote,opacity=0.2] at ([pos33]5,-3) {\tangoicon[height=0.3cm]{mimetypes_application-certificate}};
        \node [votelabel,below] at (5,-3) {$\frac{2n}{3}+1$};

        \node [block] (bJ) at (6,-1) {\blockLabel{J}};
        \draw [link] (bJ) -- (bI);

        \node [vote] at ([pos11]6,-3) {\tangoicon[height=0.3cm]{mimetypes_application-certificate}};
        \node [vote] at ([pos12]6,-3) {\tangoicon[height=0.3cm]{mimetypes_application-certificate}};
        \node [vote] at ([pos13]6,-3) {\tangoicon[height=0.3cm]{mimetypes_application-certificate}};
        \node [vote] at ([pos21]6,-3) {\tangoicon[height=0.3cm]{mimetypes_application-certificate}};
        \node [vote] at ([pos22]6,-3) {\tangoicon[height=0.3cm]{mimetypes_application-certificate}};
        \node [vote] at ([pos23]6,-3) {\tangoicon[height=0.3cm]{mimetypes_application-certificate}};
        \node [vote] at ([pos31]6,-3) {\tangoicon[height=0.3cm]{mimetypes_application-certificate}};
        \node [vote,opacity=0.2] at ([pos32]6,-3) {\tangoicon[height=0.3cm]{mimetypes_application-certificate}};
        \node [vote,opacity=0.2] at ([pos33]6,-3) {\tangoicon[height=0.3cm]{mimetypes_application-certificate}};
        \node [votelabel,below] at (6,-3) {$\frac{2n}{3}+1$};

        \node [block] (bK) at (7,-1) {\blockLabel{K}};
        \draw [link] (bK) -- (bJ);

        \node [vote] at ([pos11]7,-3) {\tangoicon[height=0.3cm]{mimetypes_application-certificate}};
        \node [vote] at ([pos12]7,-3) {\tangoicon[height=0.3cm]{mimetypes_application-certificate}};
        \node [vote] (voteNo) at ([pos13]7,-3) {\tangoicon[height=0.3cm]{mimetypes_application-certificate}};
        \node [vote] at ([pos21]7,-3) {\tangoicon[height=0.3cm]{mimetypes_application-certificate}};
        \node [vote] at ([pos22]7,-3) {\tangoicon[height=0.3cm]{mimetypes_application-certificate}};
        \node [vote] at ([pos23]7,-3) {\tangoicon[height=0.3cm]{mimetypes_application-certificate}};
        \node [vote] at ([pos31]7,-3) {\tangoicon[height=0.3cm]{mimetypes_application-certificate}};
        \node [vote] at ([pos32]7,-3) {\tangoicon[height=0.3cm]{mimetypes_application-certificate}};
        \node [vote] at ([pos33]7,-3) {\tangoicon[height=0.3cm]{mimetypes_application-certificate}};
        \node [votelabel,below] at (7,-3) {$n$};

        \node [locked] (lockb0) at (b0) {\notoemoji[height=0.4cm]{lock}};
        \node [locked] (lockbA) at (bA) {\notoemoji[height=0.4cm]{lock}};
        \node [locked] (lockbB) at (bB) {\notoemoji[height=0.4cm]{lock}};
        \node at (7,-3) {\notoemoji[height=0.7cm]{prohibited}};
        \node [votelabel,below] at (7,-3) {\notoemoji[height=0.3cm]{prohibited}};

        \node (validator) at (8.5,0) {\notoemoji[height=0.6cm]{technologist}};
        \draw [-latex,shorten >=-5pt,dashed] (validator) -- (voteYes) node [midway] {\DiagramStep{2}};
        \draw [-latex,shorten >=-3pt,dashed] (validator) -- ([yshift=5pt]lockbB.south east) node [midway] {\DiagramStep{1}};
        \draw [-latex,shorten >=-5pt,dashed] (validator) -- (voteNo) node [midway] {\notoemoji[height=1em]{prohibited}} node [midway,align=left,right,xshift=4pt,red] {\DiagramStep{3}};

        \node [cloud callout,cloud puffs=15,aspect=2,cloud puff arc=120,draw,callout absolute pointer={(8.3,1)}] at (8,2.5) {\DiagramStep{1} \notoemoji[height=0.9em]{lock}\blockLabel{B}};

    \end{tikzpicture}%
    \caption[]{%
        Illustration of \Oflex-Streamlet (\cf \cref{fig:protocol-views-oflex-generic}).
        Each \tangoicon{mimetypes_application-certificate} represents a vote,
        and the number of votes for each block is indicated.
        * \emph{Protocol rules:}
        \DiagramStep{1}~If a replica sees a block $B$
        that would have been confirmed according to
        Streamlet's original confirmation rule $\PiC$
        (\ie, adjacent blocks $A,B,C$ from consecutive epochs in notarized chain),
        then the replica \olocks $B$ if $B$ extends its current \olock.
        \DiagramStep{2}~%
        From a replica's vote for $D$, clients can infer that the replica,
        if honest, 
        must be \olocked on $B$.
        \DiagramStep{3}~%
        This, in turn, means the replica, if honest, will never vote for blocks
        inconsistent with $B$.
        * \emph{Safety intuition:}
        A client $k$ with high quorum $q_k=n$ confirms $B$ when it sees $q_k$ votes on $D$.
        Due to steps \DiagramStep{1} to \DiagramStep{3},
        no block $K$ conflicting with $B$ can receive $n$ votes
        and get confirmed by another client $k'$ with $q_{k'}=n$,
        \emph{unless all replicas are adversary},
        implying $\tS_k = \tS_{k'} = n-1$.
    }%
    \label{fig:protocol-views-oflex-streamlet}%
\end{figure}

%% file: figures/protocol-views-fbft.tex
\import{./}{protocol-views-_utils.tex}%
\begin{figure}[tb]
    \centering
    \begin{tikzpicture}[fig_protocol_views]%
        \footnotesize%
        \scriptspacing%

        \node [block] (b0) at (0,0) {};
        \draw [link,densely dotted] (b0) -- ++([xshift=-0.65cm]0,0);

        \node [block] (bA) at (1,+1) {\blockLabel{A}};
        \draw [link] (bA) -- (b0);

        \node [vote] at ([pos11]1,+3) {\tangoicon[height=0.3cm]{mimetypes_application-certificate}};
        \node [vote] at ([pos12]1,+3) {\tangoicon[height=0.3cm]{mimetypes_application-certificate}};
        \node [vote] at ([pos13]1,+3) {\tangoicon[height=0.3cm]{mimetypes_application-certificate}};
        \node [vote] at ([pos21]1,+3) {\tangoicon[height=0.3cm]{mimetypes_application-certificate}};
        \node [vote] at ([pos22]1,+3) {\tangoicon[height=0.3cm]{mimetypes_application-certificate}};
        \node [vote] at ([pos23]1,+3) {\tangoicon[height=0.3cm]{mimetypes_application-certificate}};
        \node [vote] at ([pos31]1,+3) {\tangoicon[height=0.3cm]{mimetypes_application-certificate}};
        \node [vote] at ([pos32]1,+3) {\tangoicon[height=0.3cm]{mimetypes_application-certificate}};
        \node [vote] at ([pos33]1,+3) {\tangoicon[height=0.3cm]{mimetypes_application-certificate}};
        \node [votelabel,above] at (1,+3) {$n$};

        \node [block] (bB) at (2,+1) {\blockLabel{B}};
        \draw [link] (bB) -- (bA);

        \node [vote] at ([pos11]2,+3) {\tangoicon[height=0.3cm]{mimetypes_application-certificate}};
        \node [vote] at ([pos12]2,+3) {\tangoicon[height=0.3cm]{mimetypes_application-certificate}};
        \node [vote] at ([pos13]2,+3) {\tangoicon[height=0.3cm]{mimetypes_application-certificate}};
        \node [vote] at ([pos21]2,+3) {\tangoicon[height=0.3cm]{mimetypes_application-certificate}};
        \node [vote] at ([pos22]2,+3) {\tangoicon[height=0.3cm]{mimetypes_application-certificate}};
        \node [vote] at ([pos23]2,+3) {\tangoicon[height=0.3cm]{mimetypes_application-certificate}};
        \node [vote] at ([pos31]2,+3) {\tangoicon[height=0.3cm]{mimetypes_application-certificate}};
        \node [vote] at ([pos32]2,+3) {\tangoicon[height=0.3cm]{mimetypes_application-certificate}};
        \node [vote] at ([pos33]2,+3) {\tangoicon[height=0.3cm]{mimetypes_application-certificate}};
        \node [votelabel,above] (qiTop) at (2,+3) {$n$};

        \node [block] (bC) at (3,+1) {\blockLabel{C}};
        \draw [link] (bC) -- (bB);

        \node [vote] at ([pos11]3,+3) {\tangoicon[height=0.3cm]{mimetypes_application-certificate}};
        \node [vote] at ([pos12]3,+3) {\tangoicon[height=0.3cm]{mimetypes_application-certificate}};
        \node [vote] at ([pos13]3,+3) {\tangoicon[height=0.3cm]{mimetypes_application-certificate}};
        \node [vote] at ([pos21]3,+3) {\tangoicon[height=0.3cm]{mimetypes_application-certificate}};
        \node [vote] at ([pos22]3,+3) {\tangoicon[height=0.3cm]{mimetypes_application-certificate}};
        \node [vote] at ([pos23]3,+3) {\tangoicon[height=0.3cm]{mimetypes_application-certificate}};
        \node [vote] at ([pos31]3,+3) {\tangoicon[height=0.3cm]{mimetypes_application-certificate}};
        \node [vote] at ([pos32]3,+3) {\tangoicon[height=0.3cm]{mimetypes_application-certificate}};
        \node [vote] at ([pos33]3,+3) {\tangoicon[height=0.3cm]{mimetypes_application-certificate}};
        \node [votelabel,above] at (3,+3) {$n$};

        \node [block] (bE) at (1,-1) {\blockLabel{E}};
        \draw [link] (bE) -- (b0);

        \node [vote] at ([pos11]1,-3) {\tangoicon[height=0.3cm]{mimetypes_application-certificate}};
        \node [vote] at ([pos12]1,-3) {\tangoicon[height=0.3cm]{mimetypes_application-certificate}};
        \node [vote] at ([pos13]1,-3) {\tangoicon[height=0.3cm]{mimetypes_application-certificate}};
        \node [vote] at ([pos21]1,-3) {\tangoicon[height=0.3cm]{mimetypes_application-certificate}};
        \node [vote] at ([pos22]1,-3) {\tangoicon[height=0.3cm]{mimetypes_application-certificate}};
        \node [vote] at ([pos23]1,-3) {\tangoicon[height=0.3cm]{mimetypes_application-certificate}};
        \node [vote] at ([pos31]1,-3) {\tangoicon[height=0.3cm]{mimetypes_application-certificate}};
        \node [vote,opacity=0.2] at ([pos32]1,-3) {\tangoicon[height=0.3cm]{mimetypes_application-certificate}};
        \node [vote,opacity=0.2] at ([pos33]1,-3) {\tangoicon[height=0.3cm]{mimetypes_application-certificate}};
        \node [votelabel,below] at (1,-3) {$\frac{2n}{3}+1$};

        \node [block] (bF) at (2,-1) {\blockLabel{F}};
        \draw [link] (bF) -- (bE);

        \node [vote] at ([pos11]2,-3) {\tangoicon[height=0.3cm]{mimetypes_application-certificate}};
        \node [vote] at ([pos12]2,-3) {\tangoicon[height=0.3cm]{mimetypes_application-certificate}};
        \node [vote] at ([pos13]2,-3) {\tangoicon[height=0.3cm]{mimetypes_application-certificate}};
        \node [vote] at ([pos21]2,-3) {\tangoicon[height=0.3cm]{mimetypes_application-certificate}};
        \node [vote] at ([pos22]2,-3) {\tangoicon[height=0.3cm]{mimetypes_application-certificate}};
        \node [vote] at ([pos23]2,-3) {\tangoicon[height=0.3cm]{mimetypes_application-certificate}};
        \node [vote] at ([pos31]2,-3) {\tangoicon[height=0.3cm]{mimetypes_application-certificate}};
        \node [vote,opacity=0.2] at ([pos32]2,-3) {\tangoicon[height=0.3cm]{mimetypes_application-certificate}};
        \node [vote,opacity=0.2] at ([pos33]2,-3) {\tangoicon[height=0.3cm]{mimetypes_application-certificate}};
        \node [votelabel,below] (qiBottom) at (2,-3) {$\frac{2n}{3}+1$};

        \node [block] (bG) at (3,-1) {\blockLabel{G}};
        \draw [link] (bG) -- (bF);

        \node [vote] at ([pos11]3,-3) {\tangoicon[height=0.3cm]{mimetypes_application-certificate}};
        \node [vote] at ([pos12]3,-3) {\tangoicon[height=0.3cm]{mimetypes_application-certificate}};
        \node [vote] at ([pos13]3,-3) {\tangoicon[height=0.3cm]{mimetypes_application-certificate}};
        \node [vote] at ([pos21]3,-3) {\tangoicon[height=0.3cm]{mimetypes_application-certificate}};
        \node [vote] at ([pos22]3,-3) {\tangoicon[height=0.3cm]{mimetypes_application-certificate}};
        \node [vote] at ([pos23]3,-3) {\tangoicon[height=0.3cm]{mimetypes_application-certificate}};
        \node [vote] at ([pos31]3,-3) {\tangoicon[height=0.3cm]{mimetypes_application-certificate}};
        \node [vote,opacity=0.2] at ([pos32]3,-3) {\tangoicon[height=0.3cm]{mimetypes_application-certificate}};
        \node [vote,opacity=0.2] at ([pos33]3,-3) {\tangoicon[height=0.3cm]{mimetypes_application-certificate}};
        \node [votelabel,below] at (3,-3) {$\frac{2n}{3}+1$};

        \node [block] (bH) at (4,-1) {\blockLabel{H}};
        \draw [link] (bH) -- (bG);

        \node [vote] at ([pos11]4,-3) {\tangoicon[height=0.3cm]{mimetypes_application-certificate}};
        \node [vote] at ([pos12]4,-3) {\tangoicon[height=0.3cm]{mimetypes_application-certificate}};
        \node [vote] at ([pos13]4,-3) {\tangoicon[height=0.3cm]{mimetypes_application-certificate}};
        \node [vote] at ([pos21]4,-3) {\tangoicon[height=0.3cm]{mimetypes_application-certificate}};
        \node [vote] at ([pos22]4,-3) {\tangoicon[height=0.3cm]{mimetypes_application-certificate}};
        \node [vote] at ([pos23]4,-3) {\tangoicon[height=0.3cm]{mimetypes_application-certificate}};
        \node [vote] at ([pos31]4,-3) {\tangoicon[height=0.3cm]{mimetypes_application-certificate}};
        \node [vote] at ([pos32]4,-3) {\tangoicon[height=0.3cm]{mimetypes_application-certificate}};
        \node [vote] at ([pos33]4,-3) {\tangoicon[height=0.3cm]{mimetypes_application-certificate}};
        \node [votelabel,below] at (4,-3) {$n$};

        \node [block] (bI) at (5,-1) {\blockLabel{I}};
        \draw [link] (bI) -- (bH);

        \node [vote] at ([pos11]5,-3) {\tangoicon[height=0.3cm]{mimetypes_application-certificate}};
        \node [vote] at ([pos12]5,-3) {\tangoicon[height=0.3cm]{mimetypes_application-certificate}};
        \node [vote] at ([pos13]5,-3) {\tangoicon[height=0.3cm]{mimetypes_application-certificate}};
        \node [vote] at ([pos21]5,-3) {\tangoicon[height=0.3cm]{mimetypes_application-certificate}};
        \node [vote] at ([pos22]5,-3) {\tangoicon[height=0.3cm]{mimetypes_application-certificate}};
        \node [vote] at ([pos23]5,-3) {\tangoicon[height=0.3cm]{mimetypes_application-certificate}};
        \node [vote] at ([pos31]5,-3) {\tangoicon[height=0.3cm]{mimetypes_application-certificate}};
        \node [vote] at ([pos32]5,-3) {\tangoicon[height=0.3cm]{mimetypes_application-certificate}};
        \node [vote] at ([pos33]5,-3) {\tangoicon[height=0.3cm]{mimetypes_application-certificate}};
        \node [votelabel,below] at (5,-3) {$n$};

        \node [block] (bJ) at (6,-1) {\blockLabel{J}};
        \draw [link] (bJ) -- (bI);

        \node [vote] at ([pos11]6,-3) {\tangoicon[height=0.3cm]{mimetypes_application-certificate}};
        \node [vote] at ([pos12]6,-3) {\tangoicon[height=0.3cm]{mimetypes_application-certificate}};
        \node [vote] at ([pos13]6,-3) {\tangoicon[height=0.3cm]{mimetypes_application-certificate}};
        \node [vote] at ([pos21]6,-3) {\tangoicon[height=0.3cm]{mimetypes_application-certificate}};
        \node [vote] at ([pos22]6,-3) {\tangoicon[height=0.3cm]{mimetypes_application-certificate}};
        \node [vote] at ([pos23]6,-3) {\tangoicon[height=0.3cm]{mimetypes_application-certificate}};
        \node [vote] at ([pos31]6,-3) {\tangoicon[height=0.3cm]{mimetypes_application-certificate}};
        \node [vote] at ([pos32]6,-3) {\tangoicon[height=0.3cm]{mimetypes_application-certificate}};
        \node [vote] at ([pos33]6,-3) {\tangoicon[height=0.3cm]{mimetypes_application-certificate}};
        \node [votelabel,below] at (6,-3) {$n$};

        \draw [red,dashed] ([xshift=-0.45cm]qiBottom.south) rectangle ([xshift=0.45cm]qiTop.north);

    \end{tikzpicture}%
    \caption[]{%
        An execution showing that in \FBFT-Streamlet
        with $q = \frac{2n}{3} + 1$
        clients with $q_{k} = n$
        are not safe if $f = \frac{2n}{3} + 1$.
        Blocks $A,B,C$ receive $n$ votes each, which leads a client $k$ with `heavy quorum' $q_k = n$ to confirm the block $B$. Since the adversary controls $\frac{2n}{3}+1$ replicas, it can notarize (with `light quorum' $q = \frac{2n}{3}+1$) inconsistent blocks $E,F,G$.
        Subsequently,
        leaders propose blocks $H,I,J$ and all honest replicas vote for these blocks,
        which is in accordance with the protocol rules. Including the adversary's votes, these blocks meet a `heavy quorum' $q_{k'} = n$
        needed for another client $k'$ to confirm, leading to a safety violation.
    }%
    \label{fig:protocol-views-fbft}%
\end{figure}
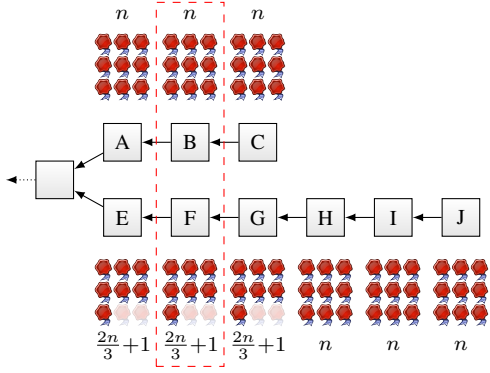

%% file: 05_application_eth.tex
\section{\Oflex Confirmation Rules for Ethereum}
\label{sec:eth}

Ethereum is a proof-of-stake blockchain
built~\cite{gasper} on Casper~\cite{casper}, a PBFT-style consensus protocol very similar~\cite{ethresearch-streamletffg} to Streamlet.\footnote{Casper~\cite{casper} is only proven to satisfy a weaker liveness notion called `plausible liveness'~\cite[Thm.~2]{casper}. Indeed, \cite{ebbandflow,2attacks,3attacks} show liveness attacks on
the Casper component of
Ethereum.
The \Oflex confirmation rules for Ethereum make the liveness and safety notions guaranteed by Casper optimally flexible. Once Ethereum upgrades to ensure standard liveness for Casper, the \Oflex rule 
will guarantee
standard liveness accordingly.}
Ethereum allows participants to lock up 
32~ETH tokens to
participate as replicas (called `validators' in Ethereum) in Casper.

In the generic \Oflex construction, we added extra \olocking and \ovoting
to the replica logic and designed a new confirmation rule.
In \Oflex-Streamlet, we reused Streamlet's votes as \ovotes,
and
only required modifying the replica voting rule to introduce the \olock constraint, and designing a new confirmation rule.
Due to Casper's similarity with Streamlet, we could do the same
for Casper.
However, 
we observe that Ethereum's \emph{implementation} of Casper already has an unrelated performance optimization that de-facto implements the required \olock and voting constraint.\footnote{In Ethereum's implementation,
validators 
ignore blocks that conflict with a block that they previously considered irreversible (`finalized'), thus effectively \olocking that block.
This improves computational efficiency.
Furthermore, this \olock is not present in the \emph{textbook} version
of Casper~\cite{casper}.
We therefore believe that this \olocking mechanism
was added as an unrelated performance optimization and was previously
not considered an integral part of Ethereum's consensus protocol.
It seems plausible that other PBFT-style consensus deployments
feature a similar \olock for performance optimization,
and therefore
can also be upgraded to \Oflex confirmation rules
with only client-local changes.}
As a result, to provide optimal flexible consensus on top of Ethereum as-is today, we do not require \emph{any} modifications to the validators, but only
new
client-local confirmation rules.

We describe how Ethereum's implementation provides the \olock and voting constraint and state our flexible \Oflex confirmation rules in \cref{sec:eth-confrule}.
We describe our implementation of the confirmation rule in \cref{sec:eth-implementation} and show experimental results in \cref{sec:eth-experiments}.

\subsection{Confirmation Rules}
\label{sec:eth-confrule}

Due to Casper's similarity~\cite{ethresearch-streamletffg}
to Streamlet, \Oflex confirmation rules for Casper
closely follow \Oflex-Streamlet.
We first briefly describe the features already implemented by Ethereum validators which enable designing client-side \Oflex confirmation rules without having to change the validator logic. 
A detailed description of Casper and of the Ethereum protocol can be found in \cite{casper,gasper,eth_specs}.
\begin{enumerate}
    \item \label{item:eth-finalized-checkpoint}
    Validators maintain a `finalized checkpoint',
    which is a block that satisfies the `default confirmation rule' called `finality' of Casper
    (\cf $\PiC$ in \cref{fig:blockdiagrams-oflex-generic}).
    The finalized checkpoint is safe up to $n/3$ adversary validators, \ie, 
    no other block inconsistent with a finalized checkpoint will ever be finalized if $\leq n/3$ validators are adversary.

    \item \label{item:eth-finalized-checkpoint-update}
    When the finalized checkpoint is updated, the new finalized checkpoint must extend the old one.

    \item \label{item:eth-voting-rule}
    Validators only vote for blocks that extend the finalized checkpoint in their view.

    \item \label{item:eth-votes-on-chain}
    Votes for a block are included on-chain in descendant blocks. 
    Validators consider only votes 
    included on-chain
    to update their finalized checkpoint.
    Blocks with votes deemed invalid are deemed invalid.
\end{enumerate}

We see below
how the above features realize the required \olock and voting constraint, just as in \Oflex-Streamlet (\cref{sec:optflex-streamlet-protocol}).
\Cref{item:eth-finalized-checkpoint,item:eth-finalized-checkpoint-update} show that a validator's finalized checkpoint behaves as its \olock, \ie, the finalized checkpoint satisfies Casper's original confirmation rule, and once a block is finalized, further finalized checkpoints must extend that block.
The voting rule in \cref{item:eth-voting-rule} shows that validators never vote for a block that is inconsistent with their finalized checkpoint, \ie, \olock.
Just as in \Oflex-Streamlet, the safety of Casper's finality when $f \leq n/3$ 
(\cref{item:eth-finalized-checkpoint}) guarantees that the voting constraint in \cref{item:eth-voting-rule} is never active when $f \leq n/3$ (recall that an analogous property was required for liveness of \Oflex-Streamlet).%
\footnote{This justifies why \cref{item:eth-voting-rule} is a correct performance optimization: it does not cause the optimized system to behave differently from the un-optimized system under normal conditions when $f \leq n/3$.}

Recall from \Oflex-Streamlet, that we also needed 
clients to be able to infer how many validators have \olocked a certain block and when it is thus safe to confirm that block.
In Ethereum's Casper implementation, having votes on chain (\cref{item:eth-votes-on-chain}) provides the evidence for the client to know when a validator must have \olocked,
because if a validator votes for a block,
it must have seen (and deemed valid)
all votes contained in the chain
leading up to its vote target.

\import{./figures/}{protocol-views-oflex-eth.tex}%

Since \olocking and \ovoting are de-facto already
implemented in Ethereum's Casper,
the following suffices for optimal flexible consensus
in Ethereum:

\emph{\Oflex confirmation rules $\PiCmod(q_{k})$ for Ethereum:}~
A client $k$ confirms block $A$ iff:
(1) it sees a block $C$ descending from $A$ that contains votes (in $C$ or its prefix) to finalize block $A$,
and
(2) it sees votes from
$q_{k}$
validators for $C$ (included on-chain or received otherwise).

\Cref{fig:protocol-views-oflex-eth} illustrates
how this rule provides safety
in an example with clients
with
$\tS_{k} = n-1$.
The example proceeds analogously to \Oflex-Streamlet (\cref{fig:protocol-views-oflex-streamlet}). 
If a client sees $n$ validators vote for a block $C$ which contains enough votes to finalize a previous block $A$, then the client confirms $A$.
When the client sees a validator vote for $C$ (\cref{fig:protocol-views-oflex-eth}~\DiagramStep{2}), it knows that this validator, if honest, must have seen the votes included in the chain of $C$ resulting in the finalization of $A$, and therefore must have set $A$ as its finalized checkpoint (\cref{fig:protocol-views-oflex-eth}~\DiagramStep{1}).
This validator will thus never vote for a block inconsistent with $A$ (\cref{fig:protocol-views-oflex-eth}~\DiagramStep{3}).
Thus, 
if a client sees all $n$ replicas vote for $C$,
then it knows that, unless all replicas are adversary, a block inconsistent with $A$ cannot obtain votes from all replicas.
Thus, clients that confirm blocks with $q_k=n$
remain safe even when $f = n-1$.

Note that clients using \Oflex rules can operate alongside clients that continue to use Casper finality as their confirmation rule.
Specifically, all clients simultaneously enjoy the flexible consensus guarantees (\cref{def:flex-consensus}) with their respective resiliences ($\tL_{k}=\frac{n}{3}-1, \tS_{k} = \frac{n}{3}$ for clients using finality).
Although Casper finality and the \Oflex rule with $q_{k} = \frac{2n}{3} + 1$ enjoy the same resiliences, the logs confirmed by them are not identical. The \Oflex rule requires an additional block ($C$ in \cref{fig:protocol-views-oflex-eth}).

\subsection{Implementation}
\label{sec:eth-implementation}

\import{./figures/}{ethcode-arch.tex}

Before describing our implementation of the \Oflex rule, we first briefly explain the software stack that a user runs in order to interact with the Ethereum blockchain, and how our implementation fits in the current system. There are two essential pieces of software: an Ethereum \emph{consensus} client, and an \emph{execution} client, as shown in \cref{fig:ethcode-arch-before}. The consensus client connects to the Ethereum peer-to-peer network to obtain latest blocks, and tries to confirm blocks according to the consensus protocol (\ie, Casper finality). It feeds confirmed blocks
to the execution client, which executes the transactions inside to obtain the system state (\eg, account balances) with regard to the tip of the confirmed log. User applications such as wallets can then query the execution client for latest system state through the JSON-RPC API.

We notice that existing Ethereum consensus clients already expose all the data required to run the \Oflex confirmation rule, namely the finalized checkpoint according to votes in the chain leading up to any block $C$, and the fraction of validators voting for $C$. As a result, the \Oflex rule can be implemented as a standalone program that runs alongside an existing, unmodified consensus client, subscribes to the client for chain data, and outputs the tip of the chain confirmed by the \Oflex rule with any desired safety resilience. \Cref{fig:ethcode-arch-after} shows one way to integrate such a program into the existing system. Instead of connecting directly to the execution client (which by default answers queries with regard to the chain tip confirmed by Casper finality), user applications connect to Patronum~\cite{patronum,popos}, an RPC proxy. Patronum learns the confirmed tip from the \Oflex rule, and rewrites user queries so that the execution client always answers them with regard to \emph{this} tip. This process is transparent to the user as well as the consensus and the execution clients, so that applications can benefit from the new rule without any change to their code.

We implemented the \Oflex confirmation rule following this design in 1{,}100 lines of Rust code.\footnote{Source code: \gitSourceUrl} For any block $C$, we use the \texttt{/states/finality\_checkpoints} endpoint of the Ethereum Beacon API~\cite{beaconapi} to query the latest block finalized by Casper according to votes in the chain leading to $C$, and use the \texttt{/states/committees} endpoint to query the set of validators selected to vote for $C$. We then use the \texttt{/blocks} endpoint to fetch subsequent blocks and examine the votes included to count the number of validators who actually vote for $C$. We tested our implementation against Prysm and Lighthouse, the two most popular Ethereum consensus clients as of now~\cite{clientdiversity}.

\subsection{Experiments}
\label{sec:eth-experiments}

\import{./figures/}{eth-experiment-latency.tex}

\import{./figures/}{eth-experiment-intro.tex}

\import{./figures/}{eth-experiment-shanghai.tex}

To evaluate the behavior of \Oflex rules in the real world, we use our implementation to apply the rule with various quorum sizes on a section of Ethereum mainnet between slots $5{,}970{,}000$ (March 10th, 2023) and $6{,}970{,}000$ (July 27th, 2023), covering roughly the most recent $1/7$ of the Ethereum PoS mainnet history as of when this paper is written. We choose this section because it is recent enough to reflect current statistics, and it covers two rare but interesting events: the finality outage incident that happened on May 12th, 2023~\cite{finality_outage_prysm}, and the Shanghai hard fork that happened on April 12th, 2023. During both events, participation of validators dipped, allowing us to demonstrate the behavior of  \Oflex rules under non-standard conditions.

\Cref{fig:eth-experiment-latency} shows how the distribution of confirmation latency changes with confirmation rule and quorum. Compared to Ethereum's Casper finality, \Oflex rules incurs an extra latency of approximately one epoch due to the extra round of voting. This extra latency is proportional to the quorum size, because a quorum of $q_k$ requires waiting for $q_k$ validators to vote, which takes roughly $(q_k/n)$-fraction of an epoch to happen. This effect is shown by the curves shifting towards right as quorum size increases. The key takeaway is that for Ethereum mainnet, adopting \Oflex rules results in only modest increase of confirmation latency, even when opting for extremely high quorum (\eg, $99\%$). For example, the $95^\text{th}$-percentile tail latency to confirm blocks with $99\%$ quorum is $30$ minutes, an increase of $60\%$ compared to Casper finality. We argue that this is a cost well worth paying, because confirming with $99\%$ quorum offers resilience against an extremely powerful $98\%$ adversary, while Casper finality is safe only against a $33\%$ adversary. Similarly, once adopting our rule, increasing the quorum size results in only slight further increase of latency. For example, adjusting the quorum size from $80\%$ to $99\%$ increases the tail latency by $25\%$, while boosting the safety resilience from $60\%$ to $98\%$. Operationally, we expect users adopting our rule to use very high quorums to maximize the safety benefit.

Ethereum mainnet typically runs in a healthy steady-state with close to $100\%$ of validators 
correctly
participating, allowing our rule to achieve good confirmation latency 
even with high quorums, as just shown. Now, we zoom into two events when participation dipped, and examine how our rules behave under such 
abnormal
conditions. For each event, we plot over time the growth of the logs confirmed by our rules with different quorums, as well as that by Casper finality. We also plot the participation rate (\ie, 
fraction of validators 
online and actively voting) for reference.

\Cref{fig:eth-experiment-intro} shows an incident that happened on May 12th, 2023, when a 
bug in some validators' software was triggered and brought them offline~\cite{finality_outage_prysm}. Before the incident, the logs confirmed by \Oflex rules closely tracked Casper finality, as shown on the left side of the plot. At around 17:20, the bug was triggered and participation quickly dropped below the threshold required by Casper ($67\%$) and \Oflex rules ($67\%$ and $99\%$, respectively), causing all three logs to stop growing (the horizontal stretch of the lines). At around 18:10, validators started to patch their software, and participation surpassed $67\%$ at around 18:30, allowing the Casper log and the $67\%$-quorum log to recover. The $99\%$-quorum log stalled for a while longer as there were not enough validators to form a $99\%$ quorum, and recovered at around 20:50 when participation surpassed $99\%$.

\Cref{fig:eth-experiment-shanghai} shows the Shanghai hard fork, when validators had to upgrade their software in order to keep participating post-fork. Unlike in the other incident, since this was a scheduled transition, participation rate after the hard fork never dropped below $90\%$. As a result, logs confirmed by  \Oflex rules with quorum sizes no larger than  $90\%$ experienced no pause and kept tracking Casper finality. The other logs except for the one confirmed by $99\%$-quorum quickly recovered as shown in the scope. It took significantly longer for the participation rate to recover above $99\%$ and for the $99\%$-quorum log to resume growth.

We remark that each user can choose to confirm with different quorums at different times and for different transactions, while maintaining high safety for all transactions that were confirmed with a high quorum. For instance, a user can choose to confirm with a $99\%$ quorum most of the time but switch to confirming with a lower quorum during incidents such as the ones described above.

%% file: figures/protocol-views-oflex-eth.tex
\import{./}{protocol-views-_utils.tex}%
\begin{figure}[tb]
    \centering
    \begin{tikzpicture}[fig_protocol_views,x=1.25cm]%
        \footnotesize%
        \scriptspacing%

        \pgfdeclarelayer{background}
        \pgfdeclarelayer{foreground}
        \pgfsetlayers{background,main,foreground}

        \node [block] (b0) at (0,0) {};
        \draw [link,densely dotted] (b0) -- ++([xshift=-0.65cm]0,0);

        \node [block] (bA) at (1,+1) {\blockLabel{A}};
        \draw [link] (bA) -- (b0);

        \node [vote] at ([pos11]1,+3) {\tangoicon[height=0.3cm]{mimetypes_application-certificate}};
        \node [vote,opacity=0.2] at ([pos12]1,+3) {\tangoicon[height=0.3cm]{mimetypes_application-certificate}};
        \node [vote,opacity=0.2] at ([pos13]1,+3) {\tangoicon[height=0.3cm]{mimetypes_application-certificate}};
        \node [vote] at ([pos21]1,+3) {\tangoicon[height=0.3cm]{mimetypes_application-certificate}};
        \node [vote] at ([pos22]1,+3) {\tangoicon[height=0.3cm]{mimetypes_application-certificate}};
        \node [vote] at ([pos23]1,+3) {\tangoicon[height=0.3cm]{mimetypes_application-certificate}};
        \node [vote] at ([pos31]1,+3) {\tangoicon[height=0.3cm]{mimetypes_application-certificate}};
        \node [vote] at ([pos32]1,+3) {\tangoicon[height=0.3cm]{mimetypes_application-certificate}};
        \node [vote] at ([pos33]1,+3) {\tangoicon[height=0.3cm]{mimetypes_application-certificate}};
        \node [votelabel,above] at (1,+3) {$\frac{2n}{3}+1$};

        \node [block] (bC) at (3,+1) {\blockLabel{C}};

        \node [vote] at ([pos11]3,+3) {\tangoicon[height=0.3cm]{mimetypes_application-certificate}};
        \node [vote] at ([pos12]3,+3) {\tangoicon[height=0.3cm]{mimetypes_application-certificate}};
        \node [vote] (voteYes) at ([pos13]3,+3) {\tangoicon[height=0.3cm]{mimetypes_application-certificate}};
        \node [vote] at ([pos21]3,+3) {\tangoicon[height=0.3cm]{mimetypes_application-certificate}};
        \node [vote] at ([pos22]3,+3) {\tangoicon[height=0.3cm]{mimetypes_application-certificate}};
        \node [vote] at ([pos23]3,+3) {\tangoicon[height=0.3cm]{mimetypes_application-certificate}};
        \node [vote] at ([pos31]3,+3) {\tangoicon[height=0.3cm]{mimetypes_application-certificate}};
        \node [vote] at ([pos32]3,+3) {\tangoicon[height=0.3cm]{mimetypes_application-certificate}};
        \node [vote] at ([pos33]3,+3) {\tangoicon[height=0.3cm]{mimetypes_application-certificate}};
        \node [votelabel,above] at (3,+3) {$n$};

        \draw [link,densely dotted] (bC) -- (bA);
        \node [align=left,anchor=west,xshift=0.4cm] at (1,3) {Votes fi-\\nalizing \blockLabel{A}};

        \begin{pgfonlayer}{background}
            \draw [draw=black!30,fill=black!3] ([xshift=-0.5cm,yshift=1.75cm]bA.north) -- ([xshift=1.9cm,yshift=1.75cm]bA.north) -- ([xshift=1.9cm,yshift=0.2cm]bA.north) -- ([xshift=0.1cm,yshift=-0.1cm]bC.north west) -- ([xshift=1.5cm,yshift=0.15cm]bA.north) -- ([xshift=-0.5cm,yshift=0.15cm]bA.north) -- cycle;
        \end{pgfonlayer}

        \coordinate (bE) at (1,-1);

        \node [block] (bF) at (2,-1) {\blockLabel{F}};
        \draw [link,densely dotted] (bF) -- (bE) -- (b0);

        \node [vote] at ([pos11]2,-3) {\tangoicon[height=0.3cm]{mimetypes_application-certificate}};
        \node [vote] at ([pos12]2,-3) {\tangoicon[height=0.3cm]{mimetypes_application-certificate}};
        \node [vote] at ([pos13]2,-3) {\tangoicon[height=0.3cm]{mimetypes_application-certificate}};
        \node [vote] at ([pos21]2,-3) {\tangoicon[height=0.3cm]{mimetypes_application-certificate}};
        \node [vote] at ([pos22]2,-3) {\tangoicon[height=0.3cm]{mimetypes_application-certificate}};
        \node [vote] at ([pos23]2,-3) {\tangoicon[height=0.3cm]{mimetypes_application-certificate}};
        \node [vote] at ([pos31]2,-3) {\tangoicon[height=0.3cm]{mimetypes_application-certificate}};
        \node [vote,opacity=0.2] at ([pos32]2,-3) {\tangoicon[height=0.3cm]{mimetypes_application-certificate}};
        \node [vote,opacity=0.2] at ([pos33]2,-3) {\tangoicon[height=0.3cm]{mimetypes_application-certificate}};
        \node [votelabel,below] at (2,-3) {$\frac{2n}{3}+1$};

        \node [block] (bH) at (4,-1) {\blockLabel{H}};

        \node [vote] at ([pos11]4,-3) {\tangoicon[height=0.3cm]{mimetypes_application-certificate}};
        \node [vote] at ([pos12]4,-3) {\tangoicon[height=0.3cm]{mimetypes_application-certificate}};
        \node [vote] (voteNo) at ([pos13]4,-3) {\tangoicon[height=0.3cm]{mimetypes_application-certificate}};
        \node [vote] at ([pos21]4,-3) {\tangoicon[height=0.3cm]{mimetypes_application-certificate}};
        \node [vote] at ([pos22]4,-3) {\tangoicon[height=0.3cm]{mimetypes_application-certificate}};
        \node [vote] at ([pos23]4,-3) {\tangoicon[height=0.3cm]{mimetypes_application-certificate}};
        \node [vote] at ([pos31]4,-3) {\tangoicon[height=0.3cm]{mimetypes_application-certificate}};
        \node [vote] at ([pos32]4,-3) {\tangoicon[height=0.3cm]{mimetypes_application-certificate}};
        \node [vote] at ([pos33]4,-3) {\tangoicon[height=0.3cm]{mimetypes_application-certificate}};
        \node [votelabel,below] at (4,-3) {$n$};

        \draw [link,densely dotted] (bH) -- (bF);
        \node [align=left,anchor=west,xshift=0.4cm] at (2,-3) {Votes fi-\\nalizing \blockLabel{F}};

        \begin{pgfonlayer}{background}
            \draw [draw=black!30,fill=black!3] ([xshift=-0.5cm,yshift=-1.75cm]bF.south) -- ([xshift=1.9cm,yshift=-1.75cm]bF.south) -- ([xshift=1.9cm,yshift=-0.2cm]bF.south) -- ([xshift=0.1cm,yshift=0.1cm]bH.south west) -- ([xshift=1.5cm,yshift=-0.15cm]bF.south) -- ([xshift=-0.5cm,yshift=-0.15cm]bF.south) -- cycle;
        \end{pgfonlayer}

        \node [locked] (lockb0) at (b0) {\notoemoji[height=0.4cm]{lock}};
        \node [locked] (lockbA) at (bA) {\notoemoji[height=0.4cm]{lock}};
        \node at (4,-3) {\notoemoji[height=0.7cm]{prohibited}};
        \node [votelabel,below] at (4,-3) {\notoemoji[height=0.3cm]{prohibited}};

        \node (validator) at (5.5,0) {\notoemoji[height=0.6cm]{technologist}};
        \draw [-latex,shorten >=-5pt,dashed] (validator) -- (voteYes) node [midway] {\DiagramStep{2}};
        \draw [-latex,shorten >=-3pt,dashed] (validator) -- ([yshift=5pt]lockbA.south east) node [midway] {\DiagramStep{1}};
        \draw [-latex,shorten >=-5pt,dashed] (validator) -- (voteNo) node [midway] {\notoemoji[height=1em]{prohibited}} node [midway,align=left,right,xshift=4pt,red] {\DiagramStep{3}};

        \node [cloud callout,cloud puffs=15,aspect=2,cloud puff arc=120,draw,callout absolute pointer={(5.5,1)}] at (5.5,2.5) {\DiagramStep{1} \notoemoji[height=0.9em]{lock}\blockLabel{A}};

    \end{tikzpicture}%
    \caption[]{%
        Illustration of \Oflex confirmation rules for Ethereum.
        * \emph{Protocol rules:}
        \DiagramStep{1}~If a validator sees a block $C$ that contains enough votes to finalize block $A$, then the validator sets $A$ as its finalized checkpoint.
        \DiagramStep{2}~%
        From a validator's vote on block $C$, clients can infer that the validator, if honest, must have set $A$ as its finalized checkpoint.
        \DiagramStep{3}~An honest validator that voted for $C$ will never vote for $F,H$ that are inconsistent with its finalized checkpoint $A$. 
        * \emph{Safety intuition:}
        Seeing votes from $n$ validators on block $C$, a client with $q_{k} = n$ confirms block $A$.
        Due to steps \DiagramStep{1} to \DiagramStep{3}, when $\frac{2n}{3}+1 \leq f < n$, blocks $F,H$ may be finalized (by adversarial votes alone), but
        will never obtain votes from all $n$ validators.
        Therefore, no client $k'$ with $q_{k'} = n$ confirms a block inconsistent with $A$, unless \emph{all} validators are adversary, implying $\tS_{k} = \tS_{k'} = n-1$.
    }%
    \label{fig:protocol-views-oflex-eth}%
\end{figure}
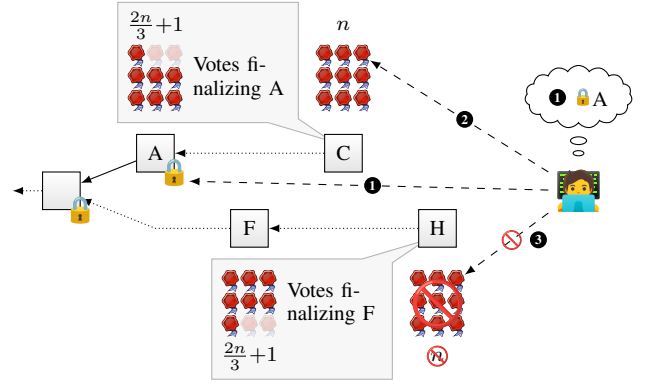

%% file: figures/ethcode-arch.tex
\tikzset{fig_ethcode_arch/.style={
            x=2.6cm,
            y=1.6cm,
        }}
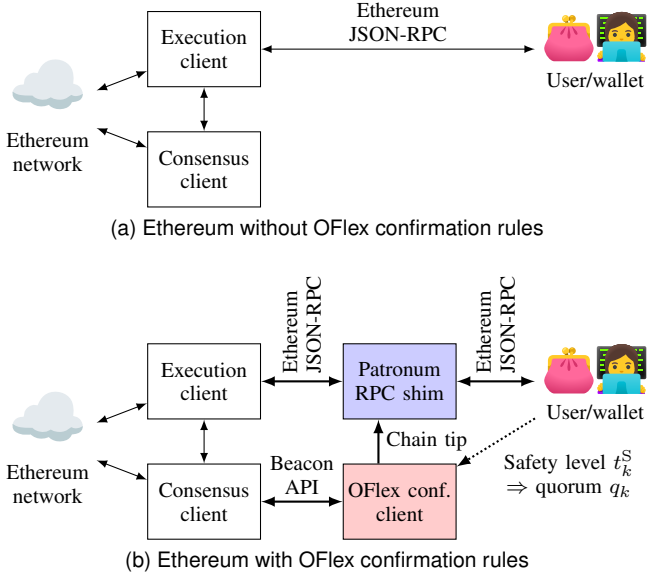
\begin{figure}[tb]%
    \centering%
    \subfloat[Ethereum without \Oflex confirmation rules\label{fig:ethcode-arch-before}]{%
        \begin{tikzpicture}[fig_ethcode_arch]%
            \footnotesize%
            \scriptspacing%

            \node [align=center,draw,minimum width=1.5cm,minimum height=1cm,inner sep=0pt] (el) at (0,0) {Execution\\client};
            \node [align=center,draw,minimum width=1.5cm,minimum height=1cm,inner sep=0pt] (cl) at (0,-1) {Consensus\\client};

            \node [align=center] (network) at (-0.8,-0.5) {\notoemoji[height=1cm]{cloud}\\Ethereum\\network};

            \node [align=center] (wallet) at (2,0) {\notoemoji[height=0.7cm]{purse}\notoemoji[height=0.7cm]{woman technologist}\\User/wallet};

            \draw [latex-latex] (network) -- (el);
            \draw [latex-latex] (network) -- (cl);

            \draw [latex-latex] (el) -- (cl) node [midway,right] {};

            \draw [latex-latex] (wallet) -- (el) node [midway,above,align=center] {Ethereum\\JSON-RPC};

        \end{tikzpicture}%
    }%
    \\%
    \subfloat[Ethereum with \Oflex confirmation rules\label{fig:ethcode-arch-after}]{%
        \begin{tikzpicture}[fig_ethcode_arch]%
            \footnotesize%
            \scriptspacing%

            \node [align=center,draw,minimum width=1.5cm,minimum height=1cm,inner sep=0pt] (el) at (0,0) {Execution\\client};
            \node [align=center,draw,minimum width=1.5cm,minimum height=1cm,inner sep=0pt] (cl) at (0,-1) {Consensus\\client};

            \node [align=center,draw,minimum width=1.5cm,minimum height=1cm,inner sep=0pt,fill=blue!20] (proxy) at (1,0) {Patronum\\RPC shim};
            \node [align=center,draw,minimum width=1.5cm,minimum height=1cm,inner sep=0pt,fill=red!20] (superfin) at (1,-1) {\Oflex conf.\\client};

            \node [align=center] (network) at (-0.8,-0.5) {\notoemoji[height=1cm]{cloud}\\Ethereum\\network};

            \node [align=center] (wallet) at (2,0) {\notoemoji[height=0.7cm]{purse}\notoemoji[height=0.7cm]{woman technologist}\\User/wallet};

            \draw [latex-latex] (network) -- (el);
            \draw [latex-latex] (network) -- (cl);

            \draw [latex-latex] (el) -- (cl) node [midway,right] {};

            \draw [-latex,thick] ([xshift=-1em]superfin.north) -- ([xshift=-1em]proxy.south) node [midway,right] {Chain tip};

            \draw [-latex,densely dotted,thick] (wallet) -- (superfin) node [midway,below right,align=center] {Safety level $\tS_k$\\$\Rightarrow$ quorum $q_k$};

            \draw [latex-latex,thick] (wallet) -- (proxy) node [midway,above,align=left,anchor=west,rotate=90] {Ethereum\\JSON-RPC};
            \draw [latex-latex,thick] (proxy) -- (el) node [midway,above,align=left,anchor=west,rotate=90] {Ethereum\\JSON-RPC};
            \draw [latex-latex,thick] (superfin) -- (cl) node [midway,above,align=center] {Beacon\\API};

        \end{tikzpicture}%
    }%
    \caption[]{%
        \Oflex confirmation rules can be adopted by Ethereum users unilaterally without any changes to internals or interfaces of execution clients, consensus clients, or wallets.
        The \Oflex confirmation client (this work) determines
        a chain tip satisfying the user-provided safety level.
        The Patronum RPC shim \cite{patronum,popos} can provide responses to queries from the wallet based on that chain tip.
    }%
    \label{fig:ethcode-arch}%
\end{figure}%

%% file: figures/eth-experiment-latency.tex
\begin{figure}[tb]
    \centering
    \begin{tikzpicture}[spy using outlines={rectangle, magnification=4, connect spies}]%
        \footnotesize
        \begin{axis}[
            mysimpleplot,
            name=plot1,
            xlabel={Latency [seconds \& epochs]},
            ylabel={Percentage},
            xmin=0, xmax=2688,
            ymin=-0.05, ymax=1.05,
            height=0.48\linewidth,
            width=\linewidth,
            legend columns=8,
            yticklabel={\pgfmathparse{\tick*100}\pgfmathprintnumber{\pgfmathresult}\%},
            ytick={0,0.2,0.4,0.6,0.8,1},
            xtick={0, 384, 768, 1152, 1536, 1920, 2304, 2688}, %
            xticklabels={{0\,s\\0\,ep.}, {384\,s\\1\,ep.}, {768\,s\\2\,ep.}, {1152\,s\\3\,ep.}, {1536\,s\\4\,ep.}, {1920\,s\\5\,ep.}, {2304\,s\\6\,ep.}, {2688\,s\\7\,ep.}}, %
            legend style={
                legend image post style={xscale=0.5},
                xshift=-1.7em,
            },
        ]

            \addplot+[black,dashed,mark=none]%
            table[ignore chars={(,)},col sep=comma, x expr=\thisrowno{0}*12,y expr=\thisrowno{2}] {figures/eth-experiment-latency/latency-finalized.txt};
            \addlegendentry{Casper};
        
            \addplot+[myparula11,mark=none]%
            table[ignore chars={(,)},col sep=comma, x expr=\thisrowno{0}*12,y expr=\thisrowno{2}] {figures/eth-experiment-latency/latency-q67.txt};
            \addlegendentry{67\%};    %
        
            \addplot+[myparula21,mark=none]%
            table[ignore chars={(,)},col sep=comma, x expr=\thisrowno{0}*12,y expr=\thisrowno{2}] {figures/eth-experiment-latency/latency-q80.txt};
            \addlegendentry{80\%};    %
        
            \addplot+[myparula31,mark=none]%
            table[ignore chars={(,)},col sep=comma, x expr=\thisrowno{0}*12,y expr=\thisrowno{2}] {figures/eth-experiment-latency/latency-q90.txt};
            \addlegendentry{90\%};    %
        
            \addplot+[myparula41,mark=none]%
            table[ignore chars={(,)},col sep=comma, x expr=\thisrowno{0}*12,y expr=\thisrowno{2}] {figures/eth-experiment-latency/latency-q95.txt};
            \addlegendentry{95\%};    %
        
            \addplot+[myparula51,mark=none]%
            table[ignore chars={(,)},col sep=comma, x expr=\thisrowno{0}*12,y expr=\thisrowno{2}] {figures/eth-experiment-latency/latency-q97.txt};
            \addlegendentry{97\%};    %
        
            \addplot+[myparula61,mark=none]%
            table[ignore chars={(,)},col sep=comma, x expr=\thisrowno{0}*12,y expr=\thisrowno{2}] {figures/eth-experiment-latency/latency-q98.txt};
            \addlegendentry{98\%};    %
        
            \addplot+[myparula71,mark=none]%
            table[ignore chars={(,)},col sep=comma, x expr=\thisrowno{0}*12,y expr=\thisrowno{2}] {figures/eth-experiment-latency/latency-q99.txt};
            \addlegendentry{99\%};    %

        \end{axis}
    \end{tikzpicture}%
    \caption[]{%
        Empirical cumulative distribution function of latency to confirm Ethereum mainnet blocks between slots $5{,}970{,}000$ and $6{,}970{,}000$ by the \Oflex rule with different quorums, and by Casper finality.}
    \label{fig:eth-experiment-latency}
\end{figure}
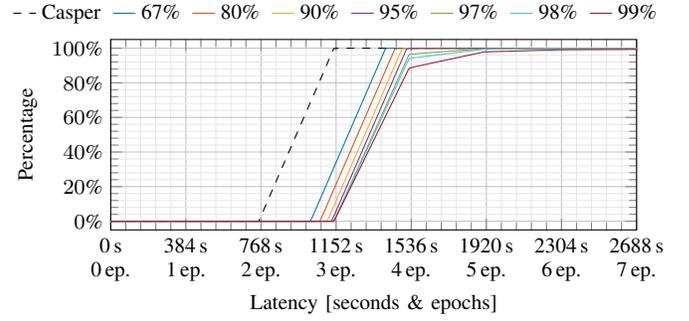

%% file: figures/eth-experiment-intro.tex
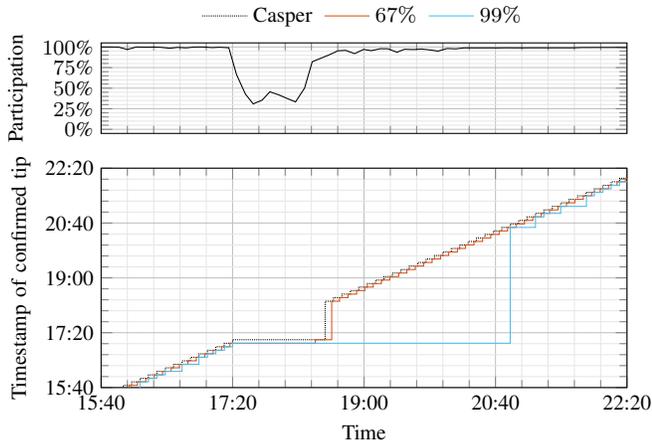
\begin{figure}[tb]
    \centering
    \begin{tikzpicture}[spy using outlines={rectangle, magnification=4, connect spies}]%
        \footnotesize
        \begin{axis}[
            mysimpleplot,
            name=plot1,
            ylabel={Participation},
            xmin=6423500, xmax=6425500,
            ymin=-0.05, ymax=1.05,
            height=0.325\linewidth,
            width=\linewidth,
            legend columns=3,
            xmajorticks=false,
            yticklabel={\pgfmathparse{\tick*100}\pgfmathprintnumber{\pgfmathresult}\%},
            ytick={0,0.25,0.5,0.75,1},
        ]

            \addplot [black,mark=none] table[ignore chars={(,)},col sep=comma, x expr=\thisrowno{0}*32+16,forget plot]
            {figures/eth-experiment-intro/participation.txt};

            \addlegendimage{black,dash pattern=on \pgflinewidth off \pgflinewidth,mark=none};
            \addlegendentry{Casper};
            \label{leg:eth-experiment-intro-ffg}
        
            \addlegendimage{myparula21,mark=none};
            \addlegendentry{$67\%$};   %
            \label{leg:eth-experiment-intro-lowQ}
        
            \addlegendimage{myparula61,mark=none};
            \addlegendentry{$99\%$};   %
            \label{leg:eth-experiment-intro-highQ}

        \end{axis}
        \begin{axis}[
            mysimpleplot,
            name=plot2,
            at=(plot1.below south), anchor=above north,
            xlabel={Time},
            ylabel={Timestamp of confirmed tip},
            xmin=6423500, xmax=6425500,
            ymin=6423500, ymax=6425500,
            height=0.525\linewidth,
            width=\linewidth,
            xtick={6423000, 6423500, 6424000, 6424500, 6425000, 6425500, 6426000},
            xticklabels={{14:00}, {15:40}, {17:20}, {19:00}, {20:40}, {22:20}, {00:00}},
            ytick={6423000, 6423500, 6424000, 6424500, 6425000, 6425500, 6426000},
            yticklabels={{14:00}, {15:40}, {17:20}, {19:00}, {20:40}, {22:20}, {00:00}},
            scaled ticks=false,
        ]

            \addplot+[black,dash pattern=on \pgflinewidth off \pgflinewidth,mark=none,const plot]
            table[ignore chars={(,)},col sep=comma,y expr=\thisrowno{1}] {figures/eth-experiment-intro/finalized.txt};
        
            \addplot+[myparula21,mark=none,const plot]
            table[ignore chars={(,)},col sep=comma,y expr=\thisrowno{1}] {figures/eth-experiment-intro/q67.txt};
        
            \addplot+[myparula61,mark=none,const plot]
            table[ignore chars={(,)},col sep=comma,y expr=\thisrowno{1}] {figures/eth-experiment-intro/q99.txt};

        \end{axis}
    \end{tikzpicture}%
    \caption[]{%
        Ethereum mainnet chain confirmed by Casper
        finality
        (\ref{leg:eth-experiment-intro-ffg})
        and by our
        \Oflex confirmation rule        with 
        low quorum ($67\%$, \ref{leg:eth-experiment-intro-lowQ})
        and high quorum ($99\%$, \ref{leg:eth-experiment-intro-highQ}).
        Shown are slots $6{,}423{,}500$ to $6{,}425{,}500$,
        covering a finality outage incident~\cite{finality_outage_coindesk,finality_outage_prysm}
        on May 12th, 2023 (times in UTC).
        }
    \label{fig:eth-experiment-intro}
\end{figure}

%% file: figures/eth-experiment-shanghai.tex
\begin{figure}[tb]
    \centering
    \begin{tikzpicture}[spy using outlines={rectangle, magnification=4, connect spies}]%
        \footnotesize
        \begin{axis}[
            mysimpleplot,
            name=plot1,
            ylabel={Participation},
            xmin=6209000, xmax=6215000,
            ymin=0.895, ymax=1.005,
            height=0.325\linewidth,
            width=\linewidth,
            legend columns=8,
            xmajorticks=false,
            yticklabel={\pgfmathparse{\tick*100}\pgfmathprintnumber{\pgfmathresult}\%},
            ytick={0.9,0.95,1},
            legend style={
                legend image post style={xscale=0.5},
                xshift=-1.7em,
            },
            xtick={6209000, 6210000, 6211000, 6212000, 6213000, 6214000, 6215000},
        ]
    
            \addlegendimage{black,dash pattern=on \pgflinewidth off \pgflinewidth,mark=none};
            \addlegendentry{Casper};

            \addlegendimage{myparula11,mark=none};
            \addlegendentry{67\%};   %

            \addlegendimage{myparula21,mark=none};
            \addlegendentry{80\%};   %

            \addlegendimage{myparula31,mark=none};
            \addlegendentry{90\%};   %

            \addlegendimage{myparula41,mark=none};
            \addlegendentry{95\%};   %

            \addlegendimage{myparula51,mark=none};
            \addlegendentry{97\%};   %

            \addlegendimage{myparula61,mark=none};
            \addlegendentry{98\%};   %

            \addlegendimage{myparula71,mark=none};
            \addlegendentry{99\%};   %

            \addplot [black,mark=none] table[ignore chars={(,)},col sep=comma, x expr=\thisrowno{0}*32+16]
            {figures/eth-experiment-shanghai/participation.txt};

        \end{axis}
        \begin{axis}[
            mysimpleplot,
            name=plot2,
            at=(plot1.below south), anchor=above north,
            xlabel={Time},
            ylabel={Timestamp of confirmed tip},
            xmin=6209000, xmax=6215000,
            ymin=6209000, ymax=6215000,
            height=0.60\linewidth,
            width=\linewidth,
            xtick={6209000, 6210000, 6211000, 6212000, 6213000, 6214000, 6215000},
            xticklabels={{20:40}, {00:00}, {03:20}, {06:40}, {10:00}, {13:20}, {16:40}},
            ytick={6209000, 6210000, 6211000, 6212000, 6213000, 6214000, 6215000},
            yticklabels={{20:40}, {00:00}, {03:20}, {06:40}, {10:00}, {13:20}, {16:40}},
            scaled ticks=false,
        ]

            \addplot+[black,dash pattern=on \pgflinewidth off \pgflinewidth,mark=none,const plot]
            table[ignore chars={(,)},col sep=comma,y expr=\thisrowno{1}] {figures/eth-experiment-shanghai/finalized.txt};
        
            \addplot+[myparula11,mark=none,const plot]
            table[ignore chars={(,)},col sep=comma,y expr=\thisrowno{1}] {figures/eth-experiment-shanghai/q67.txt};
        
            \addplot+[myparula21,mark=none,const plot]
            table[ignore chars={(,)},col sep=comma,y expr=\thisrowno{1}] {figures/eth-experiment-shanghai/q80.txt};
        
            \addplot+[myparula31,mark=none,const plot]
            table[ignore chars={(,)},col sep=comma,y expr=\thisrowno{1}] {figures/eth-experiment-shanghai/q90.txt};
        
            \addplot+[myparula41,mark=none,const plot]
            table[ignore chars={(,)},col sep=comma,y expr=\thisrowno{1}] {figures/eth-experiment-shanghai/q95.txt};
        
            \addplot+[myparula51,mark=none,const plot]
            table[ignore chars={(,)},col sep=comma,y expr=\thisrowno{1}] {figures/eth-experiment-shanghai/q97.txt};
        
            \addplot+[myparula61,mark=none,const plot]
            table[ignore chars={(,)},col sep=comma,y expr=\thisrowno{1}] {figures/eth-experiment-shanghai/q98.txt};
        
            \addplot+[myparula71,mark=none,const plot]
            table[ignore chars={(,)},col sep=comma,y expr=\thisrowno{1}] {figures/eth-experiment-shanghai/q99.txt};

            \coordinate (spyPoint) at (axis cs:6209820,6209750);
            \begin{scope}
                \spy [width=3.5cm,height=1.75cm] on (spyPoint) in node [fill=white,anchor=north west,xshift=3pt,yshift=-0.5em] at (0,0);
            \end{scope}

        \end{axis}
    \end{tikzpicture}%
    \caption[]{%
Ethereum mainnet chain confirmed by Casper 
        finality
        and by our \Oflex confirmation rule with different quorums (\cf legend).
        Shown are slots $6{,}209{,}000$ to $6{,}215{,}000$,
        covering the Shanghai hard fork at slot $6{,}209{,}536$ (on April 12th, 2023 at 22:27 UTC on the plot).%
        }
    \label{fig:eth-experiment-shanghai}
\end{figure}
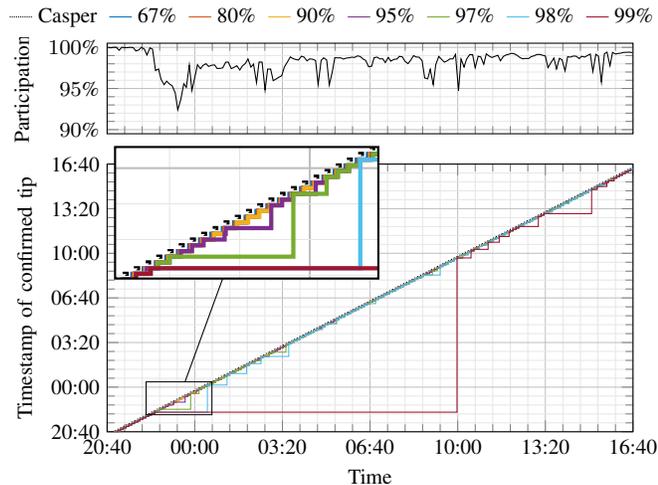

%% file: 06_relatedwork.tex
\section{Related Work}
\label{sec:relatedwork}

\smallskip\noindent\textbf{Multi-threshold consensus:}~
Multi-threshold Byzantine fault tolerance~\cite{mtbft,mtrbc,mtrbc2,aa} can be viewed as a degenerate case of flexible consensus in which all clients have the same resilience pair $(\tLrel,\tSrel)$ which may be different from $(\frac{1}{3},\frac{1}{3})$.
This is done for SMR consensus in \cite{mtbft,aa}, with a weaker notion of safety in \cite{bft2f}, and for reliable broadcast in \cite{mtbft,mtrbc,mtrbc2}.

Multi-threshold BFT~\cite{mtbft} argues that for some applications one may desire a higher safety resilience than liveness resilience ($\tS \geq \tL$).
The same problem is stated from another point of view in \cite{fbft,basilic,mtrbc2,highway,trebiz}:
`rational' adversaries may be willing to attack
safety but not liveness (\cf alive-but-corrupt faults~\cite{fbft}, deceitful faults~\cite{basilic},
Byzantine merchants~\cite{trebiz}) because 
an adversary can expect sizeable profits from double spends by attacking safety, but stands to gain little (and instead loses protocol rewards) if it were to attack liveness.
Tolerating alive-but-corrupt adversaries in addition to Byzantine adversaries is 
then 
equivalent to having a higher safety resilience than liveness resilience.
In contrast, some argue that crash faults are more common than Byzantine faults, and therefore, they design protocols to tolerate crash faults in addition to Byzantine faults~\cite{upright}.
Relatedly, \cite{sendcorruption,sendcorruption2,xft} consider other classes of adversaries.

Orthogonal to our work, protocols in \cite{mtbft,tardigrade,bkl2019,guo19} 
achieve the respective optimal safety and liveness resiliences
under different network conditions (synchrony, asynchrony),
but 
without allowing clients flexibility to choose resilience pairs.
In GearBox~\cite{gearbox}, replicas optimistically sub-sample a small committee to run the protocol, and adaptively tune the committee size and quorum threshold in order to retain worst-case safety and liveness resilience of $1/3$ of the full replica set. There are two key differences to our work. First, the goal of GearBox is not multi-threshold or flexible consensus; thus, safety and liveness of GearBox break \emph{for all clients} once $f \geq n/3$. Second, the quorum threshold is tuned in a system-wide fashion, not client-local/flexible.

\smallskip\noindent\textbf{Flexible protocols:}~
FBFT~\cite{fbft} allows clients to not only choose 
resilience pairs,
but also choose different network assumptions 
(synchronous or partially synchronous).
Protocols with an optimistic fast path~\cite{thunderella,synchotstuff,zyzzyva} allow clients to trade-off between the liveness resilience and confirmation latency, but not the safety resilience.
In this work, we focus on achieving the optimal trade-off between safety and liveness resiliences for a fixed partially synchronous network assumption.
Moreover in FBFT, the replica logic is essentially the same as in prior protocols such as Hotstuff~\cite{hotstuff} (excluding changes geared towards flexibility in the network assumptions).
In this work, we instead modify (in \Oflex -Streamlet) the replica voting rule with a restriction based on \olock, which is crucial to achieving optimal flexibility.
Highway~\cite{highway} also aims to provide flexibleresiliences but has no proven liveness guarantees when the safety resilience exceeds $\frac{1}{3}$ as per \cite[Thm.~2]{highway}.
The idea of flexible quorums, used by \cite{fbft} and our work, also appears in \cite{fpaxos}.

\smallskip\noindent\textbf{Strenghtened fault tolerance:}~
SFT~\cite{fbft2} extends \cite{fbft} from one-shot Byzantine agreement to a full SMR protocol while maintaining `linear message complexity'.
We 
discuss how
our generic \Oflex construction
can
also
be made to 
achieve linear message complexity:
In every `round' (whose duration will be determined later),
replicas should take turns being an `aggregator'.
During a round, all replicas send their \ovotes to the aggregator.
The aggregator then collects all \ovotes it receives during the round for any log that is the same as or an extension of the log that the aggregator itself \ovoted, and packages these \ovotes into a certificate.
This solution preserves safety because \ovotes cannot be forged by the aggregator.
Liveness is preserved because, after $\GST$, eventually, all replicas see the log that the aggregator \ovoted.
Moreover, in the regime where liveness needs to be provided (\ie, if $f < n/3$), the base protocol is safe, so honest replicas only \ovote logs consistent with the aggregator's \ovote.
Thus, if the round is long enough (confirmation latency of the base protocol), then all honest replicas are guaranteed to \ovote some log extending the aggregator's \ovote within that round.
With this construction, if the base protocol has linear message complexity, then so does the combined \Oflex protocol (the only additional messages are \ovotes).
Since all \ovotes collected by the aggregator extend a common prefix, the aggregator can use a SNARK to reduce the message to constant size, thus achieving linear \emph{communication} complexity.

It is equally easy to pipeline the above process in a linear-communication protocol such as DiemBFT~\cite{diembft} (which \cite{fbft2} uses) such that round leaders of the protocol also serve as aggregators.
In \FBFTtwo, the main difficulty was how to re-use votes for a block as votes for its ancestors without double-counting conflicting votes.
To do this, replicas in \FBFTtwo~\cite{fbft2} attach a `marker' to votes through which the replica signals to have not voted for an inconsistent block in the recent past.
In \Oflex, such markers are not needed since 
the \olock already provides the function of the marker (the \ovoting replica has never \ovoted and will never \ovote an inconsistent block).
By adapting the other techniques used in \cite{fbft2}, one readily obtains an `\Oflex-DiemBFT' with linear message complexity, without even the message overhead of the markers.

\import{./figures/}{blockdiagrams-snc.tex}
\smallskip\noindent\textbf{Snap-and-chat protocols:}~
Snap-and-chat protocols \cite{ebbandflow} provide a construction
of flexible consensus
where a `more live' consensus protocol
and a `more safe' one
are concatenated
to yield two confirmation rules, a `more live' one and a `more safe' one.
This construction 
can be extended to 
yield a resilience-optimal flexible consensus protocol:
a concatenation of $\frac{n}{3}$ consensus protocols, 
each with a different system-wide quorum,
starting from one with a quorum $q=\frac{2n}{3}+1$ all the way up to one with $q=n$.
A more detailed description is given in \cref{fig:blockdiagrams-snc}.
Clearly, running $\frac{n}{3}$ protocols simultaneously is impractical for the replicas.
Our generic construction (\cref{fig:blockdiagrams-oflex-generic}) points out that all these protocols except the first one ($q = \frac{2n}{3}+1$) can be collapsed into a single round of \ovote and \olock.

\smallskip\noindent\textbf{Asymmetric quorums:}~
Works~\cite{malkhibyzquorumsystems,cachinasymmetric,damgardasymmetric,cachinquorum} model 
that not all replicas may be equally trusted, thereby generalizing the resilience from a fraction of replicas to sets of replicas (called fail-prone sets).
In these works, there is no flexiblity.
Federated consensus~\cite{stellar,semitopology} adds flexibility by allowing replicas and clients to choose their own fail-prone sets.
However, clients have the same fail-prone sets for both safety and liveness (\ie, both properties fail together when replicas outside the fail-prone sets are adversary),
making this an orthogonal direction to flexible consensus.

%% file: figures/blockdiagrams-snc.tex
\tikzset{fig_blockdiagrams/.style={
            x=2cm,
            y=0.75cm,
            pir/.style={
                    fill=white,
                    draw,
                    inner sep=0pt,
                    minimum height=0.8cm,
                    minimum width=0.8cm,
                    align=center,
                },
            pic/.style={
                    fill=white,
                    draw,
                    inner sep=0pt,
                    shape=rounded rectangle,
                    minimum width=0.9cm,
                    minimum height=0.5cm,
                    xshift=0.775cm,
                    align=center,
                },
            bboxfix/.style={
                    fill=none,
                    draw=none,
                    cornerSW/.style={
                            xshift=-1cm,
                            yshift=-1em,
                        },
                    cornerNE/.style={
                            yshift=1em,
                        },
                },
            boxOverall/.style={
                    densely dotted,
                    cornerSW/.style={
                            xshift=-0.8em,
                            yshift=-1em,
                        },
                    cornerNE/.style={
                            xshift=0.8em,
                            yshift=2.75em,
                        },
                }
        }}
\begin{figure}[tb]%
    \centering%
    \begin{tikzpicture}[fig_blockdiagrams]%
        \footnotesize%
        \scriptspacing%

        \draw [bboxfix] ([cornerSW]0,0) rectangle ([cornerNE]0,0);

        \coordinate (txs) at (-0.50,0);

        \node [pic] (PiC1) at (0,0) {$\PiC^{(1)}$};
        \node [pir] (PiR1) at (0,0) {$\PiR^{(1)}$};

        \node [pic] (PiC2) at (1,0) {$\PiC^{(2)}$};
        \node [pir] (PiR2) at (1,0) {$\PiR^{(2)}$};

        \node [pic] (PiCn) at (2.35,0) {$\PiC^{(\frac{n}{3})}$};
        \node [pir] (PiRn) at (2.35,0) {$\PiR^{(\frac{n}{3})}$};

        \node (dots) at (1.9,0) {...};

        \draw [-latex] (txs) -- (PiR1) node [above,pos=0.25] {$\txs$};
        \draw [-latex] (PiC1) -- (PiR2);
        \draw [-latex] (PiC2) -- (dots);
        \draw [-latex] (dots) -- (PiRn);

        \coordinate (LOGn) at (3.25,0);
        \coordinate (LOG2) at (3.25,-1.5);
        \coordinate (LOG1) at (3.25,-2.5);

        \draw [-latex] (PiCn) -- (LOGn) node [pos=1,anchor=north west,yshift=12pt,xshift=-1.8em] {$\LOG{(0,n-1)}{}$};

        \draw [-latex] ([xshift=3pt]PiC2.east) |- (LOG2) node [pos=1,anchor=north west,yshift=12pt,xshift=-1.8em] {$\LOG{(\frac{n}{3}-2,\frac{n}{3}+2)}{}$};
        \draw [-latex] ([xshift=3pt]PiC1.east) |- (LOG1) node [pos=1,anchor=north west,yshift=12pt,xshift=-1.8em] {$\LOG{(\frac{n}{3}-1,\frac{n}{3})}{}$};

        \node at (3.2,-0.45) {\vdots};

        \draw [decorate,decoration={calligraphic brace}] ([yshift=0.75em,xshift=1pt]PiR1.north west) -- ([yshift=0.75em,xshift=-1pt]PiCn.east |- PiRn.north east) node [midway,above,anchor=base,yshift=6pt] {Concatenation of $\frac{n}{3}$ protocols};

    \end{tikzpicture}%
    \caption[]{%
        Block diagram of 
        an optimal 
        flexible consensus protocol inspired by snap-and-chat 
        (\cf \cite[Fig.~5]{ebbandflow}).
        This construction is a serial concatenation of $\frac{n}{3}$ consensus protocols (replica logic $\PiR$, confirmation rule~$\PiC$).
        The first protocol $\PI^{(1)} = (\PiR^{(1)},\PiC^{(1)})$ uses a system-wide quorum $q = \frac{2n}{3}+1$ and achieves system-wide resiliences $\tL = \frac{n}{3}-1, \tS=\frac{n}{3}$.
        At each subsequent protocol, the system-wide quorum is increased by $1$, which decreases $\tL$ by $1$ and increases $\tS$ by $2$ for that protocol, all the way up to $\PI^{(\frac{n}{3})} = (\PiR^{(\frac{n}{3})},\PiC^{(\frac{n}{3})})$ which has a system-wide quorum $q = n$ and resiliences $\tL=0, \tS=n-1$.
        The log output using $\PI^{(1)}$ is treated as input (\ie, `transactions') to be sequenced by the next protocol $\PI^{(2)}$
        (a replica `boycotts' proposals in $\PI^{(2)}$ inconsistent with its view of the log of $\PI^{(1)}$, \cf \cite[Fig.~5]{ebbandflow}).
        Thus, $\PI^{(2)}$ generates a `log of logs` which is then flattened and de-duplicated to produce the output log of $\PI^{(2)}$,
        which is then input to $\PI^{(3)}$ and so on.
        Replicas run 
        all protocols simultaneously, which makes this inefficient.
        A client $k$ with resiliences $(\tL_{k},\tS_{k})$ confirms $\LOG{(\tL_{k},\tS_{k})}{}$ output by the corresponding protocol and ignores the rest.
    }%
    \label{fig:blockdiagrams-snc}%
\end{figure}
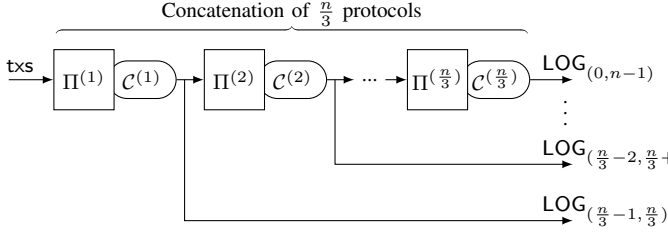%

%% file: 07_discussion.tex
\section{Discussion}
\label{sec:discussion}

\subsection{Stronger Consistency Guarantees for \Oflex}
\label{sec:discussion-strongflex}

In the flexible consensus formulation
(\cref{def:flex-consensus}),
consistency between clients $k, k'$ is guaranteed only when
$f \leq \min\{\tS_k,\tS_{k'}\}$.
This follows the definitions in~\cite{fbft,fbft2}.
The previous impossibility result $2\tL_{k} + \tS_{k} < n$ bounds the resiliences when $\tS_k = \tS_{k'}$; thus the security guarantees of \Oflex protocols are optimal for clients with equal resiliences. 
But can consistency be guaranteed when $ f > \min\{\tS_k,\tS_{k'}\}$ for clients with unequal resiliences?
In fact, a closer examination of the proofs of \cref{thm:generic-oflex-security,thm:oflex-streamlet-safety} show that the \Oflex protocols guarantee consistency between any clients $k,k'$ up to $f = (\tS_{k} + \tS_{k'})/2$.
Conversely, we show in \cref{sec:strongflex-converse} that when the resilience pairs $(\tL_{k},\tS_{k})$ and $(\tL_{k'},\tS_{k'})$ are chosen optimally (\ie, each satisfying $2\tL_{k} + \tS_{k} = n-1$), then consistency between clients $k,k'$ is impossible if $f > (\tS_{k} + \tS_{k'})/2$.

\subsection{Flexible Accountable Safety}
\label{sec:discussion-accountability}

The works
\cite{peerreview,faultdetectionproblem,casper,gasper,polygraph,blockchainisdead,aa,snapandchat,bftforensics,accimpliesfin}
strengthen consensus protocols' safety property
to \emph{accountable safety},
which 
guarantees safety if few replicas deviate from
the protocol,
\emph{and} 
if safety is ever violated, then a sizable number of replicas are identified
to have provably violated the protocol.
Our \Oflex protocols provide \emph{flexible accountable safety}: if two clients $k,k'$ confirm inconsistent logs (a safety violation), then at least $(\tS_{k}+\tS_{k'})/2$ replicas must have provably violated the protocol (see \cref{sec:discussion-strongflex} to understand the increased resilience compared to \cref{def:flex-consensus}).

The following rule identifies adversary replicas in the generic \Oflex construction:
a replica is detected adversary if
it sends two \emph{equivocating} \ovotes,
\ie, two \ovotes 
for inconsistent logs $\LOGempty$ and $\LOGempty'$.
No such \ovotes will exist for any honest replica,
due to \olocking,
and authentication using signatures.
On the other hand,
as the quorum-intersection-based safety arguments for \Oflex show
(\cf \cref{thm:generic-oflex-security,thm:oflex-streamlet-safety}),
a necessary condition for a safety violation between clients $k,k'$ is that
at least $(\tS_{k}+\tS_{k'})/2$ replicas have sent equivocating \ovotes.

This rule readily makes safety accountable in the generic \Oflex construction (\cref{sec:optflex}).
Adapting the rule to \Oflex-Streamlet (\cref{sec:optflex-streamlet-protocol}), if a replica votes for a block whose parent chain contains three adjacent blocks from consecutive epochs, it shall never vote for a block inconsistent with the second block of them (because it is \olocked on that block).
For Ethereum (\cref{sec:eth}), if a validator votes for a block whose state commits a certain block as finalized, it shall never vote for a block inconsistent with that finalized block (because it is, effectively, \olocked on that block).

\subsection{Recovering from Liveness Outages}
\label{sec:discussion-liveness}

If $f$ exceeds $\tL_{k}$, client $k$ may lose liveness.
If the adversary replicas are only crash faults, then the client can regain liveness either by temporarily increasing $\tL_{k}$ (by decreasing $q_k$), or by waiting until the crashes are restored. This is because crash faults cannot make honest replicas \olock inconsistent logs, so once the number of faults is below $\tL_{k}$, the client's confirmation quorum will eventually be fulfilled.
Similarly, if there are Byzantine faults, but
not too many
($\tL_k < f < n/3$),
then
liveness can be regained
because $f < n/3$ adversary replicas cannot make honest replicas \olock inconsistent logs.%

If $f > n/3 > \tL_k$ are Byzantine, then the base protocol may output inconsistent logs, which the OFlex gadget’s \olocking would `convert' into a (permanent) liveness violation (note that all clients’ liveness resiliences are violated in this case anyway). But in case of such a permanent stall via inconsistency of the base protocol, if the base protocol provides accountable safety (as Casper, Streamlet, and HotStuff do~\cite{casper,snapandchat,bftforensics}), then at least $n/3$ adversary replicas can be identified.
In proof-of-stake blockchains, this makes such an attack expensive, as external repair can be used to slash the stake of the identified replicas.%

\subsection{Preserving Safety after External Repair}
\label{sec:discussion-flexeth}

Clients that use \Oflex confirmation rules for Ethereum
with high $q_k$ retain safety up to high $f$,
\emph{throughout the consensus protocol's execution}.
However, 
in case inconsistencies occur in Casper finality or if liveness is lost (which may occur when $f$ exceeds $n/3$),
Ethereum intends to use a process
\emph{external to the consensus protocol} (`social consensus'~\cite{eth_attacks_and_defense,casper51})
to `repair' the protocol.
Interestingly, this intervention can be carried out in a manner
that preserves consistency of high-safety \Oflex rules
\emph{despite the intervention},
namely 
if care is taken to not revert transactions 
deemed
confirmed by \Oflex rules with high $q_k$
for which no conflicting confirmations have been observed.

%% file: A01_streamlet_proofs.tex
\section{\Oflex-Streamlet Security Proof Details}
\label{sec:oflex-streamlet-proofs}

For completeness, we provide a pseudocode of Streamlet in \cref{alg:streamlet}. The pseudocode of \Oflex-Streamlet in \cref{alg:oflex-streamlet} highlights its differences with respect to Streamlet.

\import{./figures/}{alg-streamlet.tex}

\begin{proof}[Proof of \cref{lem:no-conflict-notarize}]
    This proof follows techniques from the proof of \cite[Lem. 2]{streamlet} and is recapped here for completeness.
    Denote the heights of the blocks $A,B,C$ as $\ell,\ell+1,\ell+2$ respectively. Denote the epochs of $A,B,C$ as $e,e+1,e+2$ respectively.

    Assume for the sake of contradiction that some block $F \neq B$ with epoch $e'$, at the same height as $B$, is also notarized in the view of some honest replica. We look at three cases.

    First, if $e' \in \{e,e+1,e+2\}$, then in one of these three epochs, at least $n/3+1$ replicas must have voted for two different blocks. This cannot happen if $f \leq n/3$.

    Second, let $e' < e$.
    Since both blocks $A$ and $F$ are notarized, at least $n/3 + 1$ replicas must have voted for both blocks. Since $f \leq n/3$, at least one of these replicas must be honest.
    Note that this replica must have voted for $F$ by the end of epoch $e'$ and thus before the beginning of epoch $e$. This implies that the replica must have observed a notarized chain of length $\ell$ before the beginning of epoch $e$ (by observing $F$'s notarized parent chain, which is a prerequisite for voting). However, in epoch $e$, the honest replica voted for block $A$ which has length $\ell$. This is a contradiction because the parent chain of $A$ is not one of the longest notarized chains seen by the replica since it had already seen a notarized chain of length $\ell$.

    Third, let $e' > e+2$.
    In this case, since both blocks $C$ and $F$ are notarized, at least $n/3 + 1$ replicas must have voted for both blocks. Since $f \leq n/3$, at least one of these replicas must be honest.
    Note that this replica must have voted for $C$ by the end of epoch $e+2$ and thus before the beginning of epoch $e'$. This implies that the replica must have observed a notarized chain of length $\ell+1$ before the beginning of epoch $e'$ ($C$'s notarized parent chain). However, in epoch $e'$, the honest replica voted for block $F$ which has length $\ell+1$. This is a contradiction because the parent chain of $F$ is not one of the longest notarized chains seen by the replica since it had already seen a notarized chain of length $\ell+1$.
\end{proof}

\begin{lemma}[{\cf \cite[Fact 3]{streamlet}}]
\label{lem:two-honest-leaders}
    Suppose that $f < n/3$ and that after GST, there are two epochs $e$ and $e + 1$ both with honest leaders denoted $L_e$ and $L_{e+1}$ respectively, and suppose that $L_e$ and $L_{e+1}$ propose blocks $B_e$ and $B_{e+1}$ at heights $\ell_0$ and $\ell_1$ respectively,
    it must be that $\ell_1 \geq \ell_0 + 1$.
\end{lemma}
\begin{proof}
    The proof is similar to that of \cite[Fact 3]{streamlet}, but additionally uses the fact that the voting rule based on the \olock is never invoked in the regime of interest for liveness, \ie, $f < n/3$ (\cref{lem:voting}).
    All we need to prove is that the longest notarized chain of $L_{e+1}$ at the beginning of epoch $e+1$ is of length at least $\ell_0$.
    Since $L_e$ proposes a block whose parent chain $L_e$ saw notarized at the beginning of epoch $e$, due to synchrony, all honest replicas see this notarized chain $\Delta$ time into epoch $e$. Then, every honest replica will vote for $B_e$ unless by time $\Delta$ into epoch $e$, it had already observed a conflicting notarized chain of length $\ell_0$ (otherwise, $B_e$'s parent chain is one of the longest notarized chains see by the honest replica, and by \cref{lem:voting}, $B_e$ is also a descendant of the replica's \olock). If all honest replicas vote for $B_e$, then since $f \leq n/3$, all honest replicas see the chain ending in $B_e$ (of length $\ell_0$) notarized by the beginning of epoch $e+1$. If not, then at least one honest replica had already observed a notarized chain of length $\ell_0$ which all honest replicas observe by the beginning of epoch $e+1$. This completes the proof.
\end{proof}

\begin{lemma}[{\cf \cite[Lem.~5]{streamlet}}]
    \label{lem:three-honest-leaders}
    Suppose $f < n/3$. After GST, suppose there are three consecutive epochs $e$, $e + 1$, $e + 2$ all with honest leaders denoted $L_e$, $L_{e+1}$, and $L_{e+2}$, then the following holds (below we use $B$ to denote the block proposed by $L_{e+2}$ during epoch $e + 2$):
    \begin{enumerate}
        \item by the beginning of epoch $e + 3$, every replica/client will observe a notarized chain ending at $B$ and $n-f$ votes for $B$ (and $B$ had not received $n-f$ votes before the beginning of epoch $e$);
        \item furthermore, no conflicting block $F \neq B$ with the same height as $B$ will ever get notarized in the view of any honest replica/client.
    \end{enumerate}
\end{lemma}
\begin{proof}
    This proof is identical to that of \cite[Lem.~5]{streamlet} and is repeated for completeness.
    Let $\ell_0,\ell_1,\ell_2$ be the heights of the blocks proposed by $L_{e}, L_{e+1}, L_{e+2}$ respectively. By \cref{lem:two-honest-leaders}, $\ell_2 > \ell_1 > \ell_0$.
    Recall that $B$ is the block proposed by $L_{e+2}$ in epoch $e+2$, whose length is $\ell_2$.
    
    We now argue that by the beginning of epoch $e + 3$, no honest replica voted for a conflicting $F \neq B$ at the same height as $B$.
    No honest replica will have voted for a block $F \neq B$ at height $\ell_2$ in epochs $e, e+1$ or $e+2$ because the leaders $L_{e}$ and $L_{e+1}$ proposed blocks at heights $\ell_0$ and $\ell_1$ respectively, which are different from $\ell_2$, and the leader $L_{e+2}$ proposed the block $B \neq F$. 
    If an honest replica had voted for such a block $F$ before epoch $e$ started, at that time this replica must have observed a notarized parent chain of length $\ell_2 - 1$.
    Due to synchrony after GST, this notarized parent chain of length $\ell_2 - 1$ must have been observed by all honest replicas by the beginning of epoch $e + 1$. Therefore, $L_{e+1}$ must propose a block at length at least $\ell_2$. Thus we have reached a contradiction.
    
    Since by the beginning of epoch $e + 3$, no honest replica has signed any $F \neq B$ at length $\ell_2$, at this time there cannot be a notarization for any $F \neq B$ at length $\ell_2$. Moreover, $L_{e+2}$’s proposal and the notarized parent chain that triggered the proposal will be observed by all honest replicas at the beginning of $\Delta$ time into epoch $e + 2$. Therefore all honest replicas will vote for $B$ by $\Delta$ time into epoch $e + 2$ (since $B$'s parent is indeed one of the longest notarized chains seen by any replica). Thus by the beginning of epoch $e + 3$, all honest replicas will have seen a notarization for $B$. Thus, no honest replica will ever sign any conflicting  $F \neq B$ at length $\ell_2$ after the start of epoch $e + 3$ either; and any conflicting  $F \neq B$ at length $\ell_2$ cannot ever gain notarization.
\end{proof}

\begin{proof}[Proof of \cref{lem:liveness-intermediate}]
    Due to \cref{lem:three-honest-leaders}~(1), client $k$ observes the blocks proposed by $L_{e+2}, ..., L_{e+5}$, henceforth denoted $B_2, ..., B_5$, to have received $n - \tL_{k} \geq q_{k}$ votes. 

    Further, the blocks $B_2, B_3, B_4, B_5$ must form a chain. This is because due to \cref{lem:three-honest-leaders}~(2), the leader of epoch $e+3$ observed the chain ending in $B_2$ as notarized by the beginning of epoch $e+3$ and did not observe any other notarized chains at the same length as $B_2$. Therefore, the honest leader of epoch $e+3$ must propose block $B_3$ with the chain ending in $B_2$ as the parent chain. The same argument holds for blocks $B_3,B_4,B_5$. Thus, client $k$, based on its confirmation rule, confirms the block $B_3$ by the beginning of epoch $e+5$.
\end{proof}

%% file: figures/alg-streamlet.tex
\begin{algorithm}[tb]
    \caption{Streamlet protocol $\PI$ \cite{streamlet} (\cf \cref{sec:optflex-streamlet-recap})}
    \label{alg:streamlet}
    \begin{algorithmic}[1]
        \footnotesize
        \LineComment{Replica-side logic $\PiR$}
        \smallskip
        \On{\Call{init}{\null}}
            \State $\CB, \CV \gets \{ B_0 \}, \{ \}$
                \Comment{Background task: receive blocks and votes into $\CB$ and $\CV$, respectively, subject to the canonical validation: retain only messages with valid signatures; retain only blocks produced by respective epoch leader; add messages only once hash pointers can be resolved in $\CB$; add messages only when their epoch has come}
                \label{loc:streamlet-housekeeping}
        \EndOn
        \smallskip
        \For{each epoch $e=1,2,3,...$}
            \LineComment{\textbf{Propose} (done by epoch leader at the start of the epoch)}
            \State $B' \gets$ tip of any one longest notarized chain in $(\CB, \CV)$
            \State $h \gets \operatorname{Hash}(B')$
            \State $\txs \gets$ transactions not present in chain of $B'$
            \State Sign and broadcast block $(h, e, \txs)$
            \smallskip
            \LineComment{\textbf{Vote} (done by all replicas \emph{once} during the epoch)}
            \State $B \gets$ first block from epoch $e$ in $\CB$ signed by epoch leader
            \State $B' \gets$ parent block of $B$ in $\CB$
            \If{$B'$ is tip of any longest notarized chain in $(\CB, \CV)$}
                \State $h \gets \operatorname{Hash}(B)$
                \State Sign and broadcast vote $h$
            \EndIf
        \EndFor
        \medskip
        \LineComment{Client-side confirmation rule $\PiC$}
        \smallskip
        \On{\Call{init}{\null}}
            \State $\CB, \CV \gets \{ B_0 \}, \{ \}$
                \Comment{Background task: receive blocks and votes into $\CB$ and $\CV$, respectively, subject to the canonical validation: retain only messages with valid signatures; retain only blocks produced by respective epoch leader; add messages only once hash pointers can be resolved in $\CB$; add messages only when their epoch has come}
        \EndOn
        \smallskip
        \LineComment{\textbf{Confirmation}}
        \If{$(\CB, \CV)$ contains a notarized chain with three adjacent blocks $A, B, C$ from consecutive epochs}
            \State Choose $A, B, C$ as such blocks with maximum height
            \State $\LOG{}{} \gets$ sequence of transactions as ordered in chain of $B$
        \EndIf
    \end{algorithmic}
\end{algorithm}

%% file: A02_strongflex_converse.tex
\section{Impossibility Result for Strongly-Consistent Flexible Consensus}
\label{sec:strongflex-converse}

In this section, we prove the impossibility result referred to in \cref{sec:discussion-strongflex}: if two clients $k,k'$ choose optimal resilience pairs, that is, $2\tL_{k} + \tS_{k} = n-1$ (and similarly for $k')$, then consistency between the clients $k$ and $k'$ is impossible if $f > (\tS_{k} + \tS_{k'})/2$.
Observe that due to the optimality of the resilience pairs, this is equivalent to showing that consistency is impossible if $f > (n-1-2\tL_{k})/2 + (n-1-2\tL_{k'})/2$, \ie, for $f \geq n - \tL_{k} - \tL_{k'}$. We prove this in \cref{thm:strong-flexible-converse}.
This proof is a generalization of the
partially-synchronous resilience trade-off ($2\tL_{k} + \tS_{k'} < n$) proven in~\cite{pbft,aav1}.

\begin{theorem}
    \label{thm:strong-flexible-converse}
    In a partially synchronous network,
    there is no consensus protocol
    in which 
    for all clients $k,k'$ with $0 \leq \tL_{k},\tL_{k'} \leq n$:
    \begin{itemize}
        \item
              \textbf{Liveness:}
              For every $\tx$
              input to all honest replicas,
              eventually,
              for all clients $k$ with $f \leq \tL_k$,
              $\tx \in \LOG{k}{}$.

        \item
              \textbf{Safety:}
              For all clients $k, k'$ with 
              $f \leq n - \tL_{k} - \tL_{k'}$
              for all times $\tau, \tau'$,
              $\LOG{k}{\tau}$ and $\LOG{k'}{\tau'}$ are consistent.
    \end{itemize}
\end{theorem}
\begin{proof}%
    Suppose that there is a consensus protocol $\PI$ which provides the above mentioned liveness property. We will now prove that there is an execution of the protocol in which the safety property is violated.
    Suppose that there are two clients $k$ and $k'$.
    Let $P,Q,R$ be disjoint subsets of the $n$ replicas such that $|P|=\tL_{k}$, $|Q| = \tL_{k'}$, and $|R| = n - \tL_{k} - \tL_{k'}$.
    Consider the following three worlds:

    World 1: A high-entropy transaction $\mathsf{tx}_1$ is sent to all replicas. No other transactions are sent. Replicas in the subsets $P$ and $R$ are honest. Replicas in the subset $Q$ are adversary and do not communicate with the honest replicas and clients $k,k'$. Since $f \leq \tL_{k'}$ in this world, due to liveness, client $k'$ confirms $\mathsf{tx}_1$, \ie, $\mathsf{tx} \in \LOG{k'}{\tau'}$ at some time $\tau'$.

    World 2: A high-entropy transaction $\mathsf{tx}_2 \neq \mathsf{tx}_1$ is sent to all replicas. No other transactions are sent. Replicas in the subsets $Q$ and $R$ are honest. Replicas in the subset $P$ are adversary and do not communicate with the honest replicas and clients $k,k'$. Since $f \leq \tL_{k}$ in this world, due to liveness, client $k$ confirms $\mathsf{tx}_2$, \ie, $\mathsf{tx}_2 \in \LOG{k}{\tau}$ at some time $\tau$.

    World 3: Transaction $\mathsf{tx}_1$ is sent to replicas in $P$ and $\mathsf{tx}_2$ is sent to replicas in $Q$ and both transactions are sent to replicas in $R$. Replicas in the subsets $P$ and $Q$ are honest. Replicas in $R$ are adversary. The adversary chooses $\GST = \max\{\tau,\tau'\}$ (recall from \cref{sec:modprob} that $\GST$ is chosen by the adversary and unknown to honest replicas) and until $\GST$, replicas in $P$ and $Q$ cannot communicate with each other. The adversary replicas in $R$ perform a `split-brain' attack. One brain interacts with $P$ and the client $k'$ as if it only received input $\mathsf{tx}_1$ and this brain does not communicate with replicas in $Q$ and the client $k$. The other brain interacts with $Q$ and the client $k$ as if it only received input $\mathsf{tx}_2$ and this brain does not communicate with replicas in $P$ and the client $k$.

    For client $k'$, worlds 1 and 3 are indistinguishable, so $\mathsf{tx}_1 \in \LOG{k'}{\tau'}$ in world 3.
    For client $k$, worlds 2 and 3 are indistinguishable, so $\mathsf{tx}_2 \in \LOG{k}{\tau}$ in world 3.
    Since $\mathsf{tx}_1$ is high-entropy, $\mathsf{tx}_1 \not\in \LOG{k}{\tau}$ because no replicas saw $\mathsf{tx}_1$ in world 2. Similarly, $\mathsf{tx}_2 \not\in \LOG{k'}{\tau'}$. Thus, $\LOG{k}{\tau}$ and $\LOG{k'}{\tau'}$ are inconsistent.
    Thus, in world 3, where $f \leq n - \tL_{k} - \tL_{k'}$, there is a safety violation for clients $k$ and $k'$. This shows that there cannot be any protocol with the above safety and liveness properties.
\end{proof}